\documentclass[british]{aa}
\usepackage[T1]{fontenc}
\usepackage{CJKutf8}
\usepackage{color}
\usepackage{babel}
\usepackage{url}
\usepackage{amstext}
\usepackage{amssymb}
\usepackage{graphicx}
\usepackage{txfonts}
\usepackage[unicode=true,
 bookmarks=true,bookmarksnumbered=false,bookmarksopen=false,
 breaklinks=true,pdfborder={0 0 0},pdfborderstyle={},backref=false,colorlinks=true]
 {hyperref}
\hypersetup{
 linkcolor=magenta, urlcolor=blue, citecolor=blue}

\makeatletter
\providecommand{\tabularnewline}{\\}
\newcommand{\lyxdot}{.}

\renewcommand*\aa@pageof{, page \thepage{} of \pageref*{LastPage}}

\DeclareUnicodeCharacter{00B0}{$\degr$}
\DeclareUnicodeCharacter{00D7}{$\times$}
\DeclareUnicodeCharacter{03B1}{$\alpha$}
\DeclareUnicodeCharacter{03B2}{$\beta$}
\DeclareUnicodeCharacter{03BC}{$\mu$}
\DeclareUnicodeCharacter{2192}{$\rightarrow$}

\makeatother

\begin{document}
\begin{CJK*}{UTF8}{gbsn}
\title{ALMA imaging of the cold molecular and dusty disk in the type 2 active
nucleus of the Circinus galaxy}
\author{Konrad R. W. Tristram\inst{1}, C. M. Violette Impellizzeri\inst{2},
Zhi-Yu Zhang (张智昱)\inst{3,4}, Eric Villard\inst{5}, Christian Henkel\inst{6,7,8},
Serena Viti\inst{2}, Leonard Burtscher\inst{2}, Françoise Combes\inst{9},
Santiago García-Burillo\inst{10}, Sergio Martín\inst{1,11}, Klaus
Meisenheimer\inst{12}, Paul P. van der Werf\inst{2}}
\institute{European Southern Observatory, Alonso de Córdova 3107, Vitacura, Santiago,
Chile\\
\email{konrad.tristram@eso.org} \and Leiden Observatory, Leiden
University, PO Box 9513, 2300 RA Leiden, The Netherlands \and School
of Astronomy and Space Science, Nanjing University, Nanjing 210093,
PR China \and Laboratory of Modern Astronomy and Astrophysics (Nanjing
University), Ministry of Education, Nanjing 210093, PR China \and
European Southern Observatory, Karl-Schwarzschild-Straße 2, 85748
Garching bei München, Germany \and Max-Planck-Institut für Radioastronomie,
Auf dem Hügel 69, 53121 Bonn, Germany \and Dept. of Astronomy, Faculty
of Science, King Abdulaziz University, P.O. Box 80203, Jeddah 21589,
Saudi Arabia \and Xinjiang Astronomical Observatory, Chinese Academy
of Sciences, Urumqi 830011, China \and Observatoire de Paris, LERMA,
Collège de France, CNRS, PSL University, Sorbonne University, 75014
Paris, France \and Observatorio de Madrid, OAN-IGN, Alfonso XII,
3, 28014-Madrid, Spain\and Joint ALMA Observatory, Alonso de Córdova
3107, Vitacura, Santiago, Chile\and Max-Planck-Insitut für Astronomie,
Königstuhl 17, 69117 Heidelberg, Germany}
\date{Received 14 March 2022 / Accepted 25 May 2022}
\titlerunning{ALMA imaging of the Circinus nucleus}
\authorrunning{K. R. W. Tristram et al.}
\abstract{The central engines of many active galactic nuclei (AGNs) are thought
to be surrounded by warm molecular and dusty material in an axisymmetric
geometry, thus explaining part of the observational diversity of active
nuclei.}  {We aim to shed light on the physical properties and
kinematics of the molecular material in the nucleus of one of the
closest type 2 active galaxies.}  {To this end, we obtained high
angular resolution Atacama Large Millimeter/submillimeter Array (ALMA)
observations of the nucleus of the Circinus galaxy. The observations
map the emission at $350\,\text{GHz}$ and $690\,\text{GHz}$ with
spatial resolutions of $\sim3.8\,\mathrm{pc}$ and $\sim2.2\,\mathrm{pc}$,
respectively.}  {The continuum emission traces cold ($T\lesssim100\,\mathrm{K}$)
dust in a circumnuclear disk with spiral arms on scales of $25\,\mathrm{pc}$,
plus a marginally resolved nuclear emission peak. The latter is not
extended in polar direction as claimed based on earlier ALMA observations.
A significant amount (of the order of $40\%$) of the $350\,\text{GHz}$
emission is not related to dust, but most likely free-free emission
instead. We detect CO(3$-$2) and CO(6$-$5) as well as $\mathrm{HCO}^{+}$(4$-$3),
HCN(4$-$3), and CS(4$-$3). The CO emission is extended, showing
a spiral pattern, similar to the extended dust emission. Towards the
nucleus, CO is excited to higher transitions and its emission is self-absorbed,
leading to an apparent hole in the CO(3$-$2) but not the CO(6$-$5)
emission. On the other hand, the high gas density tracers $\textrm{HCO}^{+}$,
HCN, and CS show a strong, yet unresolved (${\lesssim}4\,\mathrm{pc}$)
concentration of the emission at the nucleus, pointing at a very small
`torus'. The kinematics are dominated by rotation and point at a
geometrically thin disk down to the resolution limit of our observations.
In contrast to several other AGNs, no HCN enhancement is found towards
the nucleus.}   {The Circinus nucleus is therefore composed of
at least two distinct components: (1) an optically thin, warm outflow
of ionised gas containing clouds of dust which are responsible for
the polar mid-infrared emission, but not seen at submillimetre wavelengths;
and (2) a cold molecular and dusty disk, traced by submillimetre continuum
and line emission. The latter is responsible for the bulk of the obscuration
of the nucleus. These findings support the most recent radiative transfer
calculations of the obscuring structures in AGNs, which find a similar
two-component structure.}

\maketitle
\end{CJK*}

\section{Introduction\label{sec:introduction}}

In an active galactic nucleus (AGN), the luminosity of at least the
inner region of a galaxy (if not the entire galaxy for a powerful
quasar) is dominated by the non-stellar continuum emission produced
by matter accretion onto the central supermassive black hole. The
central accretion disk and broad line region are surrounded by gaseous
and dusty material on parsec scales, which absorbs and reprocesses
the emission from the accretion disk. This dense material is believed
to be distributed in a toroidal distribution, the so-called dusty
or molecular torus. The toroidal structure allows for free lines of
sight towards the centre when observed face-on (type 1 source), or
it obscures the centre when seen edge-on (type 2 source). The axisymmetric
structure of the obscuring material is the basis of the unified scheme
of AGNs, which has been successful in explaining the main observational
characteristics of AGNs \citep{Antonucci1993,Urry1995,Netzer2015}.

 The covering factor of the nuclear obscuring material seems to be
relatively high, $>50\%$, even if its dependency on luminosity and
redshift is under discussion and the obtained values may be biased
by specific methods or the underlying AGNs samples \citep[for a review, see e.g.][]{RamosAlmeida2017}.
The large covering factor implies a geometrically thick torus with
$h/r\sim1$ and raises the question about the physical mechanism keeping
the torus geometrically thick and preventing it from collapsing to
a thin disk. Several mechanisms, such as radiation pressure \citep[e.g.][]{Pier1992a,Krolik2007,Chan2016},
starburst disks providing energy injection from supernovae \citep[e.g.][]{Wada2002,Schartmann2009,Wada2009},
magnetic-driven winds \citep[e.g.][]{Emmering1992,Koenigl1994,Begelman2017a},
or warped disks \citep{Phinney1989,Lawrence2007}, have been proposed
as the dominant physical mechanism of keeping the torus geometrically
thick. Interferometric observations in the mid-infrared of the thermal
emission from the centrally heated dust have revealed a predominance
of polar-elongated dust structures \citep[e.g.][]{Lopez-Gonzaga2016,Leftley2018}
and also deep single-dish observations reveal that the mid-infrared
emission is extended in polar direction out to $100\,\mathrm{pc}$
scales, rather than in the equatorial direction \citep[e.g.][]{Packham2005,Asmus2014,Asmus2016}.
This has led to new models explaining the torus as part of a wind
\citep{Hoenig2017} or a `radiation-driven fountain' \citep{Wada2016}.
The actual kinematics of the torus on parsec scales have, however,
remained essentially inaccessible to observations apart from the detection
of water maser disks and outflows in a few objects with a close to
edge-on orientation \citep[e.g. ][]{Kuo2011,Zhao2018}.

Only with the advent of the Atacama Large Millimeter/submillimeter
Array (ALMA) has it become possible to investigate the kinematics
and physical state of the material using emission lines in the submillimetre.
Not surprisingly, the best studied source is the well-known Seyfert
2 nucleus of \object{NGC~1068}. A wealth of observations have, however,
led to contradicting interpretations: Using CO(6$-$5) molecular line
emission and the $432\,\mathrm{\text{\ensuremath{\mu}m}}$ continuum,
\citet{GarciaBurillo2016b} detected the dust emission, the molecular
gas, and the kinematics of a $\sim10\,\mathrm{pc}$ disk which is
interpreted as the submillimetre counterpart of the dusty torus. The
lopsided morphology and complex kinematics of this disk (including
apparent counter-rotation) are interpreted as a possible `signature
of the Papaloizou–Pringle instability, long predicted to likely drive
the dynamical evolution of active galactic nuclei tori'. Using their
own set of CO(6$-$5) observations, \citet{Gallimore2016} resolved
the low-velocity component into a $12\times7\,\mathrm{pc}$ structure;
the higher-velocity emission is found to be consistent with a bipolar
outflow perpendicular to the nuclear disk, which is interpreted as
`compelling evidence in support of the disk-wind scenario' for the
obscuring torus. \citet{Imanishi2016,Imanishi2018a} found the dense
gas tracers HCN(3$-$2) and $\mathrm{HCO}^{+}$(3$-$2) to be morphologically
and dynamically aligned in the east-west direction (i.e.\ in the
expected torus direction, perpendicular to the radio jet), which is
seen as a sign for these molecular lines being better probes of the
rotating dense molecular gas in the torus than CO(6$-$5). More recent
observations with improved spatial resolution revealed two apparently
nested, counter-rotating disks aligned perpendicular to the radio
jet: an inner ($<1.2\,\mathrm{pc}$) disk with Keplerian rotation
consistent to an $\mathrm{H}_{2}\mathrm{O}$ megamaser disk and an
outer ($>2\,\mathrm{pc}$) counter-rotating disk \citep{Impellizzeri2019,Imanishi2020}.
Broad, blue-shifted absorption is seen as a signature of outflowing
dense molecular gas. Finally, \citet{GarciaBurillo2019} confirm strong
departures from circular motions in the torus and the circumnuclear
disk using CO(2$-$1), CO(3$-$2), and $\mathrm{HCO}^{+}$(4$-$3);
these are, however, not interpreted as counter-rotation but as signs
for a wide-angle wind reflecting AGN feedback. In summary, studying
this one emblematic object has not (yet) lead to a consensus of what
constitutes the torus and how it works. 

Apart from \object{NGC~1068}, several other nearby AGNs have been
studied on scales of the molecular and dusty torus, that is on scales
of $\sim10\,\mathrm{pc}$ and below. Among them are \object{NGC~613}
\citep{Audibert2019}, \object{NGC~1097} \citep{Izumi2017}, \object{NGC~1377}
\citep{Aalto2017}, \object{NGC~1808} \citep{Audibert2021}, \object{NGC~3772}
\citep{AlonsoHerrero2019}, and \object{NGC~5643} \citep{AlonsoHerrero2018},
as well as the NUGA (`NUclei of GAlaxies') sample of seven nuclei
\citep{Combes2019} and the GATOS (`Galaxy Activity, Torus, and Outflow
Survey') core sample of ten nuclei \citep{GarciaBurillo2021}. Compact,
disk like structures with sizes between $\sim5$ and $\sim50\,\mathrm{pc}$
are detected in many of these sources in the submillimetre continuum
and in high density gas tracers such as HCN, $\mathrm{HCO}^{+}$ and
CS. On the other hand, CO, especially CO(3$-$2) and CO(2$-$1), often
displays a deficit of emission – a `hole' – at the nucleus. This
deficit has been explained by the chemistry as a result of an X-ray
Dominated Region (XDR) caused by the AGN \citep{Izumi2020}, or as
a true nuclear gas deficit caused by AGN feedback, for example in
the form of winds \citep{GarciaBurillo2021}. It has also been recognised
that the continuum emission detected by ALMA in many cases has significant
contribution or is dominated by non-dust emission, that is free-free
or synchrotron emission. For a significant number of sources the kinematics
are clearly dominated by rotation. In addition, many sources display
molecular outflows on scales of a few (tens of) pc. A so-called submillimetre
HCN enhancement seems to be even clearer at the higher angular resolutions
provided by ALMA \citep{Izumi2016}. So despite the many faces of
the torus (or more generally the obscuration) in individual AGNs,
some of the properties such as the CO deficit, the HCN enhancement
or an outflow seem to be more common features and hence point at the
dominant physics in the tori. However it remains unclear if these
structures are directly the equivalent of the tori in the unified
scheme in the sense that they are responsible for the viewing angle
dependent obscuration.

In this paper, we present ALMA band 7 and 9 continuum and molecular
line observations of the nucleus in the \object{Circinus galaxy}.
The goal of these observations is to study the kinematics and physical
state of the dense obscuring material in this active nucleus.

At a distance of only $4\,\mathrm{Mpc}$ ($1\,\mathrm{arcsec}\sim20\,\mathrm{pc}$),
the \object{Circinus galaxy} is a prime target for the study of AGN
physics and the near-nuclear molecular material. The galaxy is regarded
as a prototypical type 2 galaxy (torus edge-on) due to its ionisation
cone \citep[e.g.][]{Marconi1994,Veilleux1997,Wilson2000,Mingozzi2019},
large-scale molecular outflow \citep{Curran1999}, broad emission
lines in polarised light \citep[e.g.][]{Oliva1998,RamosAlmeida2016},
and Compton-thick nuclear X-ray emission \citep[e.g. ][]{Arevalo2014,Uematsu2021}.
$\mathrm{H}_{2}\mathrm{O}$ masers at $22\,\mathrm{GHz}$ trace a
thin, warped disk with a radius of $r=0.4\,\mathrm{pc}$ as well as
a wide-angle outflow \citep{Greenhill2003a}. While the maser disk
matches a dense, hot ($T\sim300\,\mathrm{K}$) dust disk, the majority
of the thermal dust emission in the mid-infrared is extended in the
polar direction on scales of $\sim2.0\,\mathrm{pc}$ as revealed by
mid-infrared interferometry \citep{Tristram2007e,Tristram2014,Isbell2022}.
In fact the polar dust emission is visible out to $\sim20\,\mathrm{pc}$
\citep[e.g. ][]{Packham2005} and can be explained by a dusty hollow
cone \citep{Stalevski2017,Stalevski2019} or a magnetocentrifugal
wind \citep{Vollmer2018}. Atacama Pathfinder EXperiment (APEX) measurements
of the CO transitions up to CO(4$-$3) reveal higher excitation as
well as more violent kinematics towards the nucleus \citep{Zhang2014}.

The nucleus of the Circinus galaxy has been observed before with ALMA.
\citet{Hagiwara2013} detected the $321\,\mathrm{GHz}$ $\mathrm{H}_{2}\mathrm{O}$
maser transition at the location of the nucleus. The maser features
remain spatially unresolved for their synthesised beam of $\sim15\,\mathrm{pc}$,
but comparing the velocities of both the $22$ and $321\,\mathrm{GHz}$
maser transitions, it is concluded that both originate from a similar
region. The $321\,\mathrm{GHz}$ continuum has a total flux density
of $55\,\mathrm{mJy}$ and shows a bright core and an arm-like substructure
extending $1.5\,\mathrm{arcsec}$ to the north-east. Most recently,
\citet{Izumi2018} presented ALMA maps of the continuum as well as
the CO(3$-$2) and {[}\ion{C}{I}{]}(1$-$0) lines, tracing the molecular
and atomic phases of the obscuring structures. Both the continuum
and line emission show a $74\times34\,\mathrm{pc}$ circumnuclear
disk and spiral arms; the disk is found responsible for a significant
contribution to the Compton thickness of this nucleus, and the line
ratio found may be due to XDR chemistry at the location of the nucleus.
Decomposing the line velocity fields into rotation and dispersion,
the diffuse atomic gas is found to be more spatially extended along
the vertical direction of the disk than the dense molecular gas. This
multi-phase nature of the torus is seen as support for the validity
of the radiation-driven fountain model.

This paper is structured as follows: our observations and the data
reduction are described in Sect.~\ref{sec:observations}; the results
from both continuum and line emission are presented in Sect.~\ref{sec:results},
followed by an analysis and discussion in Sect.~\ref{sec:discussion}.
Our conclusions are summarised in Sect.~\ref{sec:conclusions}.

\section{Observations and data processing\label{sec:observations}}

\subsection{Observations\label{subsec:observations}}

We obtained ALMA band 7 and 9 observations of the nucleus of the Circinus
galaxy in programme 2012.1.00479.S (PI: K.\ Tristram), which were
mainly targeted at the CO(3$-$2) and CO(6$-$5) transitions, respectively.
For a summary of the technical setup and observational parameters,
see Table~\ref{tab:obslog}.
\begin{table}
\caption{Log of observations and observing parameters.\label{tab:obslog}}

\centering

\begin{tabular}{lrr}
\hline 
\hline ALMA band & 7 & 9\tabularnewline
\hline 
date of observations & 2015-07-16 & 2015-07-26\tabularnewline
number of antennas & 39 & 41\tabularnewline
proj. baseline lengths & $12\ldots1564\,\mathrm{m}$ & $12\ldots1569\,\mathrm{m}$\tabularnewline
integration time & $16\,\mathrm{min}$ & $15\,\mathrm{min}$\tabularnewline
pointing centre (J2000) & 14:13:09.9060 & 14:13:09.9060\tabularnewline
 & -65:20:20.4684 & -65:20:20.4684\tabularnewline
window width & $1.875\,\mathrm{GHz}$ & $1.875\,\mathrm{GHz}$\tabularnewline
number of channels & 960 & 1920 \tabularnewline
channel bandwidth & $1.953\,\mathrm{MHz}$ & $0.976\,\mathrm{MHz}$\tabularnewline
 & $\sim1.7\,\mathrm{km}\,\mathrm{s}^{-1}$ & $\sim0.42\,\mathrm{km}\,\mathrm{s}^{-1}$\tabularnewline
spectral window 0 & $344.838\,\mathrm{GHz}$ & $690.431\,\mathrm{GHz}$\tabularnewline
spectral window 1 & $342.880\,\mathrm{GHz}$ & $707.809\,\mathrm{GHz}$\tabularnewline
spectral window 2 & $356.571\,\mathrm{GHz}$ & $692.821\,\mathrm{GHz}{}^{a}$\tabularnewline
spectral window 3 & $354.880\,\mathrm{GHz}$ & $704.613\,\mathrm{GHz}{}^{a}$\tabularnewline
primary beam size & $18\,\mathrm{arcsec}$ & $9\,\mathrm{arcsec}$\tabularnewline
bandpass calibrator & J1427-4206 & J1256-0547\tabularnewline
amplitude / flux calib. & J1613-586 & J1613-586$^{b}$\tabularnewline
phase calibrator & J1424-6807 & J1147-6753\tabularnewline
delay calibrator & J1337-6509 & J1617-5848$^{b}$\tabularnewline
\hline 
\end{tabular}

\tablefoot{$^{(a)}$ Spectral windows 2 and 3 in band 9 were observed in continuum
mode, that is with only 128 channels and a channel spacing of $15.6\,\mathrm{MHz}$
for a total window width of $2\,\mathrm{GHz}$.\\
$^{(b)}$ J1617-5848 and J1613-586 are two different designations
for the same source.}
\end{table}

The observations in band 7 were carried out on 2015 July 16 with 39
antennas resulting in projected baseline lengths between $12$ and
$1564\,\mathrm{m}$. The four $1.875\,\mathrm{GHz}$ wide spectral
windows have 960 channels each and thus yield a channel spacing of
$\sim1.7\,\mathrm{km}\,\mathrm{s}^{-1}$. The spectral windows were
arranged such as to cover the CO(3$-$2) ($\nu_{\mathrm{rest}}=345.796\,\mathrm{GHz}$),
$\mathrm{HCO}^{+}$(4$-$3) ($\nu_{\mathrm{rest}}=356.734\,\mathrm{GHz}$),
HCN(4$-$3) ($\nu_{\mathrm{rest}}=354.505\,\mathrm{GHz}$) and CS(6$-$5)
($\nu_{\mathrm{rest}}=342.883\,\mathrm{GHz}$) lines as much as possible.
In order to accommodate the full expected velocity range of the CO(3$-$2)
line, the HCN(4$-$3) line was positioned at the edge of its spectral
window and only the blue-shifted part of the line up to the systemic
velocity could be covered.

\begin{figure*}
\begin{centering}
{\Large{}\includegraphics[width=0.49\textwidth]{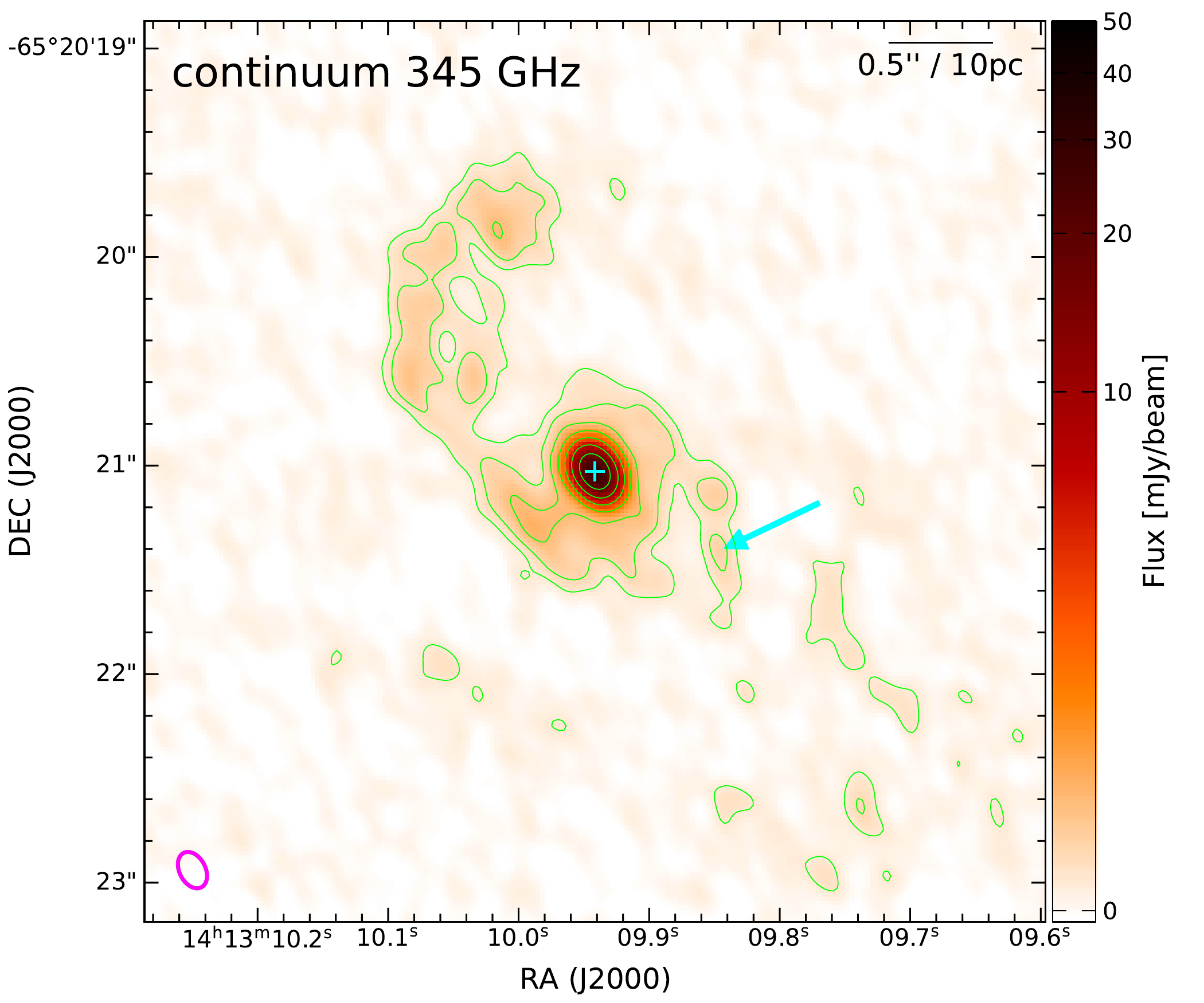}\hfill{}\includegraphics[width=0.49\textwidth]{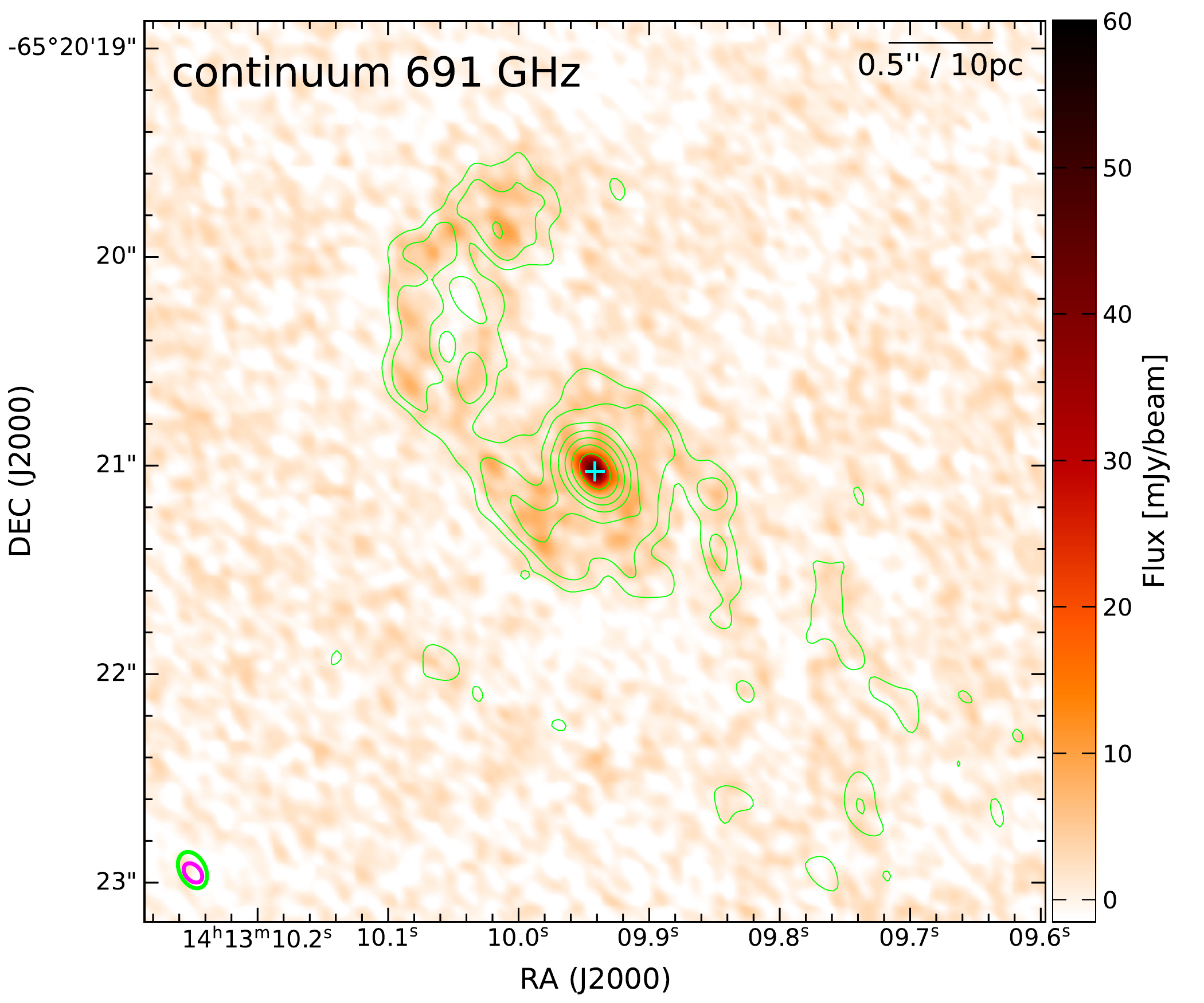}}{\Large\par}
\par\end{centering}
\caption{Continuum emission in band 7 ($\nu_{\mathrm{rest}}=345\,\mathrm{GHz}$,
$\lambda_{\mathrm{rest}}=868\,\mathrm{\mu m}$, left panel) and band
9 ($\nu_{\mathrm{rest}}=691\,\mathrm{GHz}$, $\lambda_{\mathrm{rest}}=434\,\mathrm{\mu m}$,
right panel). Contours in both panels are for the $345\,\mathrm{GHz}$
continuum at $[1,2,4,8,16,32,64,128]\times0.25\,\mathrm{mJy}/\mathrm{beam}$.
The synthesised beam is indicated in green for the contour data and
magenta for the respective image. The cyan arrow in the left panel
marks the ridge-like structure discussed in the text. \label{fig:continuum-map}}
\end{figure*}

The band 9 observations took place on 2015 July 26 using 41 antennas
resulting in very similar projected baseline lengths as for band 7,
namely between $12$ and $1569\,\mathrm{m}$. Two $1.875\,\mathrm{GHz}$
wide spectral windows with 1920 channels each and a velocity resolution
of $\sim0.42\,\mathrm{km}\,\mathrm{s}^{-1}$ were centred at $690.431$
and $707.809\,\mathrm{GHz}$ in order to cover the CO(6$-$5) ($\nu_{\mathrm{rest}}=691.473\,\mathrm{GHz}$)
and HCN(4$-$3) ($\nu_{\mathrm{rest}}=708.877\,\mathrm{GHz}$) emission
lines. The other two 2 GHz wide windows were centred at $692.821$
and $704.613\,\mathrm{GHz}$ respectively, with only 128 channels
each to trace the continuum, because they were not expected to contain
any line emission.

\subsection{Data reduction\label{subsec:datareduction}}

The data for both bands were calibrated and reduced manually with
the ALMA script generator and the Common Astronomy Software Applications
(CASA)\footnote{\url{https://casa.nrao.edu/}}. After flagging the
data, the raw phases and amplitudes were corrected by the water vapour
radiometer measurements and the system temperature ($T_{\mathrm{sys}}$),
respectively, and the positions of the antennas were updated to their
latest measurements. For band 7, phases and amplitudes were calibrated
as usual using the gain calibrator independently per spectral window.
For band 9, the signal to noise of the solutions per spectral window
is so low, that the signal from all spectral windows had to be combined
after determining the phase offsets between the windows. For the amplitude,
it is trickier to combine spectral windows because they originate
from different receiver base bands and therefore have different sensitivities.
We thus transferred the individual solutions from the band pass calibrator,
which was bright enough to provide solutions with a high signal to
noise ratio. The limitation of this approach is that the bandpass
calibrator is observed at a different elevation than the science target
and only once, at the beginning of the observation. However, the sensitivity
of the ALMA receivers change very little with elevation, and the frequent
$T_{\mathrm{sys}}$ measurements effectively compensate for differences
in opacity between the bandpass calibrator and science target. Thus
the overall error bar on the band 9 flux is probably within the ALMA
specification ($20\%$), but out of precaution, we consider a higher
absolute flux uncertainty of $30\%$. For the band 7 data we adopt
a $15\%$ uncertainty.

All imaging was carried out using the \texttt{clean} task in CASA
with Briggs weighting and a robust parameter of 0.5. For the continuum
imaging, all regions contaminated by emission lines were flagged previous
to the imaging; for the emission lines, the continuum was subtracted
from the data directly in the UV plane and then imaged. Fluxes were
not corrected for the primary beam ($18\,\mathrm{arcsec}$ for band
7, and $9\,\mathrm{arcsec}$ for band 9, respectively) as the regions
we are interested in are much smaller. 

For band 7, the continuum signal was strong enough to carry out two
iterations of phase-only self calibration before the result converges.
An attempt to also carry out amplitude self calibration did not lead
to any improvement and was therefore not used. The result of the phase-only
self calibration was applied to the main data set before continuum
subtraction and cleaning the line emission. For CO(3$-$2) and $\mathrm{HCO}^{+}$(4$-$3),
the data were imaged between $-300$ and $+300\,\mathrm{km}\,\mathrm{s}^{-1}$,
in $5\,\mathrm{km}\,\mathrm{s}^{-1}$ steps. For CS(4$-$3), which
is much fainter, the data were imaged between $-300$ and $+300\,\mathrm{km}\,\mathrm{s}^{-1}$
in $50\,\mathrm{km}\,\mathrm{s}^{-1}$ bins, the best compromise between
the low signal to noise ratio of the line and the possibility to separate
at least a few velocity bins. Finally, because the HCN(4$-$3) line
is located at the edge of the spectral window, it could only be imaged
in a range from $-300$ to $+15\,\mathrm{km}\,\mathrm{s}^{-1}$ in
$5\,\mathrm{km}\,\mathrm{s}^{-1}$ steps. All lines were directly
imaged centred on the systemic velocity of $434\,\mathrm{km}\,\mathrm{s}^{-1}$
($z=0.00145$).

The band 9 data turn out to be of much lower signal to noise ratio.
We nevertheless were able to apply one iteration of phase self calibration
on the continuum and apply it to the line data. The only emission
line detected is the CO(6$-$5) transition, which was cleaned between
$-300$ and $+300\,\mathrm{km}\,\mathrm{s}^{-1}$ in $20\,\mathrm{km}\,\mathrm{s}^{-1}$
steps.

Both the band 9 continuum and CO(6$-$5) line emission are shifted
$70\,\mathrm{mas}$ to the north-east of the peak in the band 7 continuum.
Because also the velocity map of CO(6$-$5) is offset by the same
amount, this cannot be a true offset and we therefore shifted the
band 9 data so that the continuum peaks of both bands are aligned.
This also gives a very good, although not perfect, match of the CO(3$-$2)
and CO(6$-$5) velocity maps. We could not identify an obvious phase
error like for instance \citet{Izumi2020}, who observed a similar
offset between bands in their ALMA data. However, the observed offset
is roughly aligned with the parallactic angle. Therefore, a possible
reason for the offset of the band 9 data could be a bad phase transfer
due to atmospheric refraction in combination with the low signal to
noise ratio. 

\section{Results\label{sec:results}}

In this section we present our observational results for the continuum
emission, the line emission as well as the kinematics.

\subsection{Continuum emission and astrometry}

\begin{table*}
\caption{Properties of the reconstructed maps as well as of the nuclear emission.
The errors indicated include both the statistical errors from the
fit as well as $\sim10\%$ flux calibration uncertainties. Units of
the fluxes are in mJy for the continuum and in $\mathrm{Jy}\,\mathrm{km}\,\mathrm{s}^{-1}$
for emission lines. \label{tab:nuclear-properties}}

\begin{centering}
\begin{tabular}{c|c|c|c|c|c|c|c|c}
\hline 
{\small{}line} & {\small{}rest } & \multicolumn{2}{c|}{{\small{}beam}} & \multicolumn{3}{c|}{{\small{}central peak, deconvolved}} & {\small{}integrated flux} & {\small{}integrated flux}\tabularnewline
{\small{}or} & {\small{}frequency} & {\small{}size} & {\small{}$\mathrm{PA}$} & {\small{}$\mathrm{FWHM}_{\mathrm{max}}$} & {\small{}$\mathrm{FWHM}_{\mathrm{min}}$} & {\small{}$\mathrm{PA}$} & {\small{}central peak} & {\small{}$1''$ aperture}\tabularnewline
{\small{}continuum} & {\small{}{[}GHz{]}} & {\small{}{[}$\mathrm{mas}\times\mathrm{mas}$ {]}} & {\small{}{[}$^{\circ}${]}} & {\small{}{[}mas{]}} & {\small{}{[}mas{]}} & {\small{}{[}$^{\circ}${]}} & {\small{}{[}mJy / $\mathrm{Jy}\,\mathrm{km}\,\mathrm{s}^{-1}${]}} & {\small{}{[}mJy / $\mathrm{Jy}\,\mathrm{km}\,\mathrm{s}^{-1}${]}}\tabularnewline
\hline 
\hline 
{\small{}B7 cont.} & {\small{}$345.4$} & {\small{}$190\times120$} & {\small{}$27$} & {\small{}$95\pm11$} & {\small{}$81\pm14$} & {\small{}$76\pm37$} & {\small{}$38\pm6$} & {\small{}$51\pm8$}\tabularnewline
{\small{}B9 cont.} & {\small{}$691.5$} & {\small{}$110\times70$} & {\small{}$42$} & {\small{}$160\pm20$} & {\small{}$109\pm12$} & {\small{}$45\pm15$} & {\small{}$142\pm43$} & {\small{}$383\pm115$}\tabularnewline
{\small{}CO(}3$-$2{\small{})} & {\small{}$345.8$} & {\small{}$190\times130$} & {\small{}$26$} & {\small{}–} & {\small{}–} & {\small{}–} & {\small{}–} & {\small{}$73\pm8$}\tabularnewline
{\small{}CO(}6$-$5{\small{})} & {\small{}$691.5$} & {\small{}$100\times70$} & {\small{}$34$} & {\small{}$170\pm18$} & {\small{}$94\pm7$} & {\small{}$42\pm5$} & {\small{}$17\pm5$} & {\small{}$470\pm141$}\tabularnewline
{\small{}CS}(4$-$3) & {\small{}$342.9$} & {\small{}$220\times160$} & {\small{}$29$} & {\small{}$216\pm62$} & {\small{}$102\pm81$} & {\small{}$126\pm39$} & {\small{}$0.95\pm0.17$} & {\small{}$1.14\pm0.18$}\tabularnewline
{\small{}$\textrm{HCO}^{+}$}(4$-$3) & {\small{}$356.7$} & {\small{}$180\times120$} & {\small{}$27$} & {\small{}$212\pm35$} & {\small{}$123\pm23$} & {\small{}$38\pm17$} & {\small{}$10.5\pm1.2$} & {\small{}$19\pm3$}\tabularnewline
{\small{}HCN}(4$-$3){\small{}$^{a}$} & {\small{}$354.5$} & {\small{}$190\times120$} & {\small{}$31$} & {\small{}$173\pm41$} & {\small{}$109\pm30$} & {\small{}$19\pm36$} & {\small{}$3.6\pm0.6$} & {\small{}$6\pm1$}\tabularnewline
\hline 
\end{tabular}
\par\end{centering}

\tablefoot{$^{(a)}$ Values only for the blue-shifted part of the line, see
Sect.~\ref{subsec:observations}.}
\end{table*}
The continuum maps for band 7 ($\nu_{\mathrm{rest}}=345\,\mathrm{GHz}$,
$\lambda_{\mathrm{rest}}=868\,\mathrm{\mu m}$, $\mathrm{FWHM}_{\mathrm{beam}}<0.19\,\mathrm{arcsec}$)
and band 9 ($\nu_{\mathrm{rest}}=691\,\mathrm{GHz}$, $\lambda_{\mathrm{rest}}=434\,\mathrm{\mu m}$,
$\mathrm{FWHM}_{\mathrm{beam}}<0.11\,\mathrm{arcsec}$) are presented
in Fig.\ \ref{fig:continuum-map}. Only the innermost $4\,\mathrm{arcsec}\times4\,\mathrm{arcsec}$
centred on the nucleus are shown. A central peak plus extended structure
out to a radius of $\sim2\,\mathrm{arcsec}$ ($40\,\mathrm{pc}$)
is detected.

The extended structure shows an S-shaped morphology, with a clear
spiral arm towards the north-east and a much fainter spiral structure
on the opposite side to the south-west which is only barely visible
in our band 7 data (but clearly visible in CO(3$-$2), see below).
Both spiral arms correspond to the circumnuclear disk described by
\citet{Izumi2018} and connect to their spiral arms 2 in the south
and 3 in the north. The brighter spiral arm to the north-east was
actually first seen as an extension in the $321\,\mathrm{GHz}$ continuum
map of \citet{Hagiwara2013}. There seems to be an enhancement of
the emission towards the outer, convex edge of the spiral. Towards
the south-west, at a distance of $\sim0.7\,\mathrm{arcsec}$ ($15\,\mathrm{pc}$)
from the nucleus, there is a \emph{ridge-like structure} extending
about $10\,\mathrm{pc}$ in north-south direction which will be discussed
in more detail in Sect.~\ref{subsec:results-kinematics}.

We are however mainly interested in the emission from the core which
we expect to be related to the dusty and molecular torus. We use \texttt{imfit}
in CASA to characterise this core; its main properties are summarised
in Table~\ref{tab:nuclear-properties}. The emission has very similar
apparent sizes, $206\,\mathrm{mas}\times153\,\mathrm{mas}$ and $234\,\mathrm{mas}\times173\,\mathrm{mas}$
at $345\,\mathrm{GHz}$ and $691\,\mathrm{GHz}$, respectively, despite
the difference in beam sizes by a factor of two. While at $345\,\mathrm{GHz}$,
the emission hence remains essentially unresolved (the size being
only marginally larger than the beam at this frequency), the nuclear
source is clearly resolved at $691\,\mathrm{GHz}$, with a deconvolved
size of $160\,\mathrm{mas}\times110\,\mathrm{mas}$ (i.e.\ $3.2\,\mathrm{pc}\times2.2\,\mathrm{pc}$).
We do not expect the emission to be more compact at lower frequencies
if originating from dust. Therefore the very compact, deconvolved
size of $<100\,\mathrm{mas}$ obtained with \texttt{imfit} for the
emission at $345\,\mathrm{GHz}$ may simply not be a very reliable
estimate, or it may point at an emission mechanism other than thermal
dust emission to play a significant role (see Sect.~\ref{subsec:discuss_continuum}).
The central peak appears elongated along the expected equatorial direction
of the torus as well as the major axis of the galaxy. However at $345\,\mathrm{GHz}$
this is mainly due to the elongation of the beam in this direction,
and even at $691\,\mathrm{GHz}$ the beam is most likely dominating
the observed elongation as is implied by the large uncertainty on
the position angle of the deconvolved emission components (see Table~\ref{tab:nuclear-properties}).
In conclusion, we find the nuclear emission to be very compact, possibly
elongated in the equatorial direction and with a size of $\sim3\,\mathrm{pc}$.
This is not much larger than the emission by the dust at $T\sim300\,\mathrm{K}$
traced by mid-infrared interferometry \citep{Tristram2014,Isbell2022}.

We measure fluxes of $38\pm6\,\mathrm{mJy}$ and $142\pm43\,\mathrm{mJy}$
for the central peak, at $345\,\mathrm{GHz}$ and $691\,\mathrm{GHz}$,
respectively. Additionally the peak is surrounded symmetrically by
extended emission with a diameter of $\sim1\,\mathrm{arcsec}$, corresponding
to $20\,\mathrm{pc}$. Including this emission by carrying out photometry
with an aperture of $1\,\mathrm{arcsec}$ diameter, the flux increases
by $34\%$ at $345\,\mathrm{GHz}$ and by more than a factor of two
at $691\,\mathrm{GHz}$.

The continuum peak at $345\,\mathrm{GHz}$ is located at $\mathrm{RA}=\text{14:13:09}.942$
and $\mathrm{DEC}=-\text{65:20:21}.026$ (J2000) with an uncertainty
of the fitted position of about $3\,\mathrm{mas}$. This uncertainty
is consistent with that expected following equation 10.7 in the ALMA
Technical Handbook. Comparing the locations of the amplitude and delay
calibrators associated to our observations, we find offsets of $3$
and $5\,\mathrm{mas}$, respectively, to the coordinates given in
SIMBAD. Both calibrators are further away from the phase calibrator
than the Circinus nucleus. Conservatively, this implies an uncertainty
in the position of the nucleus of $10\,\mathrm{mas}$ from the $345\,\mathrm{GHz}$
continuum map alone. As mentioned in Sect.~\ref{subsec:datareduction},
our $691\,\mathrm{GHz}$ data have an offset of $70\,\mathrm{mas}$
with respect to the $345\,\mathrm{GHz}$ data, most likely due to
the lower signal to noise ratio and a poor phase transfer at this
higher frequency. We therefore consider the position obtained at $345\,\mathrm{GHz}$
as a better estimate.

\citet{Greenhill2003a} specify the position of the $22\,\mathrm{GHz}$
maser disk\footnote{Actually, \citet{Greenhill2003a} give the position for the maser
detection at $565.2\,\mathrm{km}\,\mathrm{s}^{-1}$; the centre of
the maser disk (i.e.\ the central engine) is offset by $17.6\,\mathrm{mas}$
to the east and $13.1\,\mathrm{mas}$ to the north of this position.
This small offset is, however, much smaller than the $100\,\mathrm{mas}$
uncertainty in the absolute astrometry.} with $\mathrm{RA}=\text{14:13:09}.95\pm0.02$ and $\mathrm{DEC}=-\text{65:20:21}.2\pm0.1$.
This is about $180\,\mathrm{mas}$ to the south-east of our estimate
from the $345\,\mathrm{GHz}$ data, but within $2\sigma$ of the astrometric
uncertainties of the water masers position. An astrometric alignment
for the near-infrared emission carried out by \citet{Prieto2004}
results in a location of the infrared core at $\mathrm{RA}=\text{14:13:09}.96$
and $\mathrm{DEC}=-\text{65:20:19.91}$. Although $\sim1.1\,\mathrm{arcsec}$
to the north of our coordinates, this is in rough agreement with our
position considering the uncertainty of $900\,\mathrm{mas}$ cited
for the infrared position. Moreover, the infrared emission peak may
not be centred on the very nucleus due to obscuration effects playing
a role in the infrared. \citet{Izumi2018} and \citet{Hagiwara2013}
give coordinates for the continuum peak at $345\,\mathrm{GHz}$ and
$321\,\mathrm{GHz}$ that are located $\sim50\,\mathrm{mas}$ to the
east and $\sim200\,\mathrm{mas}$ to the north-east of our coordinate,
respectively. The reason for these relatively large differences in
the astrometry of the nucleus with ALMA remain unclear. The larger
offset to the position in \citet{Hagiwara2013} may be related to
the fact that those observations were carried out in cycle 0 with
only 18 antennas.

In conclusion, we estimate the uncertainty on the absolute astrometry
of the central engine in the Circinus galaxy to be of the order of
$100\,\mathrm{mas}$. This means we find a position of the core of
$\mathrm{RA}=\text{14:13:09}.942\pm0.016$ and $\mathrm{DEC}=-\text{65:20:21}.03\pm0.10$
from our data.

\begin{figure*}
\begin{centering}
{\Large{}\includegraphics[width=0.49\textwidth]{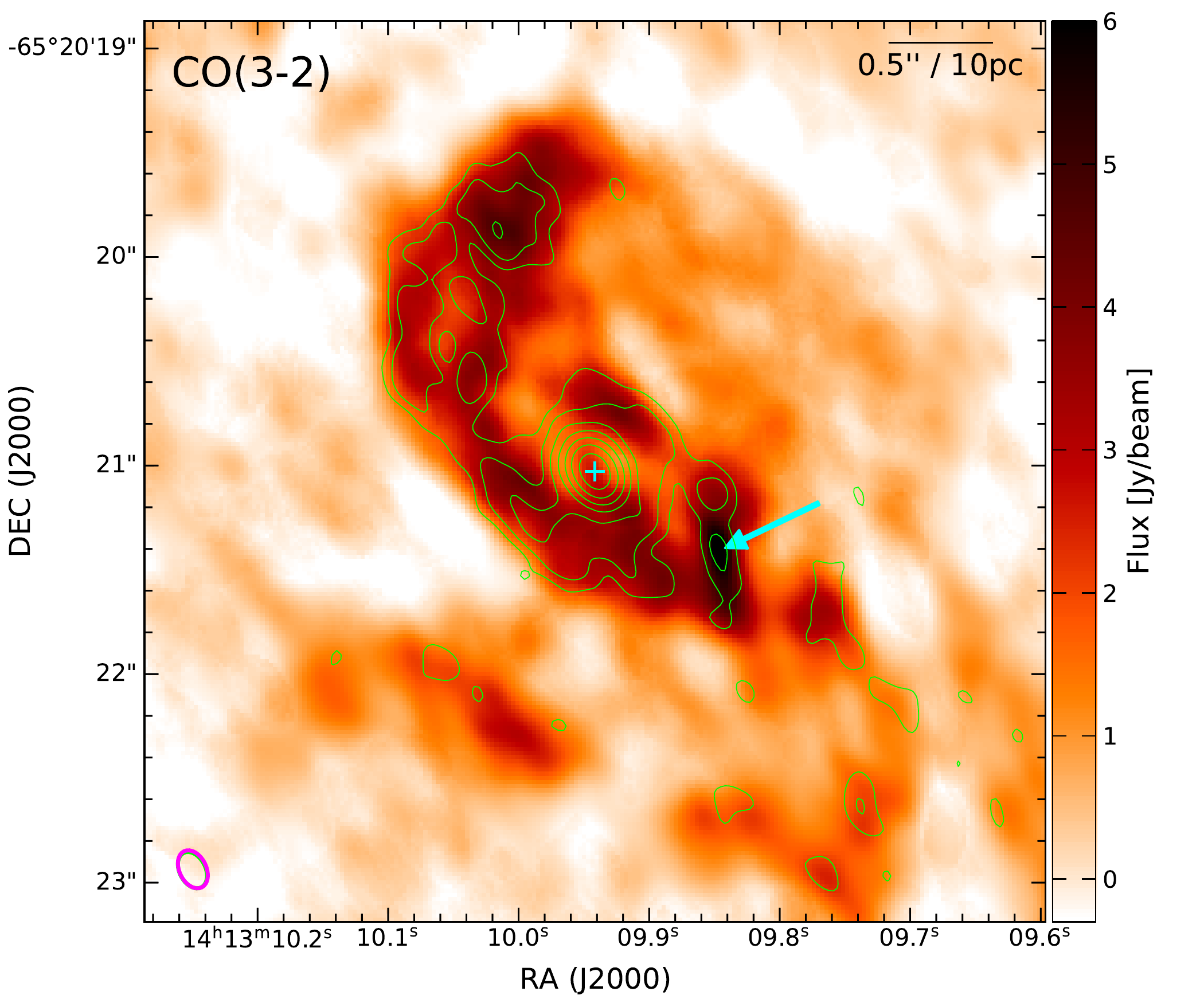}\hfill{}\includegraphics[width=0.49\textwidth]{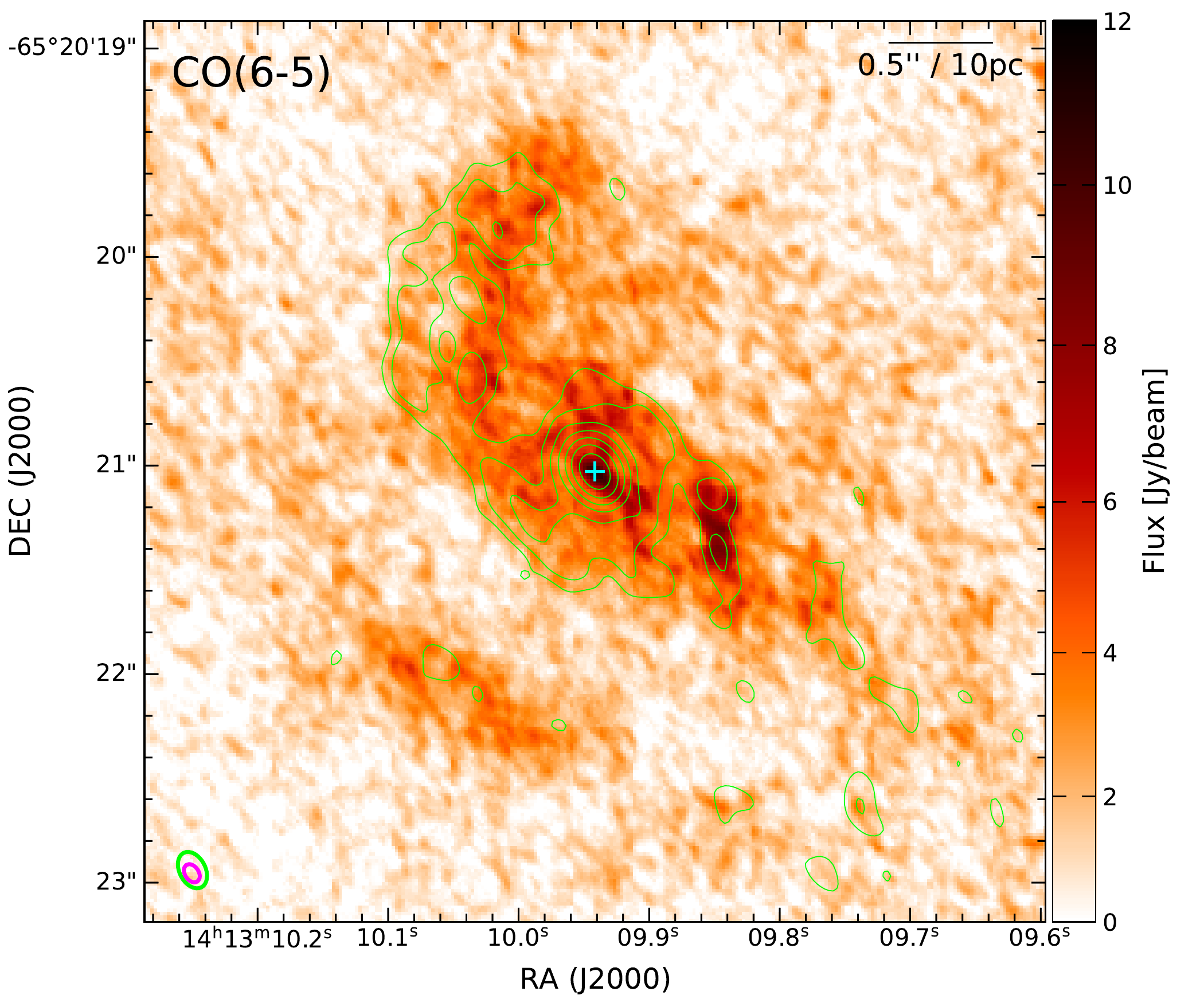}}{\Large\par}
\par\end{centering}
\begin{centering}
{\Large{}\includegraphics[width=0.49\textwidth]{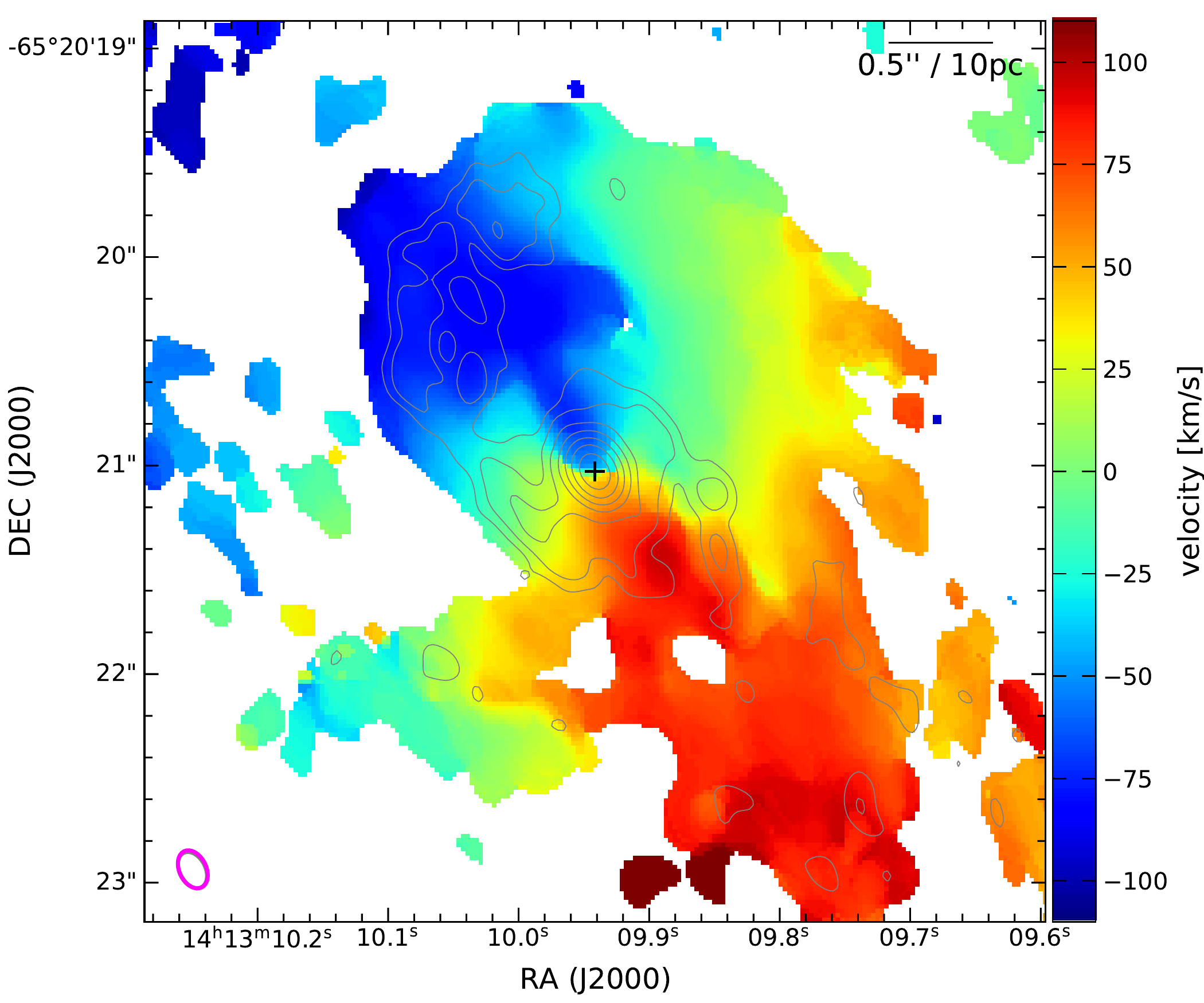}\hfill{}\includegraphics[width=0.49\textwidth]{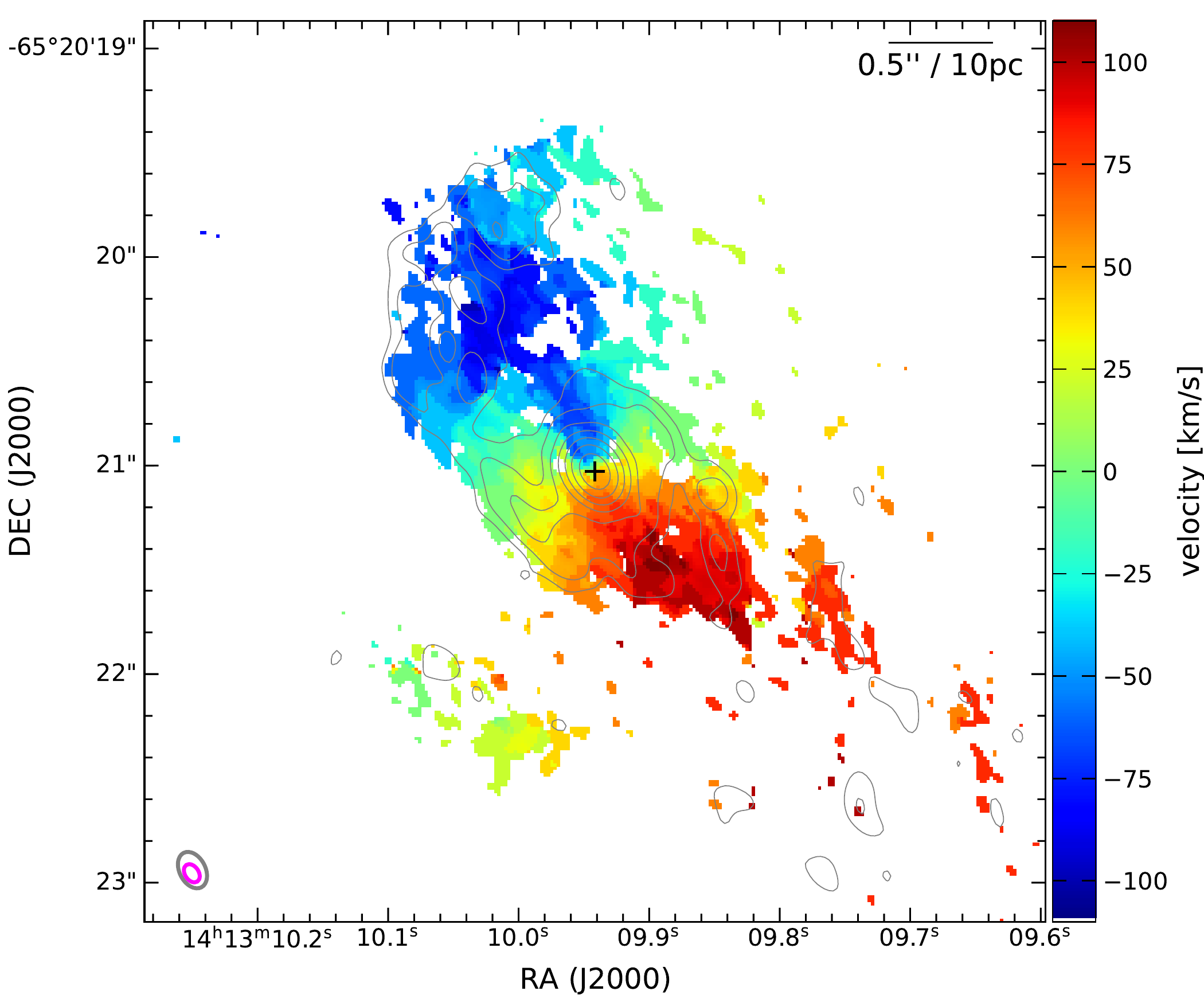}}{\Large\par}
\par\end{centering}
\begin{centering}
{\Large{}\includegraphics[width=0.49\textwidth]{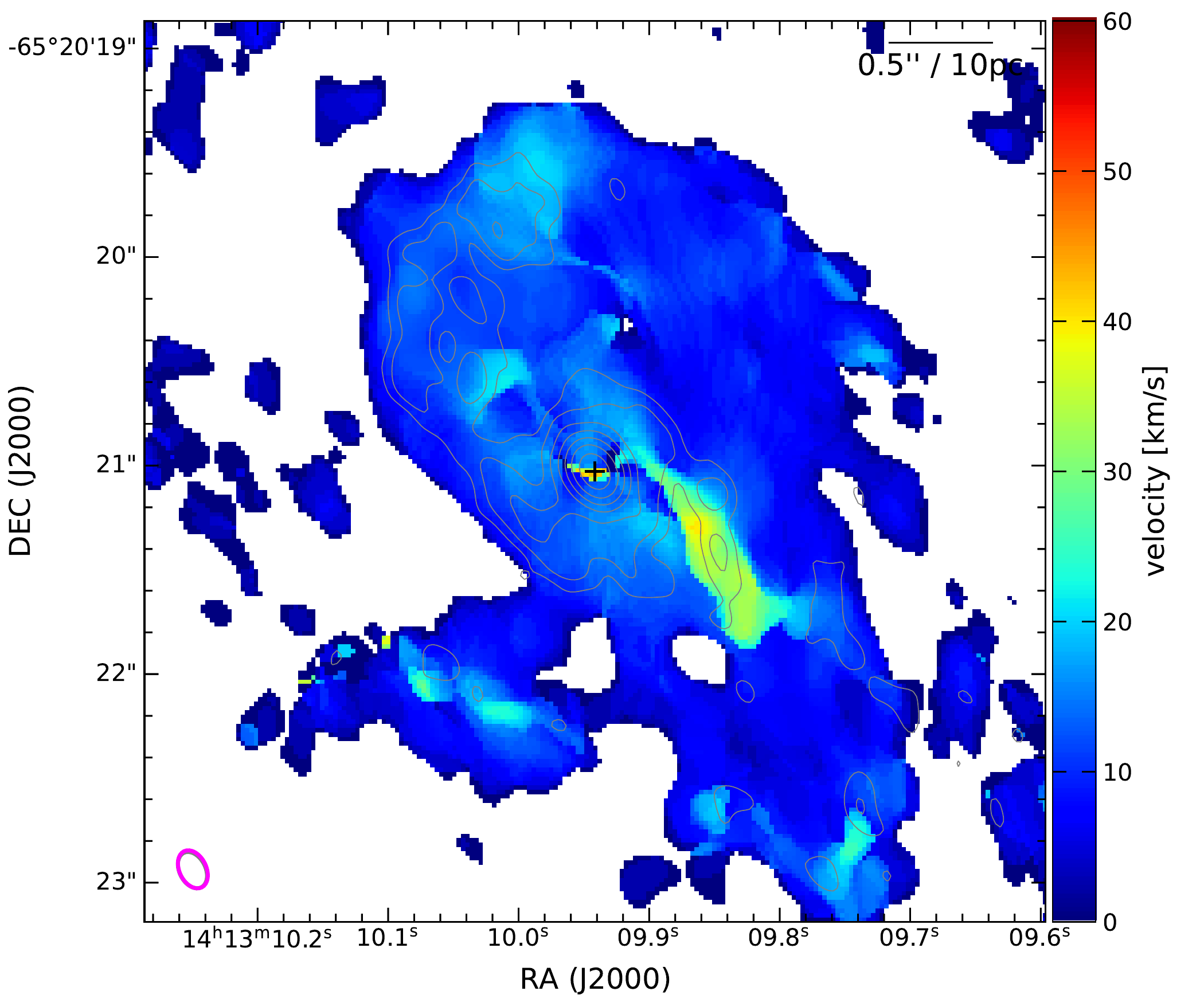}\hfill{}\includegraphics[width=0.49\textwidth]{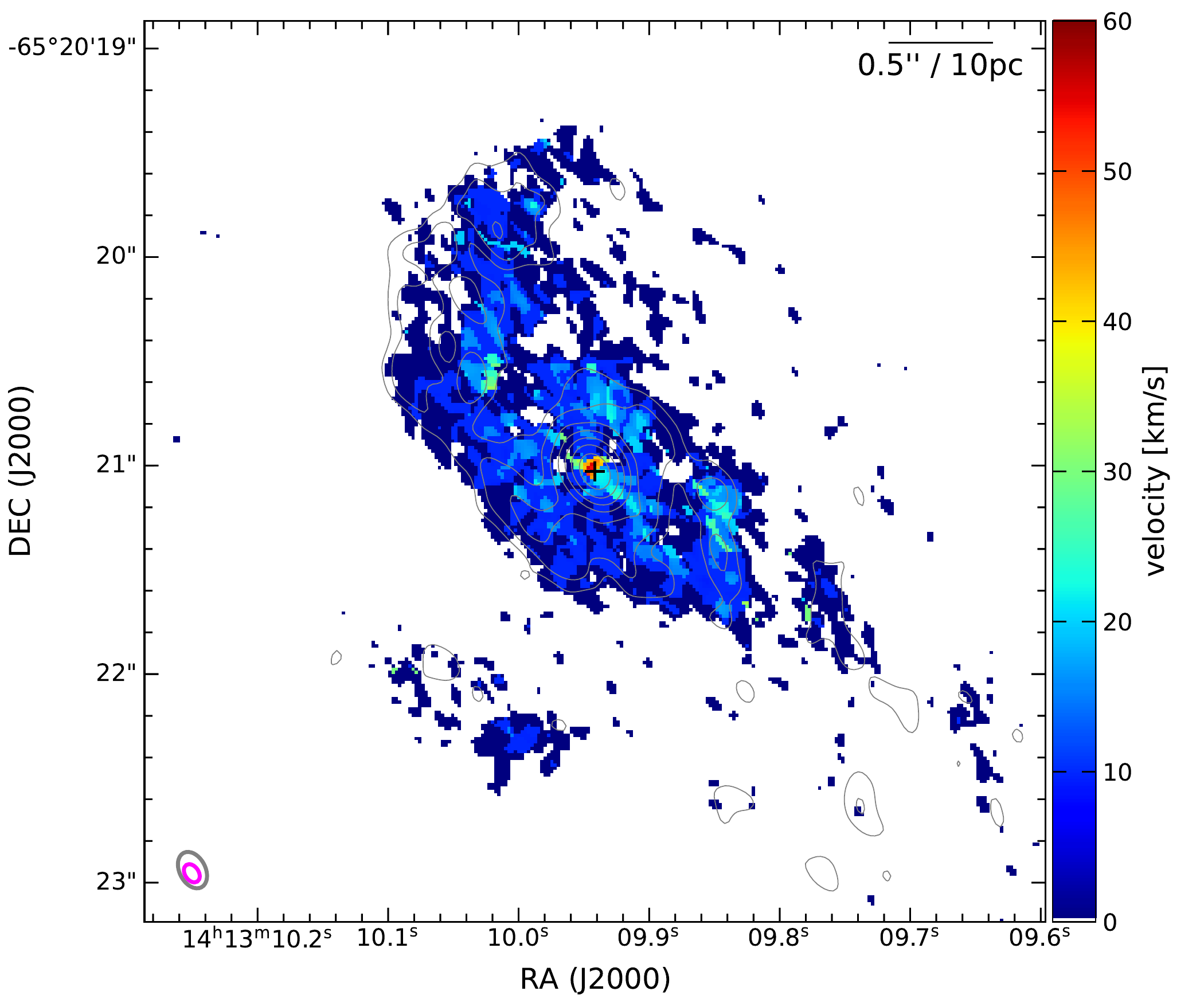}}{\Large\par}
\par\end{centering}
\caption{CO(3$-$2) ($\nu_{\mathrm{rest}}=346\,\mathrm{GHz}$, $\lambda_{\mathrm{rest}}=867\,\mathrm{\mu m}$,
left panels) and CO(6$-$5) ($\nu_{\mathrm{rest}}=691\,\mathrm{GHz}$,
$\lambda_{\mathrm{rest}}=434\,\mathrm{\mu m}$, right panels) moment
maps. In the top row, the total integrated line emission (moment 0)
maps are shown; in the middle row, the velocity (moment 1) maps are
displayed; the bottom row shows the velocity dispersion (moment 2).
The green / grey contours in all panels are for the $345\,\mathrm{GHz}$
continuum as in Fig.~\ref{fig:continuum-map}. The cyan arrow in
the top left panel marks the ridge-like structure discussed in the
text.\label{fig:CO-maps}}
\end{figure*}
\begin{figure*}
\begin{centering}
{\Large{}\includegraphics[width=0.49\textwidth]{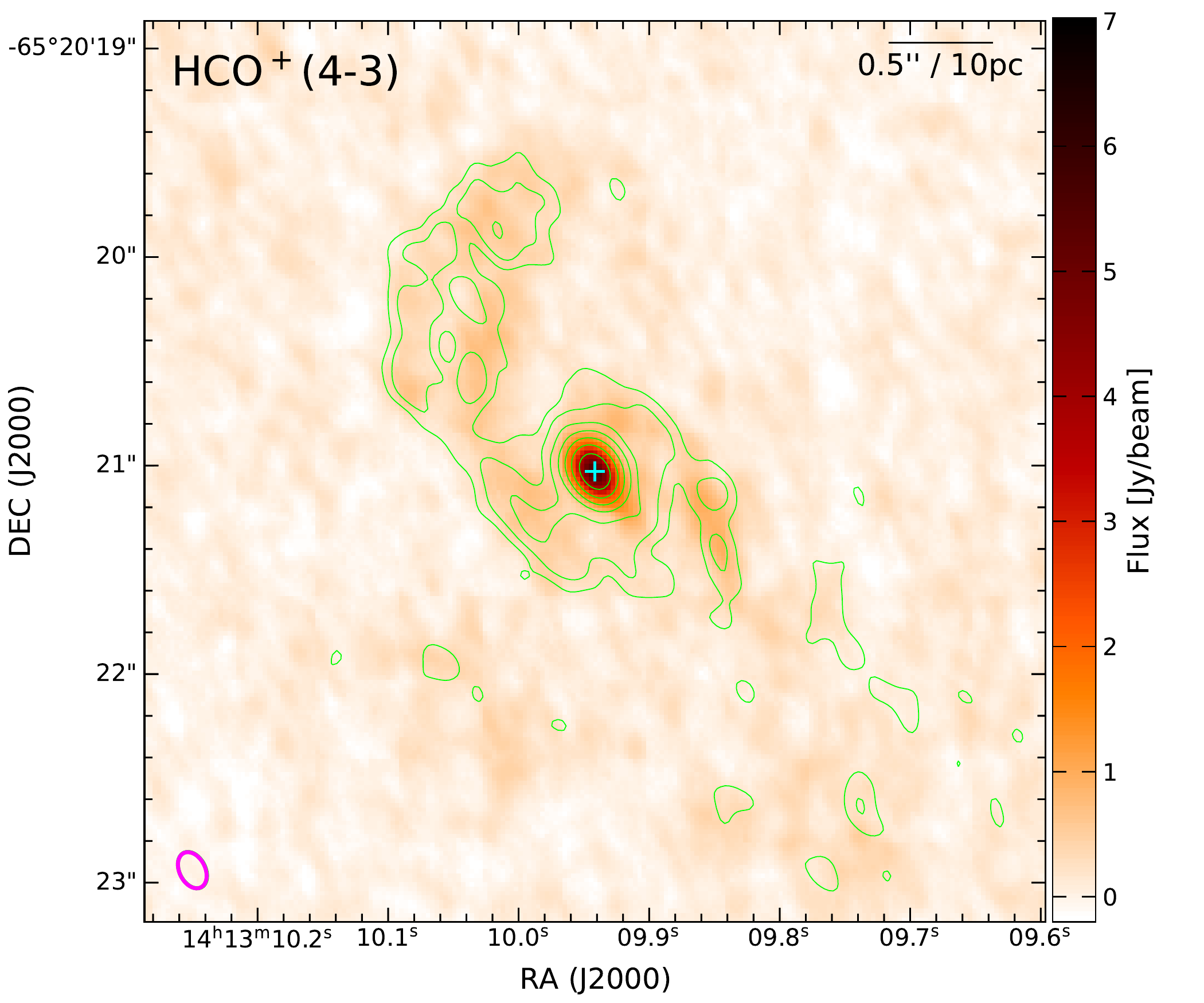}\hfill{}\includegraphics[width=0.49\textwidth]{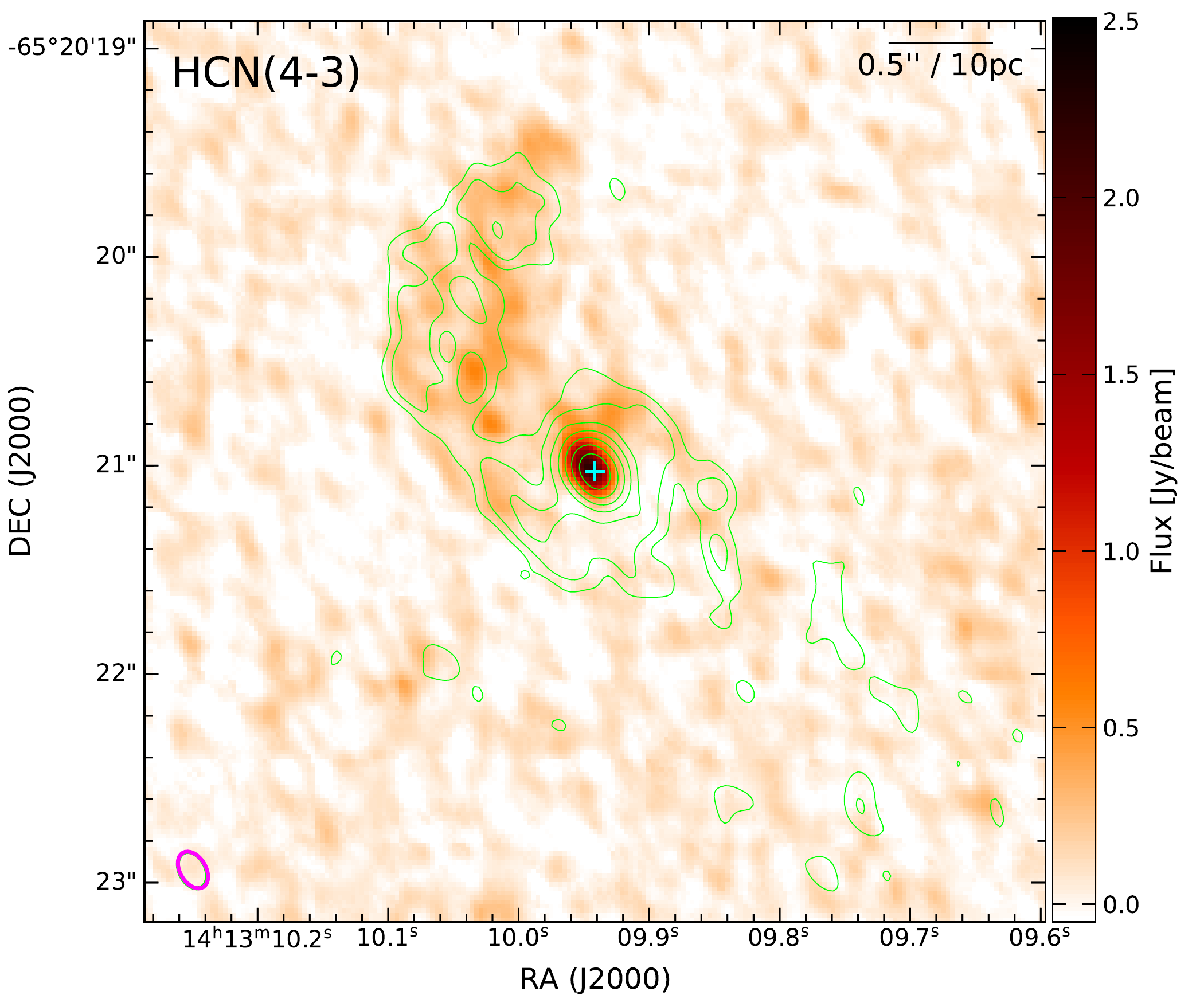}}{\Large\par}
\par\end{centering}
\begin{centering}
{\Large{}\includegraphics[width=0.49\textwidth]{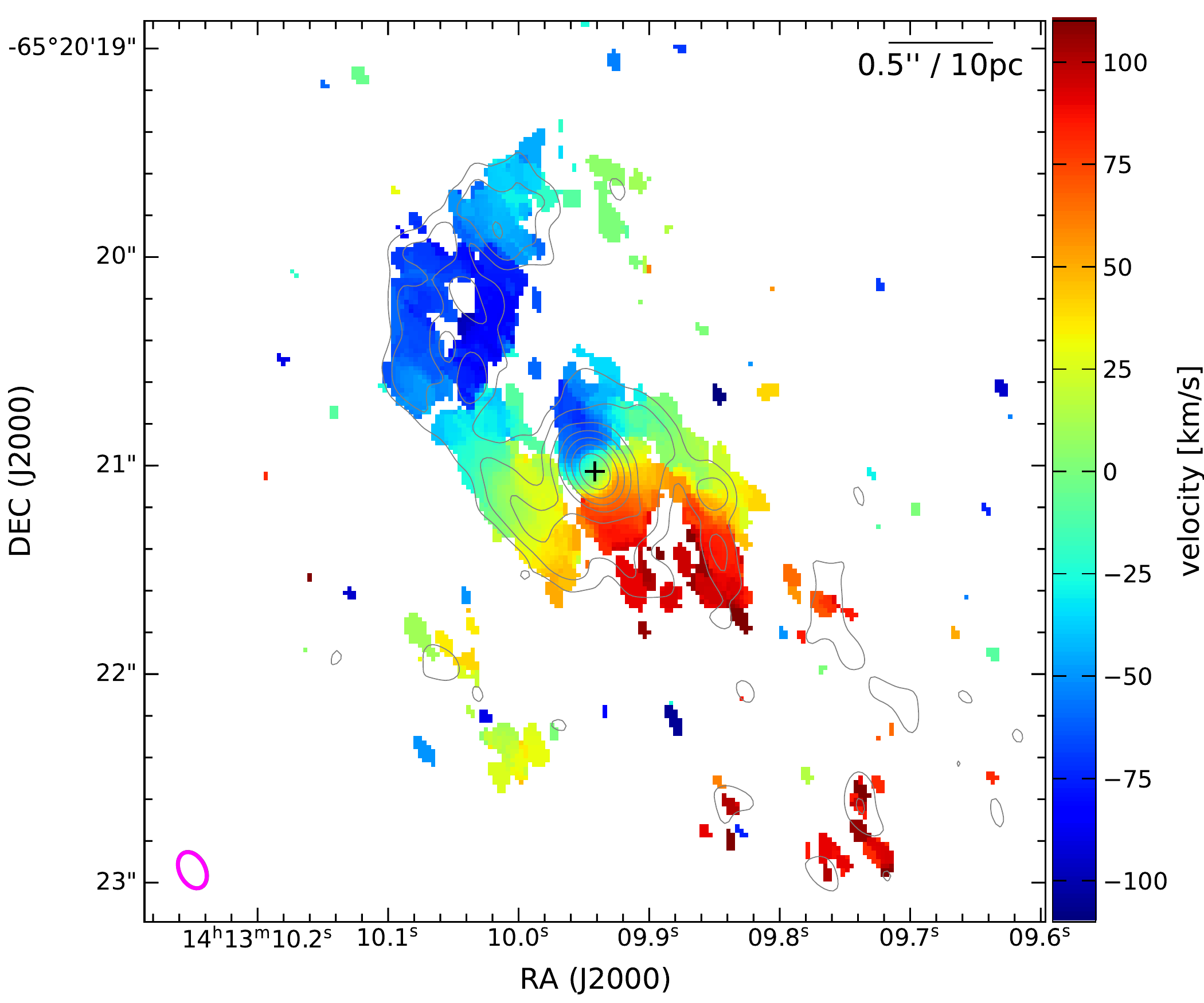}\hfill{}\includegraphics[width=0.49\textwidth]{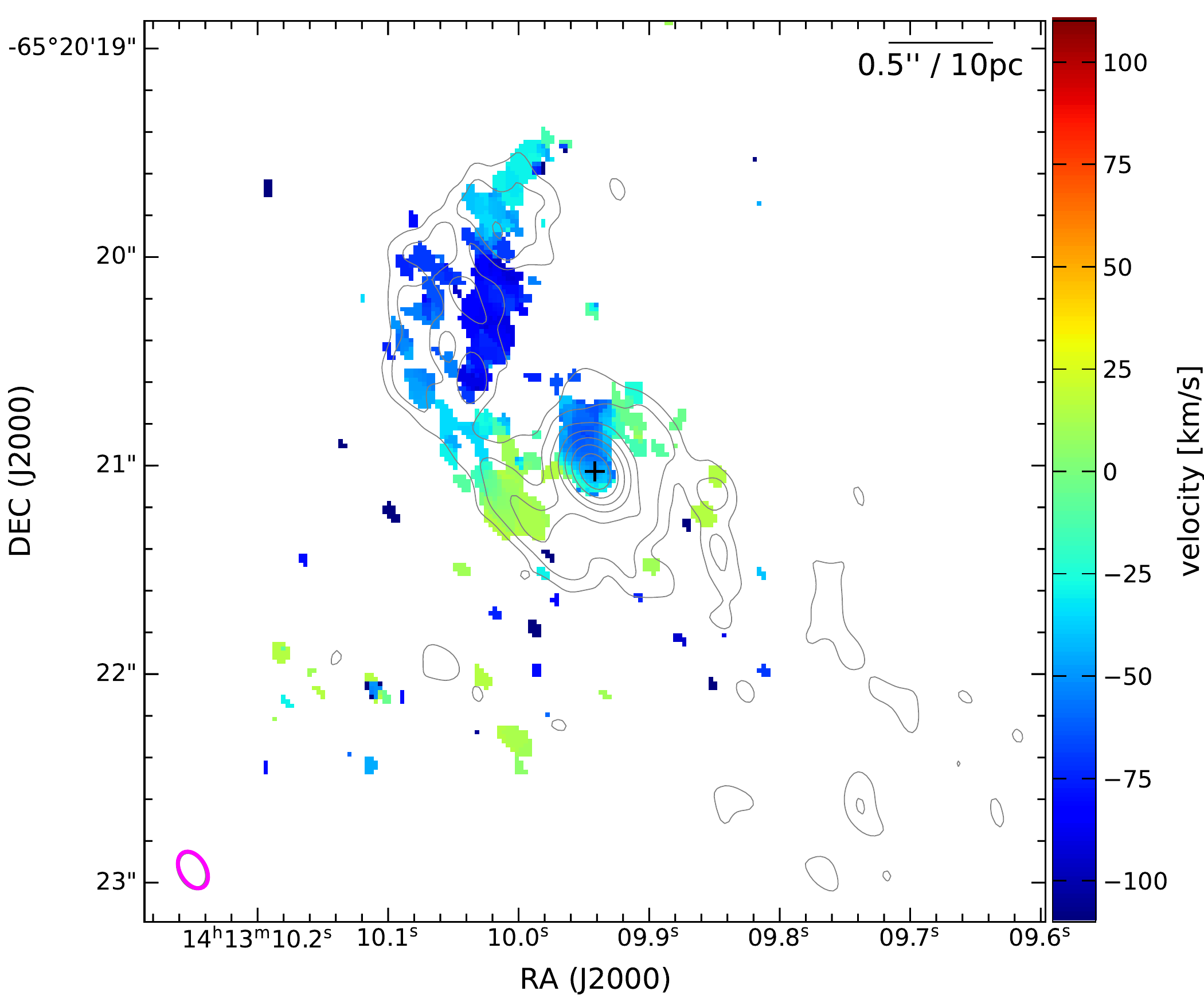}}{\Large\par}
\par\end{centering}
\begin{centering}
{\Large{}\includegraphics[width=0.49\textwidth]{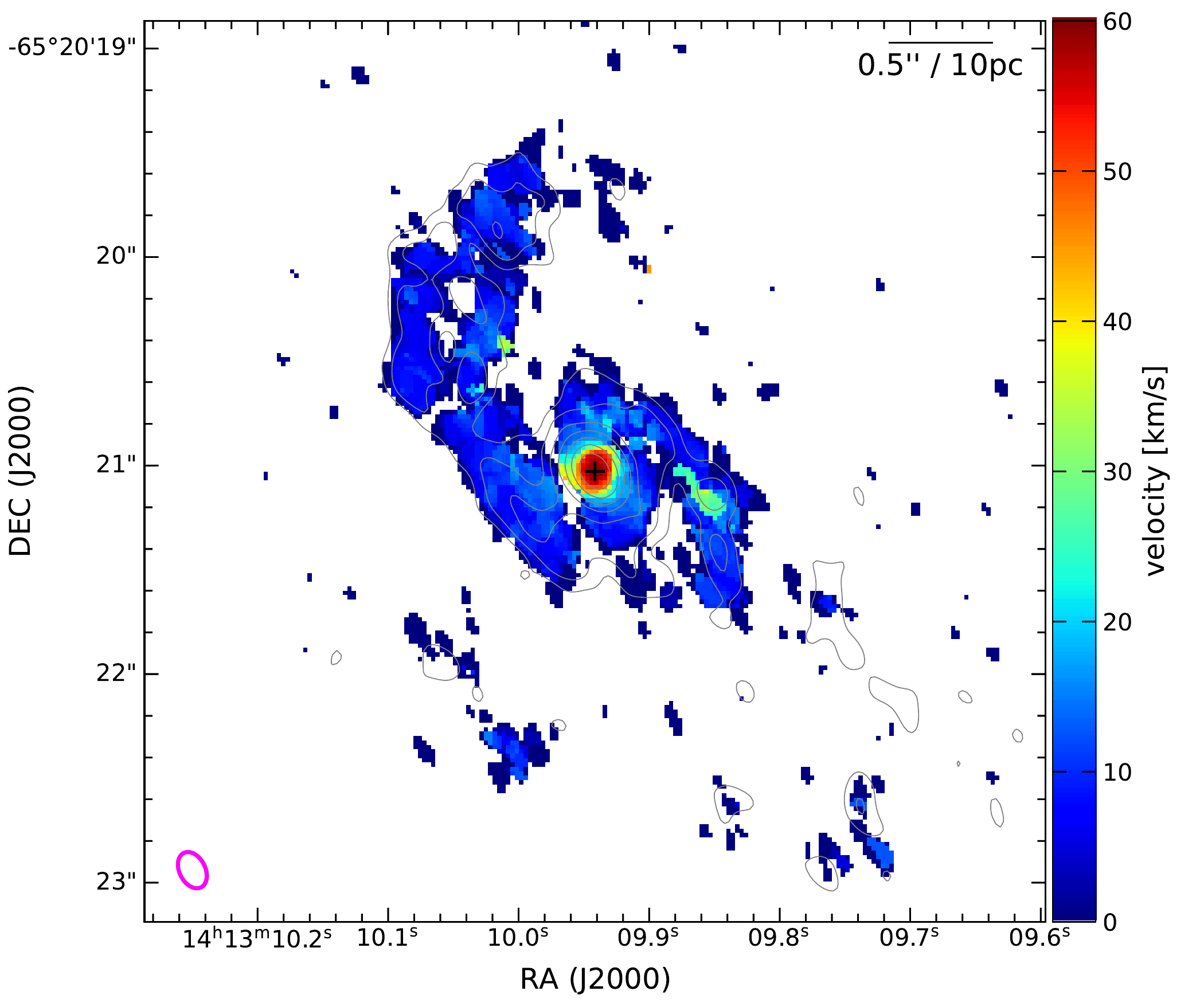}\hfill{}\includegraphics[width=0.49\textwidth]{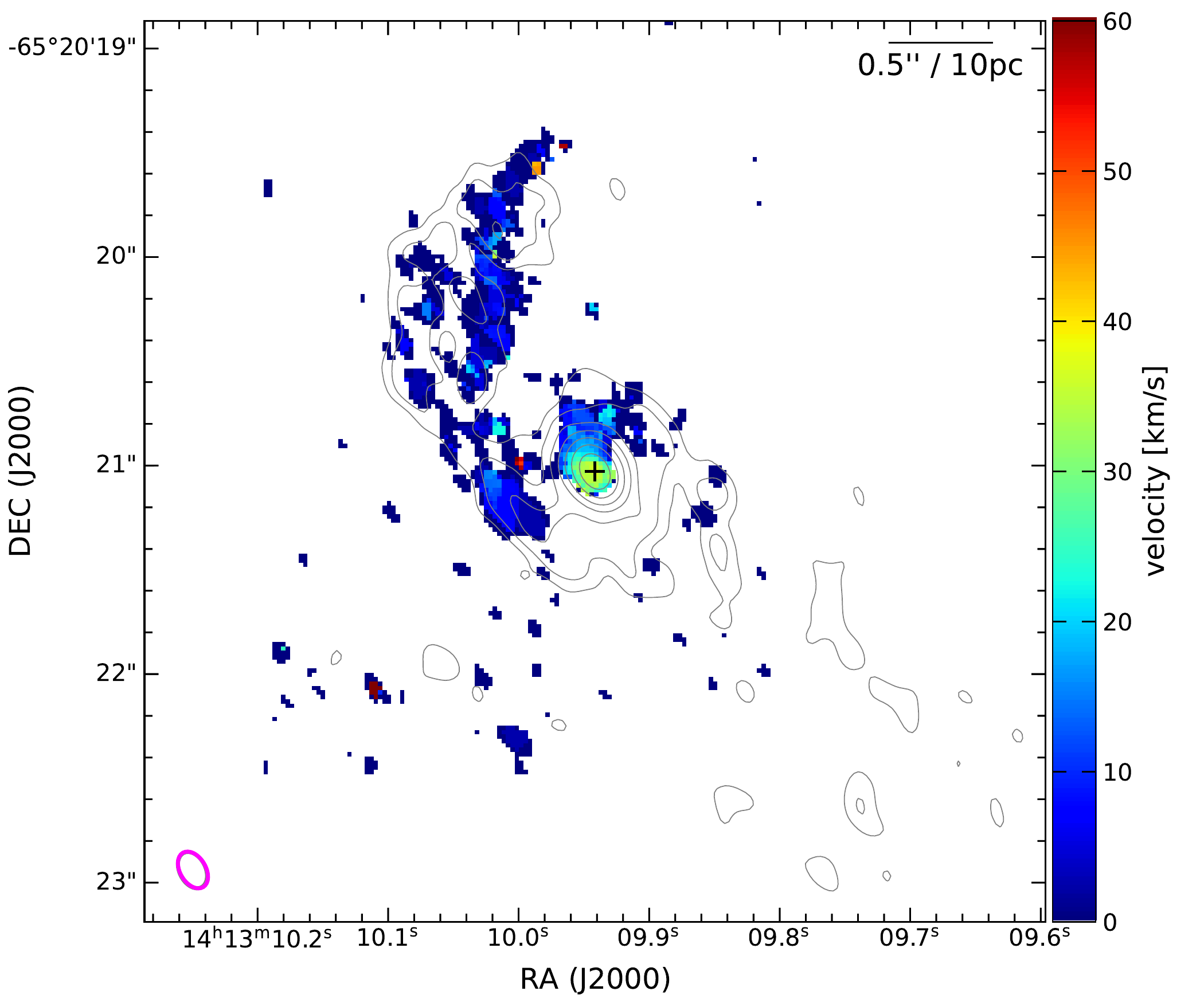}}{\Large\par}
\par\end{centering}
\caption{$\mathrm{HCO}^{+}$(4$-$3) ($\nu_{\mathrm{rest}}=357\,\mathrm{GHz}$,
left panels) and HCN(4$-$3) ($\nu_{\mathrm{rest}}=355\,\mathrm{GHz}$,
right panels) moment maps. In the top row, the total integrated line
emission (moment 0) maps are shown; in the middle row, the velocity
(moment 1) maps are displayed; the bottom row shows the velocity dispersion
(moment 2). The green / grey contours in all panels are for the $345\,\mathrm{GHz}$
continuum as in Fig.~\ref{fig:continuum-map}. For HCN(4$-$3), only
the velocity range from $-300$ to $+15\,\mathrm{km}\,\mathrm{s}^{-1}$
is covered by the observations, so that the south-western, redshifted
part of the emission is missing (see Sect.~\ref{sec:observations}).\label{fig:HCO-HCN-maps}}
\end{figure*}
\begin{figure}
\begin{centering}
{\Large{}\includegraphics[width=0.49\textwidth]{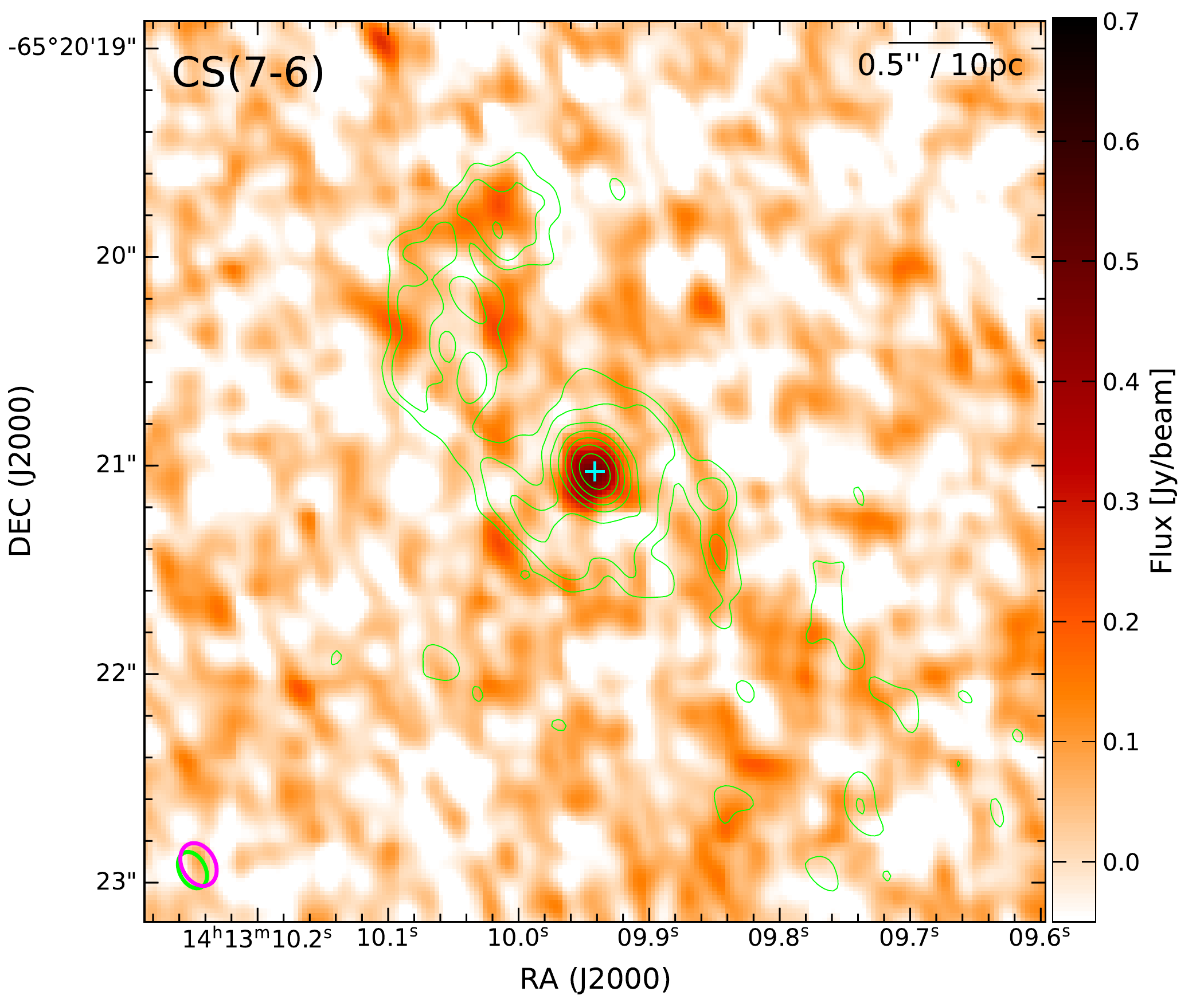}}{\Large\par}
\par\end{centering}
\begin{centering}
{\Large{}\includegraphics[width=0.49\textwidth]{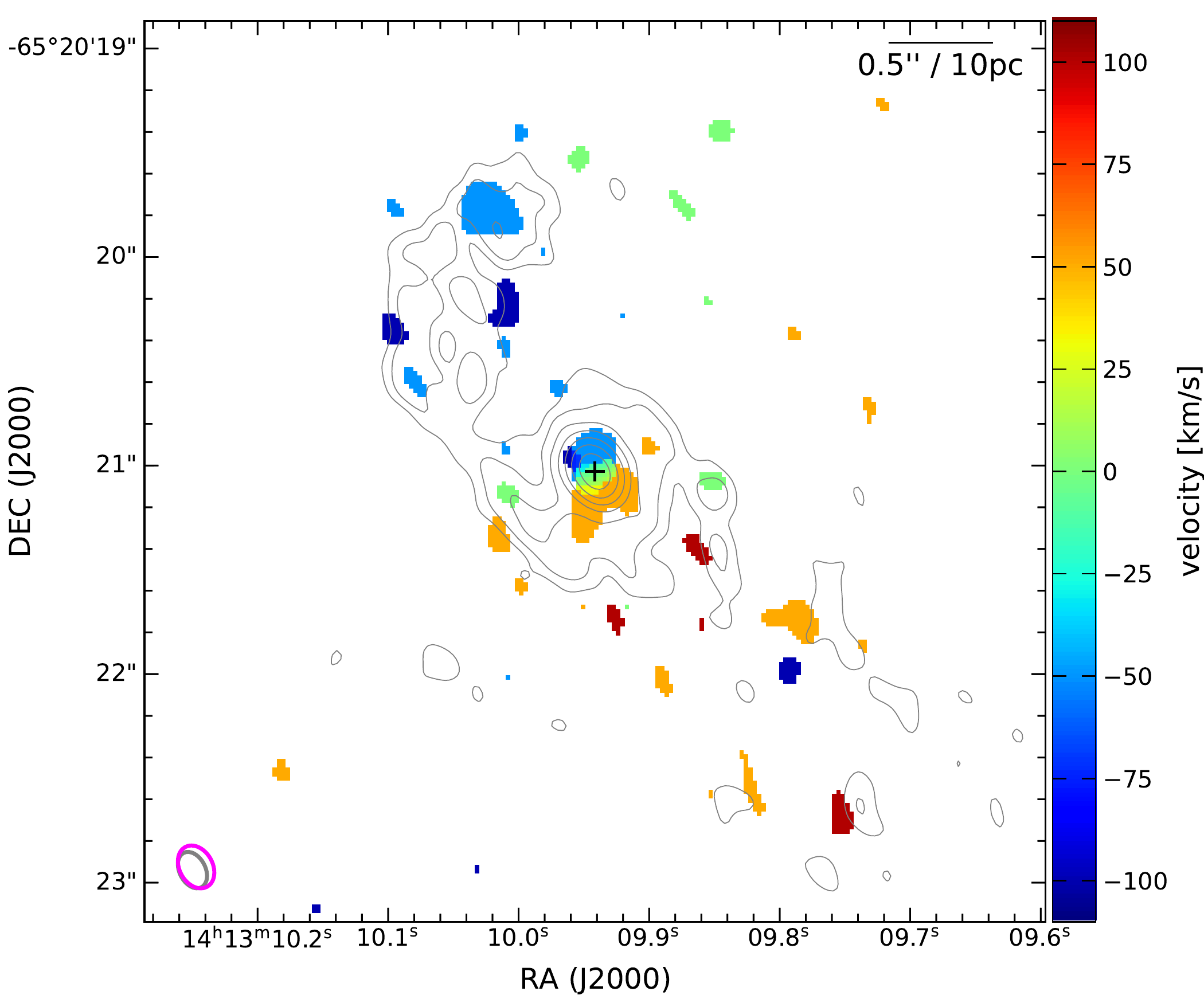}}{\Large\par}
\par\end{centering}
\begin{centering}
{\Large{}\includegraphics[width=0.49\textwidth]{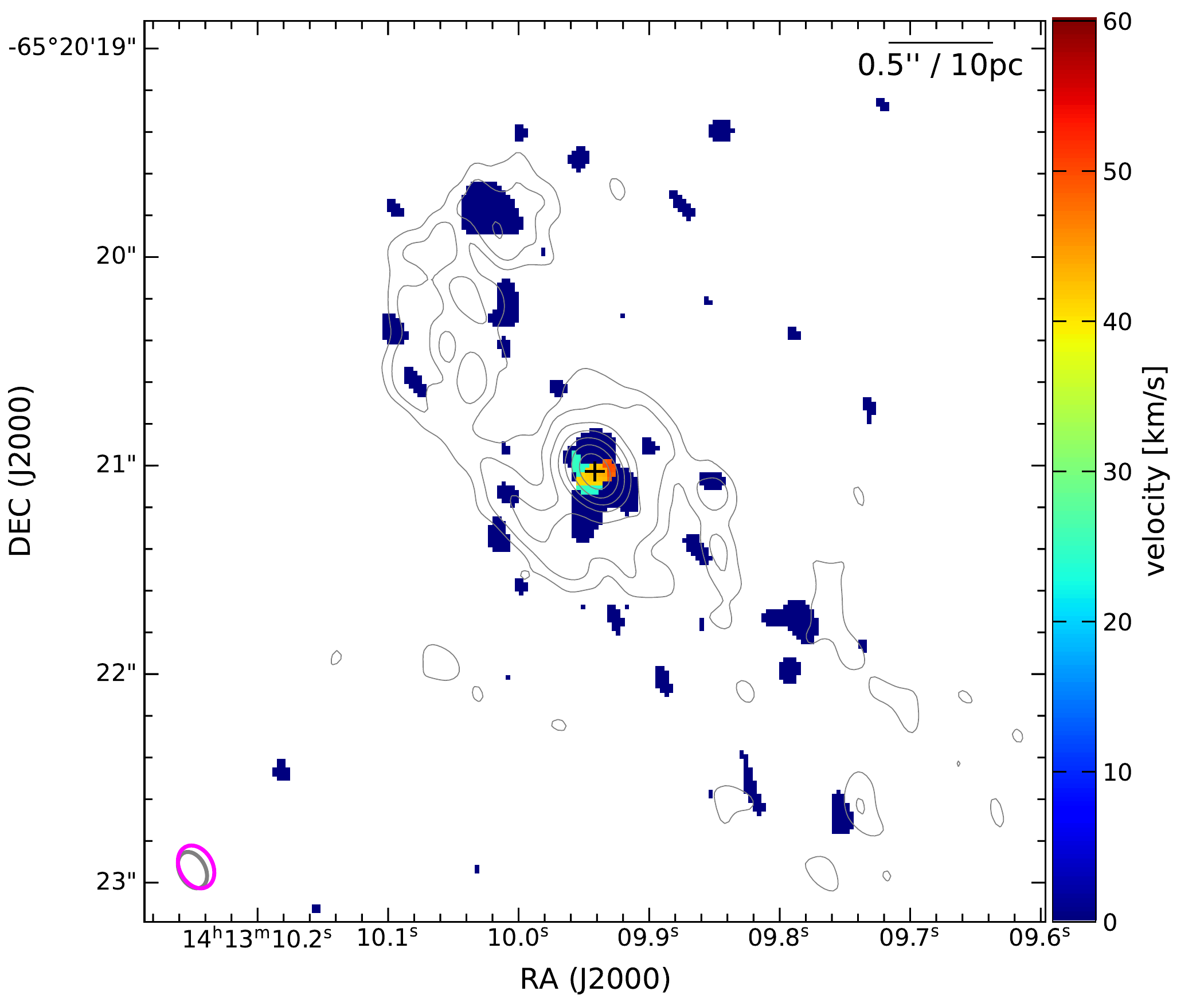}}{\Large\par}
\par\end{centering}
\caption{CS(4$-$3) ($\nu_{\mathrm{rest}}=343\,\mathrm{GHz}$) moment maps.
In the top panel, the total integrated line emission (moment 0) map
is shown; in the middle panel, the velocity (moment 1) map is displayed;
the bottom panel shows the velocity dispersion (moment 2). The green
/ grey contours in all panels are for the $345\,\mathrm{GHz}$ continuum
as in Fig.~\ref{fig:continuum-map}. \label{fig:CS-maps}}
\end{figure}

\subsection{CO(3$-$2) and CO(6$-$5) line emission\label{subsec:CO-line-emission}}

Both the CO(3$-$2) ($\nu_{\mathrm{rest}}=345.796\,\mathrm{GHz}$,
$\lambda_{\mathrm{rest}}=867\,\mathrm{\mu m}$) and CO(6$-$5) ($\nu_{\mathrm{rest}}=691.473\,\mathrm{GHz}$,
$\lambda_{\mathrm{rest}}=434\,\mathrm{\mu m}$) emission lines are
clearly detected, see Fig.~\ref{fig:CO-maps}. The CO(3$-$2) map
reveals very strong extended and patchy emission out to scales of
$10\,\mathrm{arcsec}$. The larger scale structure of the CO(3$-$2)
emission are very well described in \citet{Izumi2018}, who present
maps with slightly lower resolution but higher sensitivity. Their
data hence better trace the extended and fainter emission, revealing
three spiral arms. A large part of this large-scale emission is resolved
out in our data and we therefore concentrate only on the innermost
region. The structure on scales of $\sim1\,\mathrm{arcsec}$ largely
follows that of the continuum emission with an S-shaped morphology
composed of a strong spiral arm to the north-east of the nucleus and
a weaker spiral arm to the south-west. Intensity peaks in the CO line
emission correspond to peaks in the continuum. The line emission in
the north-eastern spiral arm is, however, more located towards the
inner side of the spiral arm, while the dust emission is stronger
along the outer edge. Surprisingly, the CO(3$-$2) emission does not
have a central peak like the continuum or the other emission lines.
In fact, there is a slight decrease of the emission levels, in the
innermost $10\:\mathrm{pc}$ ($500\,\mathrm{mas}$). The only structure
at the location of the nucleus is a $6\:\mathrm{pc}$ ($300\,\mathrm{mas}$)
large finger of emission reaching towards the nucleus from the south.
The nucleus is surrounded by a patchy `ring' of emission with peak
intensities of $\sim4\,\mathrm{Jy}\,\mathrm{beam}^{-1}\,\mathrm{km}\,\mathrm{s}^{-1}$.
Due to the absence of any core, only aperture photometry can be carried
out. We obtain a total flux of $73\pm8\,\mathrm{Jy}\,\mathrm{km}\,\mathrm{s}^{-1}$
within an aperture of $1\,\mathrm{arcsec}$ (cf.\ Table~\ref{tab:nuclear-properties}).
The brightest emission of the CO(3$-$2) line is actually located
to the south-west of the nucleus in the \emph{ridge-like structure}
extending in north-south direction ($\mathrm{PA}\sim6^{\circ}$, $F\sim30\,\mathrm{Jy}\,\mathrm{km}\,\mathrm{s}^{-1}$,
cyan arrow in Fig.~\ref{fig:CO-maps}). The structure is hence much
more prominent in CO than in the continuum or other emission lines.

The CO(6$-$5) emission has much lower signal to noise ratio but essentially
traces similar structures as the CO(3$-$2) emission, although with
significant differences. It also shows the S-shaped spiral structure,
including similar substructure such as the \emph{ridge-like structure}
to the south-west of the nucleus. The main difference is that this
transition has a clear peak at the location of the nucleus (within
a relative astrometric precision of $<40\,\mathrm{mas}$ to the continuum
peak at $691\,\mathrm{GHz}$). With a size of $170\,\mathrm{mas}\times94\,\mathrm{mas}$,
the emission is clearly resolved considering a beam size of $100\,\mathrm{mas}\times70\,\mathrm{mas}$.
The integrated fit of the central core is $17\pm5\,\mathrm{mJy}$,
while the emission within an aperture of $1\,\mathrm{arcsec}$ amounts
to $\sim470\,\mathrm{mJy}$, that is most of the CO(6$-$5) emission
comes from extended scales.

\subsection{Dense gas emission lines}

Except for the CO lines, the data contain three additional emission
lines: $\mathrm{HCO}^{+}$(4$-$3) ($\nu_{\mathrm{rest}}=356.734\,\mathrm{GHz}$,
$\lambda_{\mathrm{rest}}=840\,\mathrm{\mu m}$, Fig.~\ref{fig:HCO-HCN-maps}
left panels), HCN(4$-$3) ($\nu_{\mathrm{rest}}=354.505\,\mathrm{GHz}$,
$\lambda_{\mathrm{rest}}=846\,\mathrm{\mu m}$, Fig.~\ref{fig:HCO-HCN-maps}
right panels) and CS(4$-$3) ($\nu_{\mathrm{rest}}=342.883\,\mathrm{GHz}$,
$\lambda_{\mathrm{rest}}=874\,\mathrm{\mu m}$, Fig.~\ref{fig:CS-maps}),
all in band 7. In band 9, no other emission lines apart from CO(6$-$5)
are detected. 

$\textrm{HCO}^{+}$ is clearly detected even for a high sampling in
velocity space. The morphology of the emission in $\mathrm{HCO}^{+}$
is very similar to the continuum emission, with a strong emission
peak at the nucleus, surrounded by the S-shaped spiral structure.
Even clearer than for CO(3$-$2), $\textrm{HCO}^{+}$ is stronger
on the inside of the spiral arm in comparison to the continuum emission
which is enhanced on the outer edge of the spiral arm. Considering
the counter-clockwise rotation of the gas (see below), material is
entering the spiral arm from the concave inside of the arm, such that
the spiral arms are trailing. There might be a shock where the gas
enters the arm, and star-formation may take place. Dust and continuum
emission occurs later in the flow, towards the convex outside of the
arm. The nuclear peak contains a flux of $10.5\pm1.2\,\mathrm{Jy}\,\mathrm{km}\,\mathrm{s}^{-1}$
and is possibly resolved slightly with an elongation along a position
angle of $38\pm17^{\circ}$ (see Table~\ref{tab:nuclear-properties}).
This would imply that we start to resolve the nuclear dense gas emission
along the equatorial direction.

For the HCN(4$-$3) emission line, only the blue-shifted part is covered
by our ALMA observations (see Sect.~\ref{subsec:observations}).
The morphology of this part is very similar to that of the $\textrm{HCO}^{+}$
emission. Also HCN(4$-$3) shows an emission peak at the nucleus;
however, the peak in the moment 0 map is truncated towards the south-west
because the red-shifted velocities of the line are missing. The signal
to noise of the HCN(4$-$3) map is, however, clearly lower than for
the $\mathrm{HCO}^{+}$ line.

Finally, the emission of the CS line is faint but clearly detected
when strongly binning in velocity space (see Sect.~\ref{subsec:datareduction}).
Most of the CS emission is concentrated towards the position of the
nucleus with a peak flux of $0.57\pm0.07\,\mathrm{Jy}\,\mathrm{km}\,\mathrm{s}^{-1}$.
Faint emission on the $3\sigma$ level also traces the S-shaped spiral
arms.

\subsection{Line kinematics\label{subsec:results-kinematics}}

\begin{figure}
\includegraphics[width=1\columnwidth]{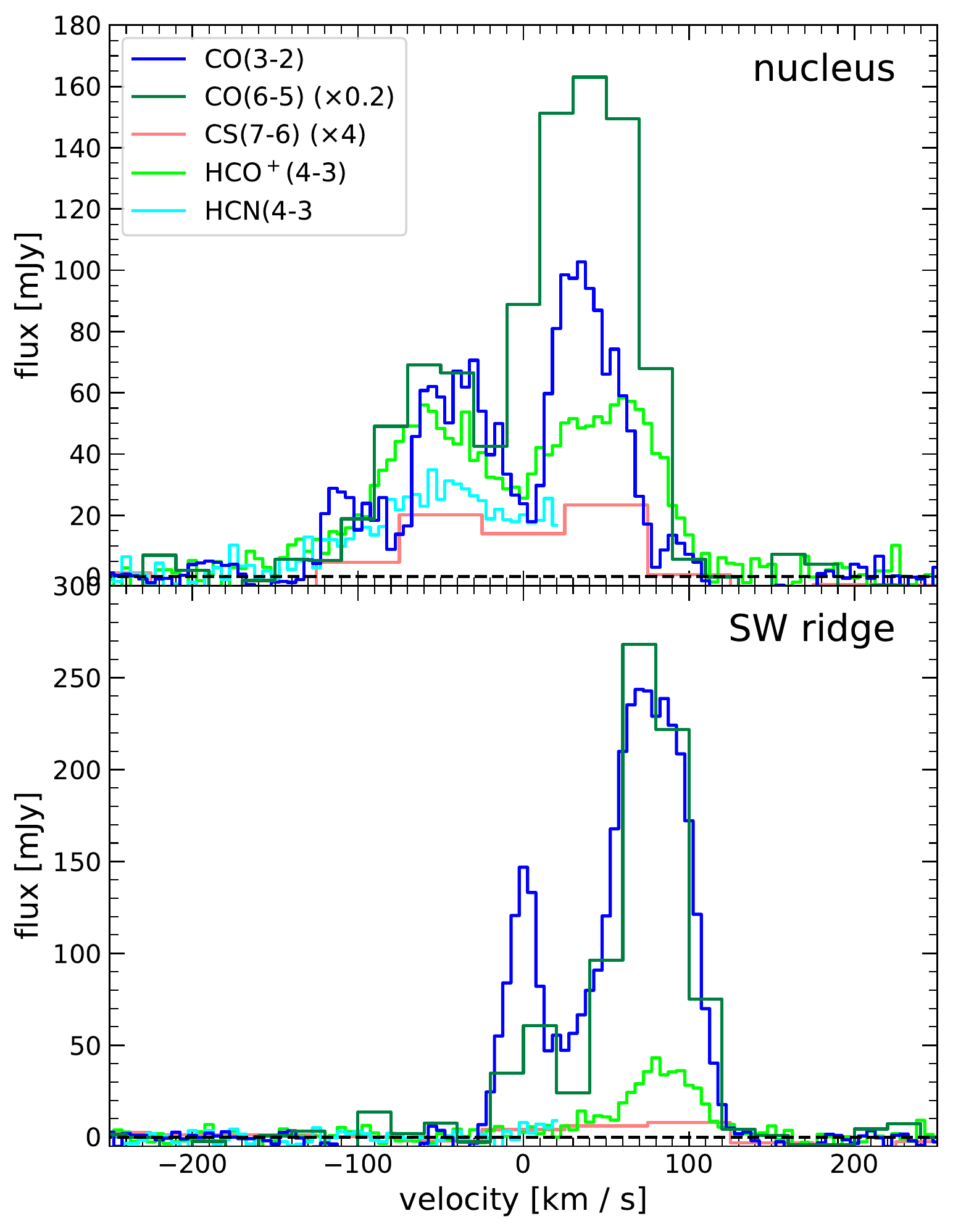}\caption{Emission line spectra for all detected emission lines, that is CO(3$-$2),
CO(6$-$5), CS(4$-$3), $\mathrm{HCO}^{+}$(4$-$3), and HCN(4$-$3),
at two locations in the nuclear region of the Circinus Galaxy: the
nucleus itself (top panel, cyan cross in Figs.~\ref{fig:CO-maps}
to \ref{fig:CS-maps}), and the ridge-like structure to the south-west
of the nucleus (lower panel, tip of the cyan arrow in the top left
panel of Fig.~\ref{fig:CO-maps}). For better visibility, the spectra
for CO(6$-$5) and CS(4$-$3) were multiplied by factors of 4 and
0.2 respectively.\label{fig:spectra}}
\end{figure}

The velocity fields of all emission lines are similar, tracing a rotating
pattern with velocities between $-130$ and $+130\,\mathrm{km}\,\mathrm{s}^{-1}$
(see Figs.~\ref{fig:CO-maps} to \ref{fig:CS-maps}). The approaching,
blue-shifted part of the rotation is located towards the north-east
and the receding, red-shifted part towards the south-west. This velocity
field is in agreement to the one found for the galactic disk on kpc
scales \citep[e.g.][]{Freeman1977,Elmouttie1998b,Curran2008,Zschaechner2016},
as well as the one on smaller, sub-parsec scales traced by the water
masers \citep{Greenhill2003a}. Within ${\sim}0.5\,\mathrm{arcsec}$
(i.e.\ $10\:\mathrm{pc}$) from the nucleus, the velocities in the
polar region to the south-east are generally more red-shifted (up
to $+40\,\mathrm{km}\,\mathrm{s}^{-1}$), while the gas towards the
north-western side has lower velocities (down to $-10\,\mathrm{km}\,\mathrm{s}^{-1}$).
This may be a sign for outflowing gas and is further discussed in
Sect.~\ref{subsec:discuss_kinematics}.

In Fig.~\ref{fig:spectra}, spectra for two locations are shown:
at the nucleus itself and for the \emph{ridge-like structure} (cyan
cross and tip of the cyan arrow in the top left panel of Fig.~\ref{fig:CO-maps},
respectively). At the nucleus itself, the spectra of all emission
lines show a double peak with the two maxima located at $v\sim\pm50\,\mathrm{km}\,\mathrm{s}^{-1}$.
For CO(3$-$2), the overall width of the emission and the separation
of the peaks is smaller than for the other lines, while the central
minimum at $v\sim0\,\mathrm{km}\,\mathrm{s}^{-1}$ is deeper. $\textrm{HCO}^{+}$
and HCN (at least on the blue-shifted side which is covered by our
observations) are displaying the highest velocities at the centre
(up to $170\,\mathrm{km}\,\mathrm{s}^{-1}$, that is significantly
more than the rotational pattern of the nuclear disk) and show a less
deep dip of the profile close to the systemic velocity. This is also
reflected by the velocity dispersion, which is the lowest for CO(3$-$2),
and the highest for $\mathrm{HCO}^{+}$(4$-$3) (cf.\ Fig.~\ref{fig:HCO-HCN-maps},
bottom left panel). The reason for these differences is discussed
in Sect.~\ref{subsec:discuss_kinematics}.

Most of the line emission outside the nucleus comes from only one
relatively narrow velocity component with $\sigma<15\,\mathrm{km}\,\mathrm{s}^{-1}$,
as can be deduced from the respective intensity-weighted velocity
dispersion (moment 2) maps. The dispersion of $\mathrm{HCO}^{+}$(4$-$3)
is the same or slightly lower as that of CO(3$-$2), while for the
other transitions the dispersions agree within uncertainties.

There are, however, several regions in the spiral arms, where multiple
distinct velocity components can be separated. The clearest example
is the \textit{ridge-like structure} to the south-west of the nucleus
(see Sect.~\ref{subsec:CO-line-emission}). At its northern end,
it has a single broad component ($v_{\mathrm{peak}}\sim30\,\mathrm{km}\,\mathrm{s}^{-1}$,
$\sigma\sim40\,\mathrm{km}\,\mathrm{s}^{-1}$), which splits up into
two velocity components towards the southern end (cf.\ bottom panel
of Fig.~\ref{fig:spectra}). This is most clearly seen in the CO(3$-$2)
data, where the main component is found at $v_{\mathrm{peak}}=90\pm5\,\mathrm{km}\,\mathrm{s}^{-1}$
with a velocity dispersion of $\sigma=18\pm2\,\mathrm{km}\,\mathrm{s}^{-1}$
and a second, fainter component is found at the systemic velocity
with a lower velocity dispersion ($v_{\mathrm{peak}}=9\pm2\,\mathrm{km}\,\mathrm{s}^{-1}$,
$\sigma=8\pm1\,\mathrm{km}\,\mathrm{s}^{-1}$). The first, red-shifted
component corresponds to the normal velocity field of a rotating disk
expected at this location. The two components lead to the increased
dispersion along the ridge clearly visible in the moment 2 map of
CO(3$-$2), see the lower left panel in Fig.~\ref{fig:CO-maps}.
CO(6$-$5) has a similar two component velocity structure as CO(3$-$2),
although the systemic peak has a much lower relative intensity (its
intensity is much less compared to the redshifted peak than for CO(3$-$2)).
For HCN, we only see a marginal increase in the spectrum at $v=0\,\mathrm{km}\,\mathrm{s}^{-1}$
at the end of the spectral coverage for this line, which could indicate
that it also traces the systemic component. The $\mathrm{HCO}^{+}$(4$-$3)
emission, on the other hand, clearly does not show the velocity component
at systemic velocity, it only shows the broader, red-shifted component
which is consistent with disk rotation. The CS line has too low signal
to noise and spectral resolution to separate any two components clearly.
We hence interpret the narrow systemic velocity component as a lower
density filament behind the disk, along which material is falling
toward the disk, that is it has a lower line-of-sight velocity than
the gas in the disk at this location. The gas is of lower density
and excitation than the gas in the disk, because it is comparatively
weaker in CO(6$-$5) and not detected in $\mathrm{HCO}^{+}$(4$-$3).
Alternatively, this could be material in front of the disk, similar
to dust and gas causing self-absorption at the nucleus, that is material
which is lifted up from the disk towards us, although the physical
mechanism responsible for this would be unclear.

\begin{figure}
\includegraphics[width=1\columnwidth]{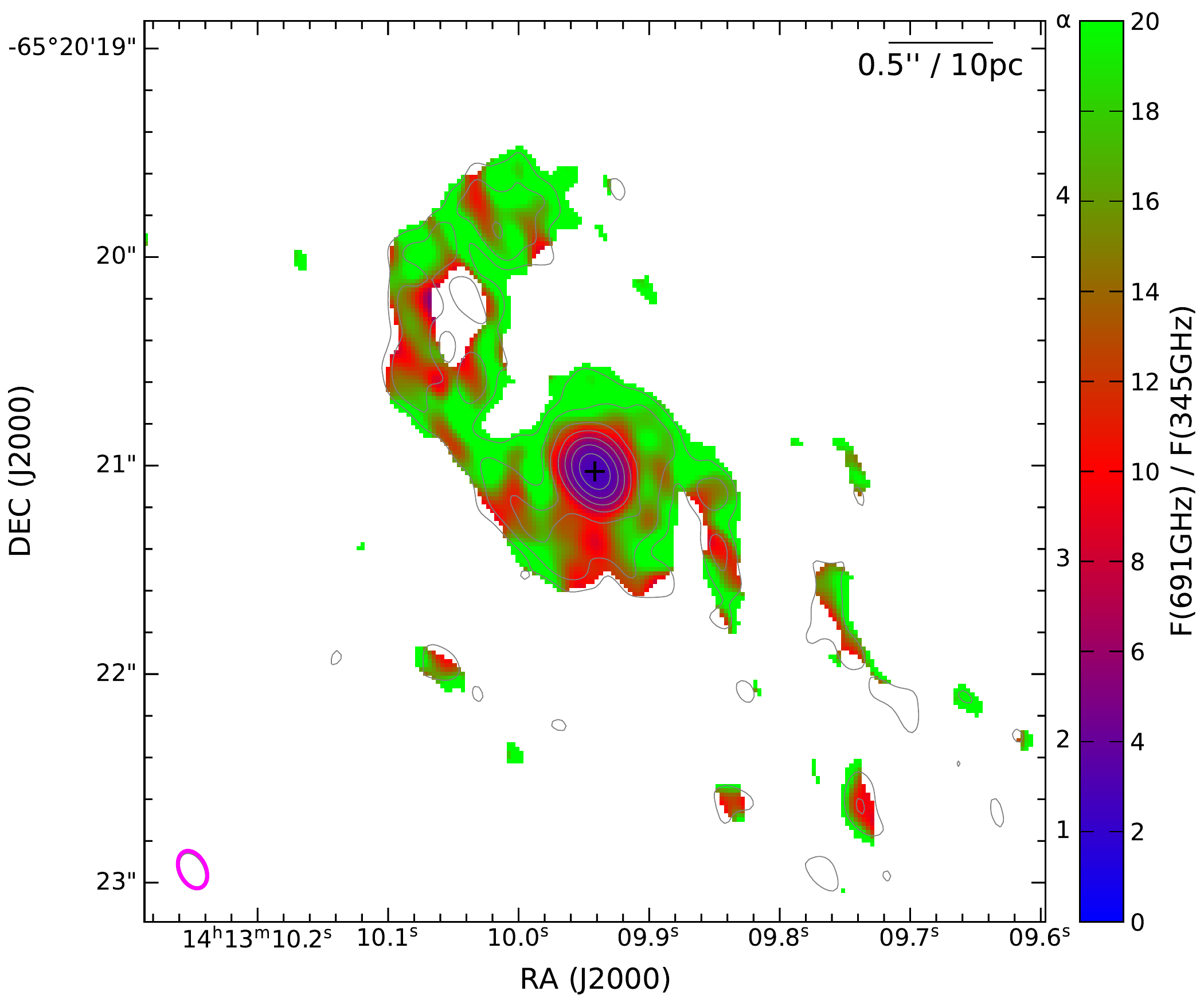}\caption{Band 9 to band 7 continuum ratio map: $F_{691\,\mathrm{GHz}}/F_{345\,\mathrm{GHz}}$.
The left axis of the colour bar is labelled with the respective spectral
index $\alpha=0.998\times\log_{2}\left(F_{691\,\mathrm{GHz}}/F_{345\,\mathrm{GHz}}\right)$.
\label{fig:band9to7ratio}}
\end{figure}
More or less symmetrically to the ridge on the other side of the nucleus
to the north-east, there also seems to be a slight enhancement of
the velocity dispersion for CO(3$-$2) and also CO(6$-$5). The spectrum
at this location in fact shows a significantly broader line emission,
with an indication of two velocity components separated by $\Delta v\sim40\,\mathrm{km}\,\mathrm{s}^{-1}$.
Most other regions with $\sigma>20\,\mathrm{km}\,\mathrm{s}^{-1}$
in the moment 2 map of CO(3$-$2) similarly show multiple velocity
components; in some cases up to three distinct components. These are
much less clear or absent in the other emission lines, indicating
that this is most likely lower density material above or below the
disk in a filamentary structure.

In conclusion we find the kinematics of the molecular material in
the nuclear region dominated by rotation, with some perturbation by
the spiral arms and by filaments of less dense material above or below
the disk with distinct velocity components. In all cases, the velocity
dispersion is much lower than the rotational velocity (see also Sect.~\ref{subsec:discuss_kinematics}).

\section{Discussion\label{sec:discussion}}

\subsection{Origin of the continuum emission\label{subsec:discuss_continuum}}

Continuum emission at submillimetre wavelengths can have its origin
in three different emission mechanisms: thermal free-free emission
(bremsstrahlung), synchrotron emission and thermal emission from dust
grains. All three mechanisms have characteristic power law spectral
indices (for $F_{\nu}\propto\nu^{\alpha}$): $\alpha=-0.1$ for (optically
thin) free-free emission, and $\alpha=-0.7$ for synchrotron emission.
For thermal dust emission, $\alpha=3\ldots4$, depending on the dust
emissivity index $\beta=\alpha-2$\footnote{The dust emission is commonly assumed to follow a modified blackbody
spectrum, $F_{\nu}=\Omega\cdot B_{\nu}\left(T\right)\cdot Q_{\nu},$
where $B_{\nu}\left(T\right)$ is the Planck function depending on
the temperature $T$, and $Q_{\nu}=Q_{0}\left(\nu/\nu_{0}\right)^{\beta}$
is the frequency dependent opacity of the dust \citep{Hildebrand1983}.
In the Rayleigh-Jeans regime, this simplifies to $F_{\nu}=\Omega\cdot2\nu^{2}k_{B}Tc^{-2}\cdot Q_{0}\left(\nu/\nu_{0}\right)^{\beta}\propto\nu^{2+\beta}.$}, which in turn depends on the dust composition and grain size distribution.
In the following we investigate which emission mechanisms are most
likely to contribute to the submillimetre continuum emission in the
nucleus of the Circinus galaxy.
\begin{table}
\caption{ALMA measurements used to complement the spectral energy distribution
in Fig.~\ref{fig:sed}, in addition to the data from \citet{Prieto2010}.
Columns are frequency $\nu$, wavelength $\lambda$, beam size of
the ALMA observations, flux per beam and reference for the measurement.\label{tab:SED-fluxes}}

\begin{tabular}{ccccc}
\hline 
$\nu$ {[}$\mathrm{GHz}${]} & $\lambda$ {[}$\mathrm{mm}${]} & beam {[}$\mathrm{mas}^{2}${]} & $F_{\nu}$ {[}mJy{]} & Reference\tabularnewline
\hline 
\hline 
134.5 & 2.23 & $56\times41\,$ & $15.0\pm1.9$ & 1\tabularnewline
145.2 & 2.06 & $52\times37$ & $14.1\pm1.6$ & 1\tabularnewline
321 & 0.93 & $700\times500$ & $40.6\pm4.1$ & 2\tabularnewline
345.4 & 0.87 & $190\times120$ & $38\pm6$ & 3\tabularnewline
351 & 0.85 & $290\times240$ & $22.4\pm2.3$ & 4\tabularnewline
485 & 0.62 & $710\times650$ & $87.8\pm8.8$ & 4\tabularnewline
691.5 & 0.43 & $110\times70$ & $142\pm43$ & 3\tabularnewline
\hline 
\end{tabular}

\tablebib{(1) Wang et al., in prep.; (2) \citet{Hagiwara2013}; (3)
this work (see Table~\ref{tab:nuclear-properties}); (4) \citet{Izumi2018}.}
\end{table}
\begin{figure*}
\begin{minipage}[b]{12cm}%
\centering

\includegraphics[clip,width=12cm]{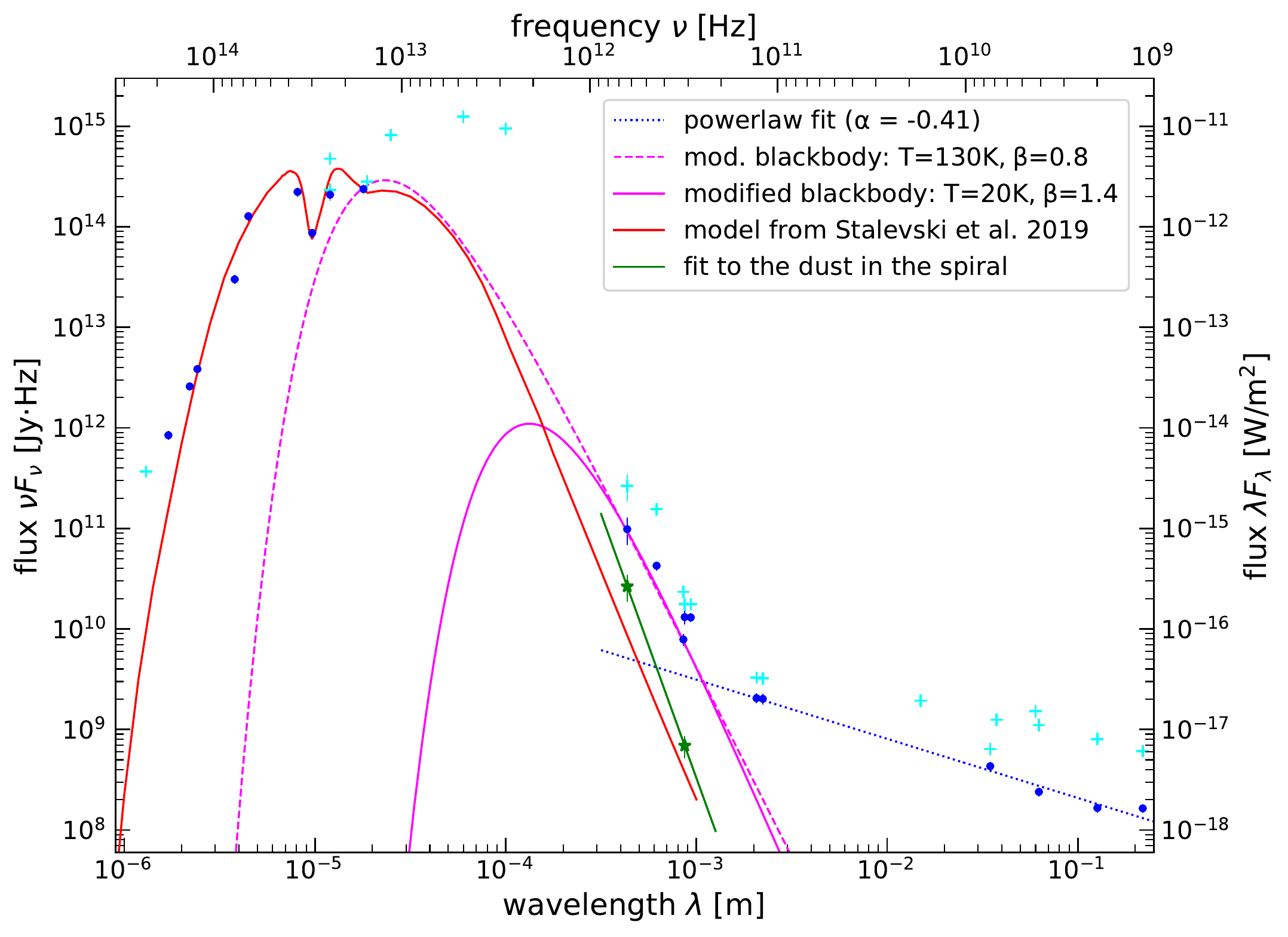}%
\end{minipage}\hfill{}%
\begin{minipage}[b]{6cm}%
\caption{Nuclear spectral energy distribution (SED) of the Circinus galaxy.
Blue points are high resolution photometry compiled by \citet{Prieto2010}
plus additional measurement with ALMA (see Table~\ref{tab:SED-fluxes}).
Also shown are low resolution measurements (cyan crosses). A power-law
fit to the radio continuum data (blue dashed line) yields a slope
of $\alpha=-0.41\pm0.04$. Modified blackbody emitters for warm dust
($T=130\,\mathrm{K}$ and $\beta=\alpha-2=0.8$, dashed line) and
cold dust ($T=20\,\mathrm{K}$ and $\beta=1.4$, continuous line)
are shown in magenta. Also plotted is the SED of the model from \citet{Stalevski2019}.
We also show for comparison flux measurements and a power-law fit
($\alpha\sim4.2$) for the emission in the head of the north-eastern
spiral arm (dark green). \label{fig:sed}}
\end{minipage}
\end{figure*}

\subsubsection{Continuum ratio map\label{subsec:continuum-ratio-map}}

To investigate the origin of the continuum emission seen by ALMA in
the nucleus of the Circinus galaxy, we create a map of the flux ratio
between the continuum emission at $691\,\mathrm{GHz}$ and $345\,\mathrm{GHz}$,
after smoothing and regridding the higher resolution $691\,\mathrm{GHz}$
map to the resolution of the $345\,\mathrm{GHz}$ map. This $F_{691\,\mathrm{GHz}}/F_{345\,\mathrm{GHz}}$
map is displayed in Fig.~\ref{fig:band9to7ratio}. For the extended
emission at $r>6\,\mathrm{pc}$ from the nucleus, we obtain a flux
ratio of $10<F_{691\,\mathrm{GHz}}/F_{345\,\mathrm{GHz}}<20$, which
corresponds to spectral indices of $3.3<\alpha<4.3$. This very steep
(inverted) slope strongly suggests that this emission is the Rayleigh-Jeans
tail of warm dust. The spectral slope corresponds to a dust emissivity
index of $\beta\sim2$. This is the `classical' value for dust in
the interstellar medium, following for example the `Chicago assumptions'
\citep{Hildebrand1983} or the calculations of \citet{Draine1984}.

In the nucleus, on the other hand, significantly lower ratios of $F_{691\,\mathrm{GHz}}/F_{345\,\mathrm{GHz}}\sim4$
are measured. This corresponds to a spectral index of $\alpha\sim2$.
The ratio and hence the spectral index might be somewhat higher, considering
that not all flux at $691\,\mathrm{GHz}$ may be concentrated at the
centre due to remaining phase residuals in the band 9 data. Nevertheless,
the very different spectral index in comparison to those at larger
distances indicates that the emission from the nucleus may be due
to a different mechanism, at least partially.

\subsubsection{Spectral energy distribution}

To investigate this possibility further, we complement the high resolution
spectral energy distribution (SED) of the nucleus compiled in \citet{Prieto2010}
with new measurements at submillimetre wavelengths with ALMA. These
new measurements are listed in Table~\ref{tab:SED-fluxes}. In all
cases we used the flux from the central beam, for which the size is
also indicated in Table~\ref{tab:SED-fluxes}. The SED is displayed
in Fig.~\ref{fig:sed}. Also given for reference (blue crosses) are
ALMA fluxes integrated in larger apertures ($>1\,\mathrm{arcsec}$)
or lower resolution data, such as the low resolution measurements
with IRAS.

The SED is dominated by two regions: For $\lambda\lesssim2\,\mathrm{mm}$
the emission is dominated by a large bump. Towards longer wavelengths
($\lambda\gtrsim2\,\mathrm{mm}$), the emission is dominated by a
power-law.

The large bump at $\lambda\lesssim2\,\mathrm{mm}$ is caused by thermal
emission from dust grains heated by the AGN. Fig.~\ref{fig:sed}
includes the SED of the dusty disk and hyperboloid cone model from
\citet[red continuous line]{Stalevski2019}. This model was created
to describe the SED and interferometric measurements in the infrared,
that is the leftmost ($\lambda<3\times10^{-5}\,\mathrm{m}$) data
points in Fig.~\ref{fig:sed}. Our ALMA measurements are well above
the Rayleigh-Jeans part of that model, indicating that additional
emission from cooler dust must be present. Fig.~\ref{fig:sed} includes
two modified blackbody emitters to represent such an additional component
of dust: first, a warm emitter with $T=130\,\mathrm{K}$ and $\beta=0.8$
(dashed magenta line); second, a cool dust component with $T=20\,\mathrm{K}$
and $\beta=1.4$ (continuous magenta line). There is a degeneracy
between the temperature $T$ and the dust emissivity index $\beta$
when fitting far-infrared and submillimetre SEDs using a modified
blackbody \citep[see e.g.][]{Shetty2009}, even more so when only
considering our two submillimetre continuum measurements. It is therefore
not possible to put tighter constraints on the temperature and dust
emissivity for such an additional dust component. However, as can
be seen from Fig.~\ref{fig:sed}, a component of dust with $T>130\,\mathrm{K}$
would be inconsistent with the measurement at $20\,\mathrm{\mu m}$
and can hence be ruled out. On the other hand, we do not expect the
dust in the vicinity of the AGN to be much cooler than $T\sim20\,\mathrm{K}$.
Therefore, we expect any additional dust to have $20\,\mathrm{K}\lesssim T\lesssim130\,\mathrm{K}$.
This implies values of $0.8\lesssim\beta\lesssim1.4$. Such low values
may either be caused by very different dust properties in the nucleus,
such as large grain sizes. However, there is no other evidence for
`abnormal' dust in the nucleus of the Circinus galaxy; on the contrary,
to the first order the dust is found to be similar to that in the
interstellar medium of the Milky Way \citep{Duy2019}. Therefore,
these values are more likely a result of contamination by emission
other than dust to the flux at $345\,\mathrm{GHz}$, leading to an
artificially low spectral index. 

At longer wavelengths ($\lambda\gtrsim2\,\mathrm{mm}$), the emission
is dominated by a power-law with a slope $\alpha=-0.41\pm0.04$. For
this fit, we used the $\lambda=2\,\mathrm{mm}$ ALMA measurement from
Wang et al., in prep., as well as the four measurements with the Australia
Telescope Compact Array (ATCA) between $\lambda=3$ and $20\,\mathrm{cm}$
by \citet{Elmouttie1998b}. Our spectral index is somewhat in between
those determined by \citet{Elmouttie1998b} for the `core' and the
`nucleus' of the Circinus galaxy using the ATCA data alone, $\alpha=-0.06\pm0.15$
and $\alpha=-0.65\pm0.01$, respectively. We note that the ATCA measurements
are from significantly larger apertures than the ALMA measurements.
Moreover, due to their increasing apertures towards longer wavelengths,
they include increasing flux components from the surrounding radio
structures. We therefore expect our fitted value to rather be a lower
limit. Both free-free absorption by thermal gas as well as a compact
synchrotron core with synchrotron self-absorption have been put forward
to explain the Circinus radio emission. Considering the spectral slope
and that Circinus does not possess a strong radio jet, we favour the
interpretation that the emission is mainly due to thermal free-free
emission, similar to what has been found for \object{NGC~1068} \citep{Gallimore2004b,Impellizzeri2019}.

\subsubsection{Non-dust contribution at 345 GHz}

Whatever the exact origin of the radio emission may be, it means that
there is a significant contribution of non-dust emission to the nuclear
emission at $345\,\mathrm{GHz}$. Using our power-law fit, we estimate
a contribution of $\gtrsim10\,\mathrm{mJy}$ to the flux at $345\,\mathrm{GHz}$,
that is of the order of $40\%$, at least $25\%$. This non-dust contribution
can explain the shallower spectral index in comparison to the dust
emission from the circumnuclear disk as well as why we measure a more
compact size for the nuclear emission at $345\,\mathrm{GHz}$ than
at $691\,\mathrm{GHz}$, as the size of this component is very compact
($<100\,\mathrm{mas}$ from the $\lambda=2\,\mathrm{mm}$ data, Wang
et al., in prep). Our result is in disagreement with the one of \citet{Izumi2018},
who estimated that the emission is dominated by dust emission alone.
These authors estimated the non-dust contribution by extrapolating
the measurement at $\lambda=3\text{\,\ensuremath{\mathrm{cm}}}$ from
\citet{Elmouttie1998b} using a power-law with $\alpha=-0.7$ assuming
the main mechanism to be synchrotron emission. However we have shown
that the slope is probably significantly flatter, leading to a stronger
contribution at $345\,\mathrm{GHz}$.

Also in several other nearby AGNs the continuum emission in the core
at $\sim350\,\mathrm{GHz}$ has been found to be either dominated
or at least significantly influenced by non-dust emission: as mentioned
above, the core of \object{NGC~1068} has been identified to be dominated
by free-free emission \citep{Gallimore2004b,Impellizzeri2019}; in
\object{NGC~7469}, the nuclear point source has a significant non-dust
contribution, probably from synchrotron emission \citep{Izumi2015,Izumi2020}.
For the NUGA sample of seven relatively low luminosity AGNs, the continuum
point source is expected to be synchrotron emission \citep{Combes2019},
and for three GATOS targets, \object{NGC~3227}, \object{NGC~5643}
and \object{NGC~7582}, spectral indices between $0.3$ and $-1.4$
are found, which is seen as evidence for free-free or a combination
of dust and synchrotron emission in the unresolved cores \citep{GarciaBurillo2021}.
The Circinus Galaxy is hence in good company.

In summary, we conclude that the nuclear continuum emission at $691\,\mathrm{GHz}$
is dominated to $\gtrsim95\%$ by dust emission, while that at $345\,\mathrm{GHz}$
has a significant contribution (of the order of $40\%$) most likely
from free-free, or possibly from synchrotron emission. This additional
non-dust emission at $345\,\mathrm{GHz}$ can explain both the significantly
shallower (inverted) spectral slope at the nucleus compared to the
pure dust emission in the circumnuclear disk and the more compact
emission at lower frequencies. We note that if the free-free interpretation
is indeed correct, the plasma should produce detectable line emission.
A detailed model is, however, outside the scope of this paper and
will be discussed in Wang et al., in prep.

\subsection{Morphology of the dust distribution}

\begin{figure}
\includegraphics[width=1\columnwidth]{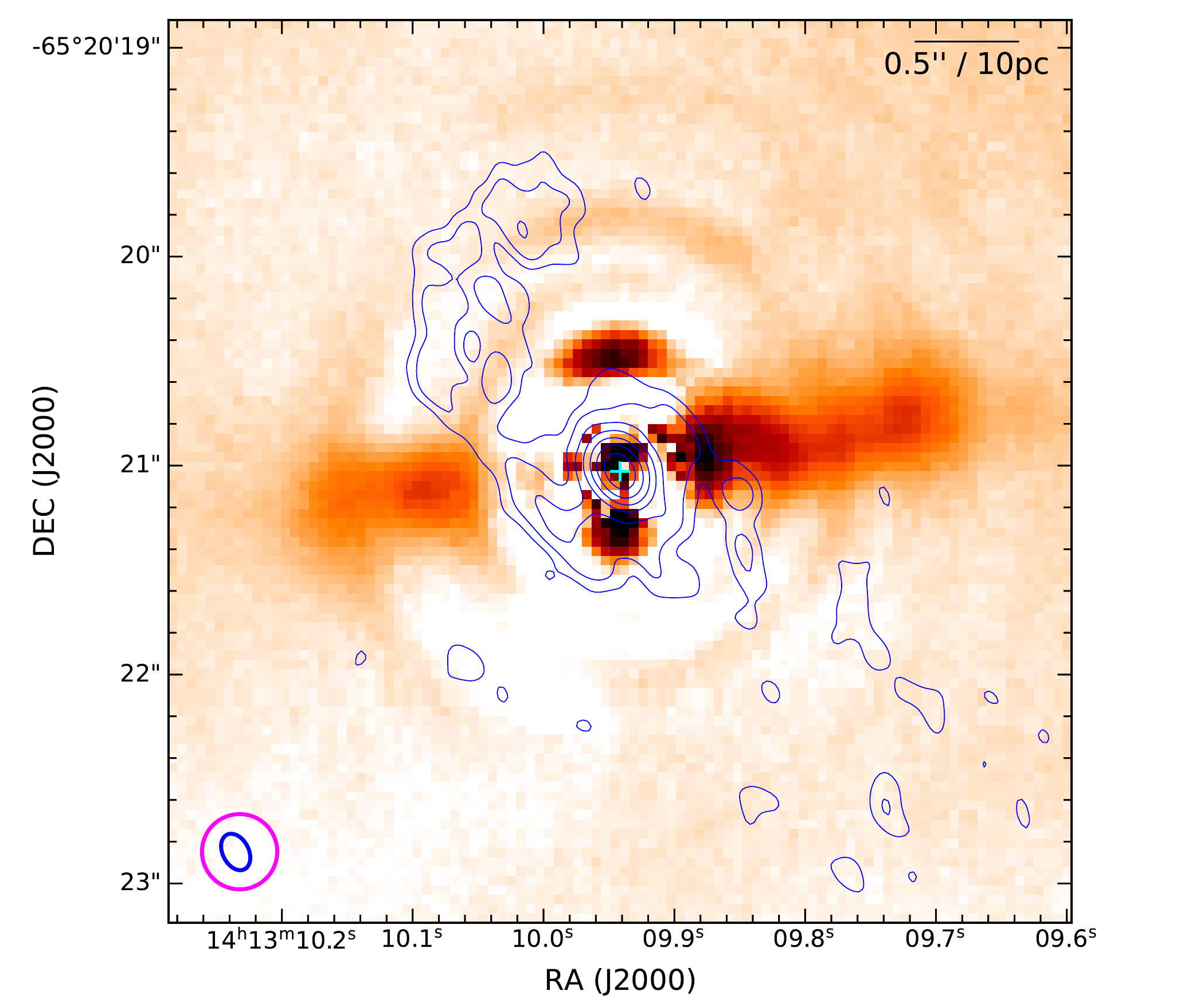}

\caption{Comparison of the polar extended warm dust emission at $\lambda=11.8\,\mathrm{\mu m}$
on scales of ten parsecs (background image with inverted colour map,
that is darker colour corresponds to brighter emission; from \citealt{Stalevski2017})
to the extended cold dust emission traced by ALMA ($345\,\mathrm{GHz}$
continuum, blue contours).\label{fig:comparison-visir-alma}}
\end{figure}
Mid-infrared interferometry and high resolution imaging has revealed
that the warm dust in AGNs on scales of several parsecs is extended
rather in the polar direction than in the equatorial direction (see
Sect.~\ref{sec:introduction}). The nucleus of the Circinus galaxy
is one of the sources showing the most prominent polar dust extensions.
With scales between $1$ and $20\,\mathrm{pc}$, the polar extension
covers the same spatial scales as our ALMA data and we therefore compare
the mid-infrared to the submillimetre continuum data. Fig.~\ref{fig:comparison-visir-alma}
shows the $\lambda=11.8\,\mathrm{\mu m}$ warm dust emission obtained
by \citet{Stalevski2017} after subtracting the point spread function
(PSF) of the strong unresolved core to better reveal the polar extension
out to $20\,\mathrm{pc}$. Also the core is to a large degree elongated
in the polar direction, as revealed by higher resolution mid-infrared
interferometric observations \citep{Tristram2014,Isbell2022}. The
polar elongation of the dust emission can be explained by a dusty
hollow cone illuminated by an inclined accretion disk \citep{Stalevski2017}.
Overplotted are the contours of the ALMA continuum emission at $345\,\mathrm{GHz}$
($868\,\mathrm{\mu m}$, the same as in Fig.~\ref{fig:continuum-map}).
Clearly there is no morphological resemblance between the two images
whatsoever and we see no indications for a polar extension of the
emission in our ALMA data. A possible polar extension of the continuum
emission at $345\,\mathrm{GHz}$ was claimed by \citet{Izumi2018}
in their lower resolution data. However we clearly do not see any
indications for a polar extension on our higher resolution data and
speculate that their fit, although done directly on the visibility
data, was affected by the beam. Moreover the polar emission extends
out to scales of $20\,\mathrm{pc}$ ($1\,\mathrm{arcsec}$) in the
mid-infrared, which would also be very well resolved in the ALMA data.

We conclude that in the submillimetre we are not just probing the
Rayleigh-Jeans tail of the dust emission peaking in the mid-infrared;
rather we must be probing two completely different dust components
in the mid-infrared and the submillimetre. We therefore paint the
following picture: The dust in the polar direction is relatively warm,
$T_{\mathrm{dust}}\sim300\,\mathrm{K}$, and its SED drops sharply
towards the submillimetre regime (cf.\ the red curve in Fig.~\ref{fig:sed}).
Furthermore, the emission has a relatively low surface filling factor
or column density so that its emission has dropped below our detection
limit with ALMA. This emission is coming from optically thin dust
in clouds within or at the edge of the outflow. However this dust
is not associated with a larger colder dust distribution (which we
would see with ALMA). The cold dust, $T_{\mathrm{dust}}\lesssim100\,\mathrm{K}$,
is mainly distributed in an equatorial plane, roughly following the
distribution of the molecular gas. This dust has a higher surface
filling factor, so that it dominates the emission at submillimetre
wavelengths. This material is most likely located in a turbulent disk
as obtained by recent hydrodynamical simulations \citep[e.g.][]{Wada2015}.
It is the material that is responsible for the bulk of the obscuration,
on scales of a few parsecs only.

\subsection{Kinematics\label{subsec:discuss_kinematics}}

\begin{figure*}
{\Large{}\includegraphics[width=0.49\textwidth]{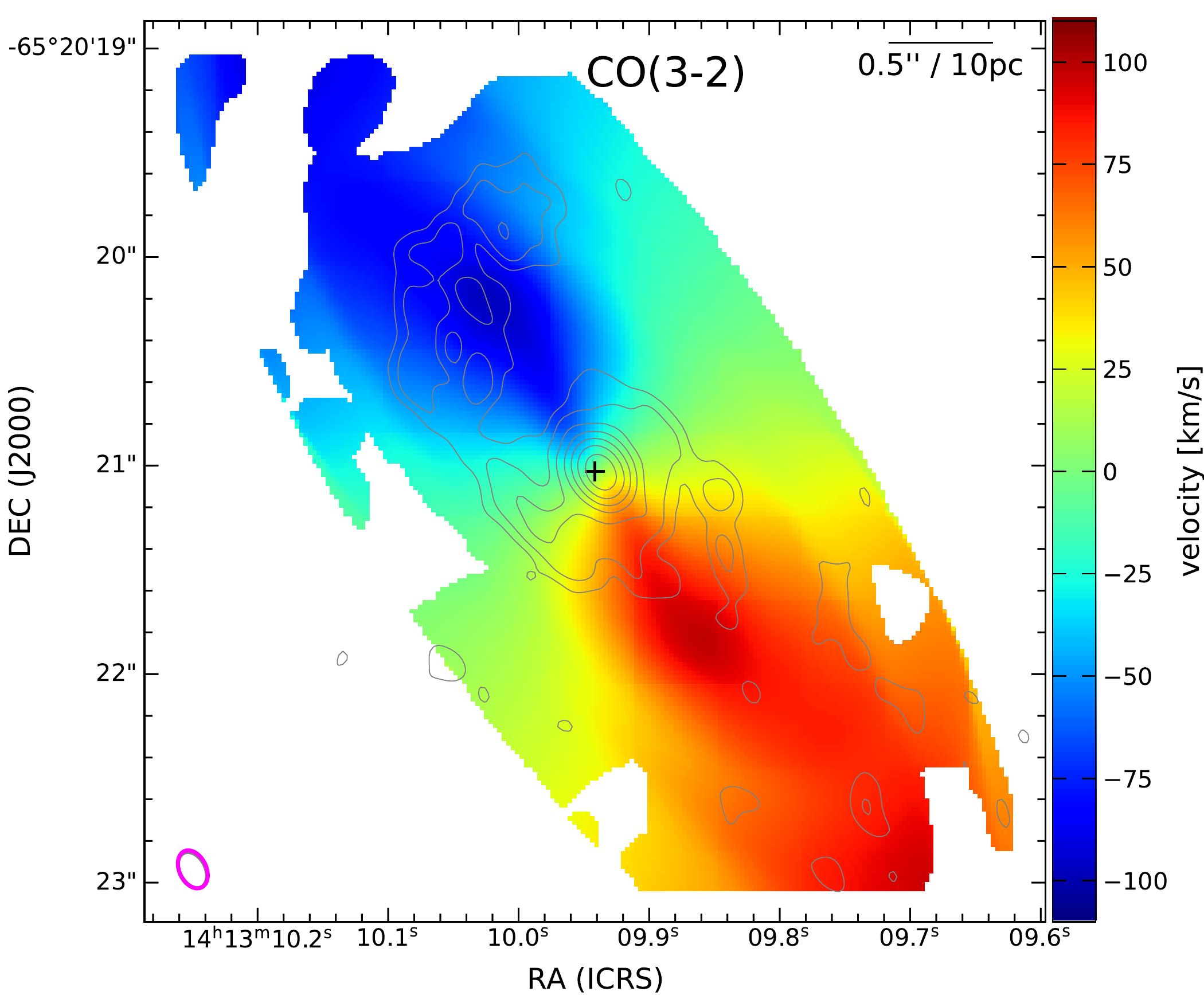}\hfill{}\includegraphics[width=0.49\textwidth]{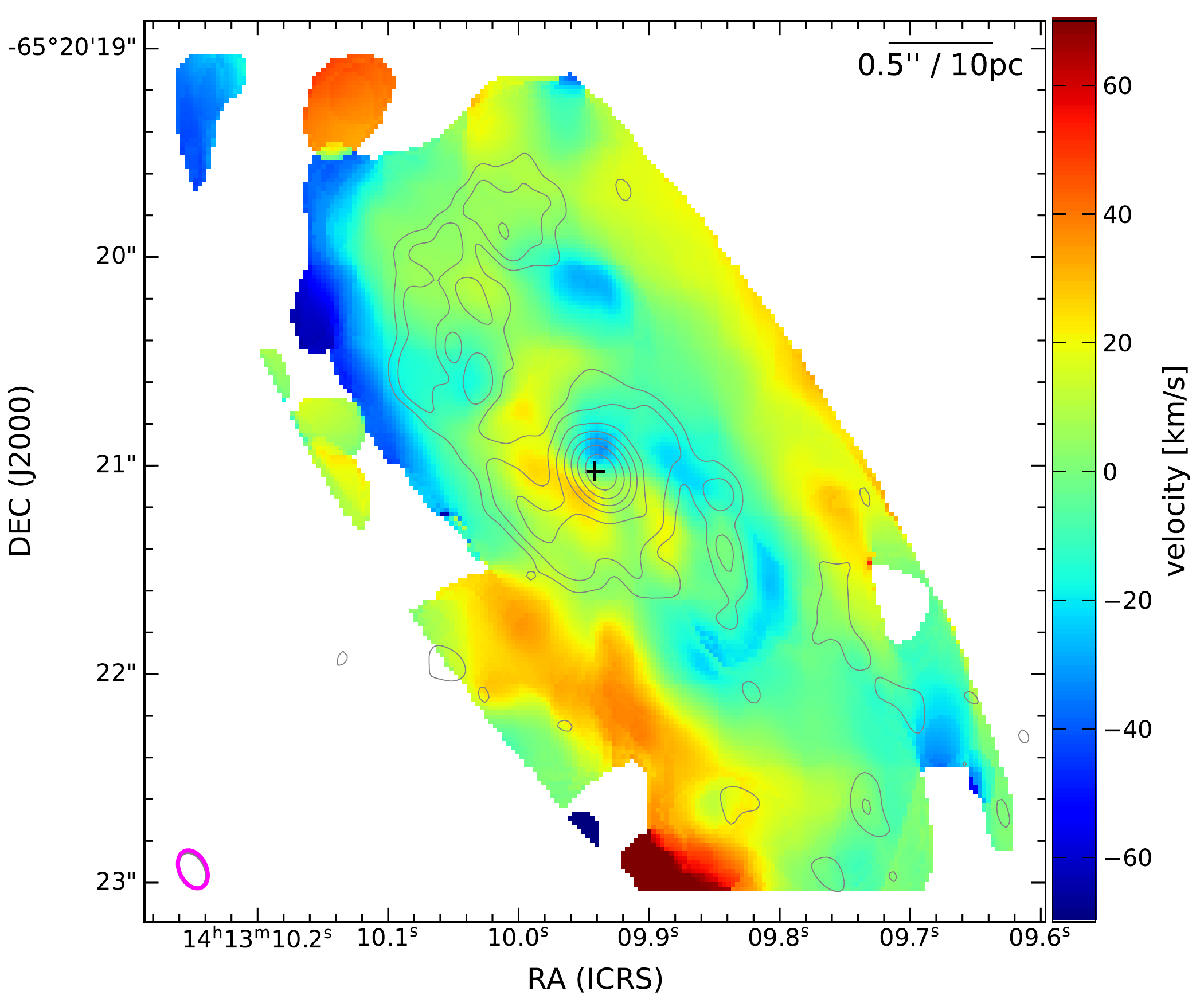}}{\Large\par}
\begin{centering}
{\Large{}\includegraphics[width=0.49\textwidth]{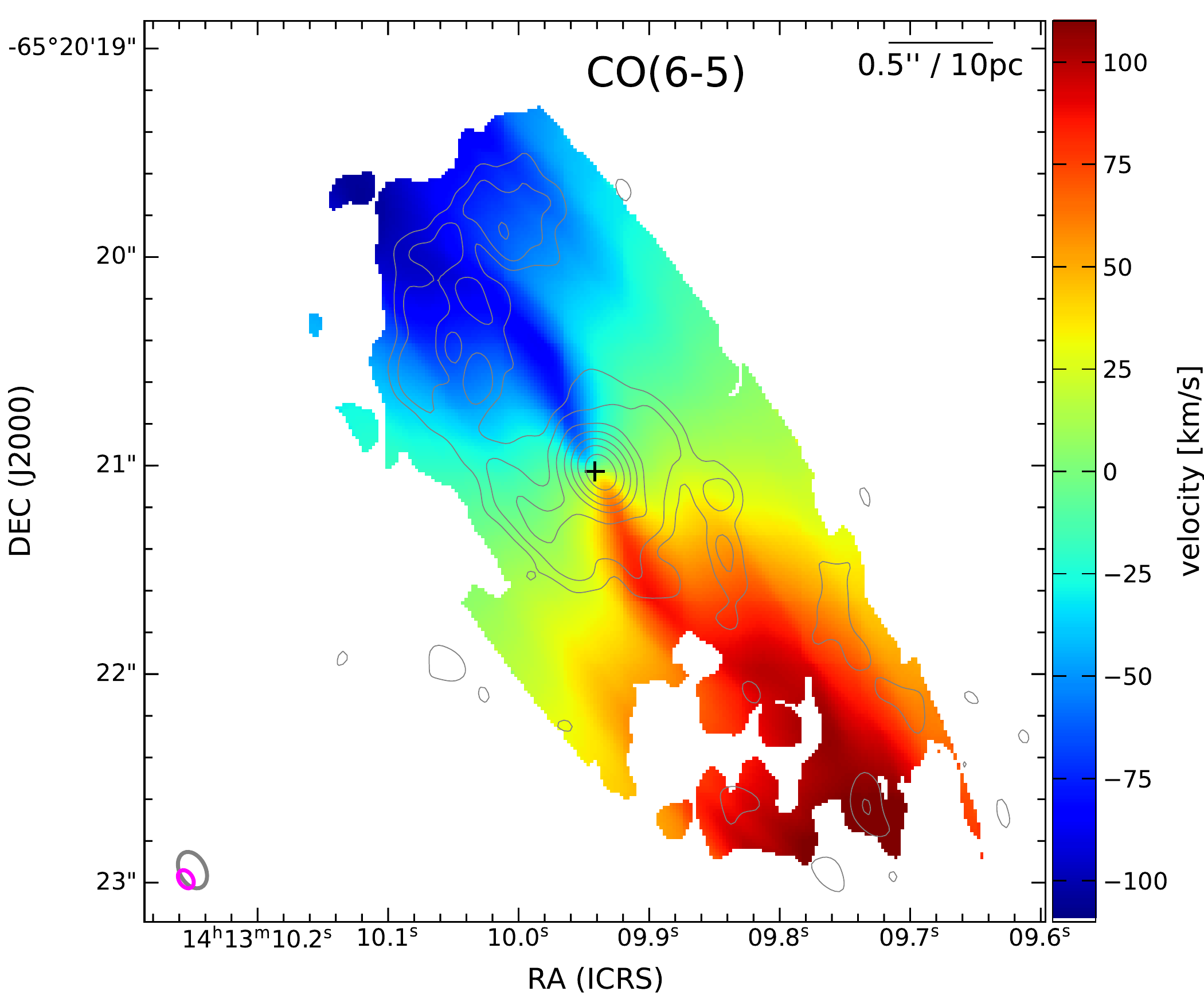}\hfill{}\includegraphics[width=0.49\textwidth]{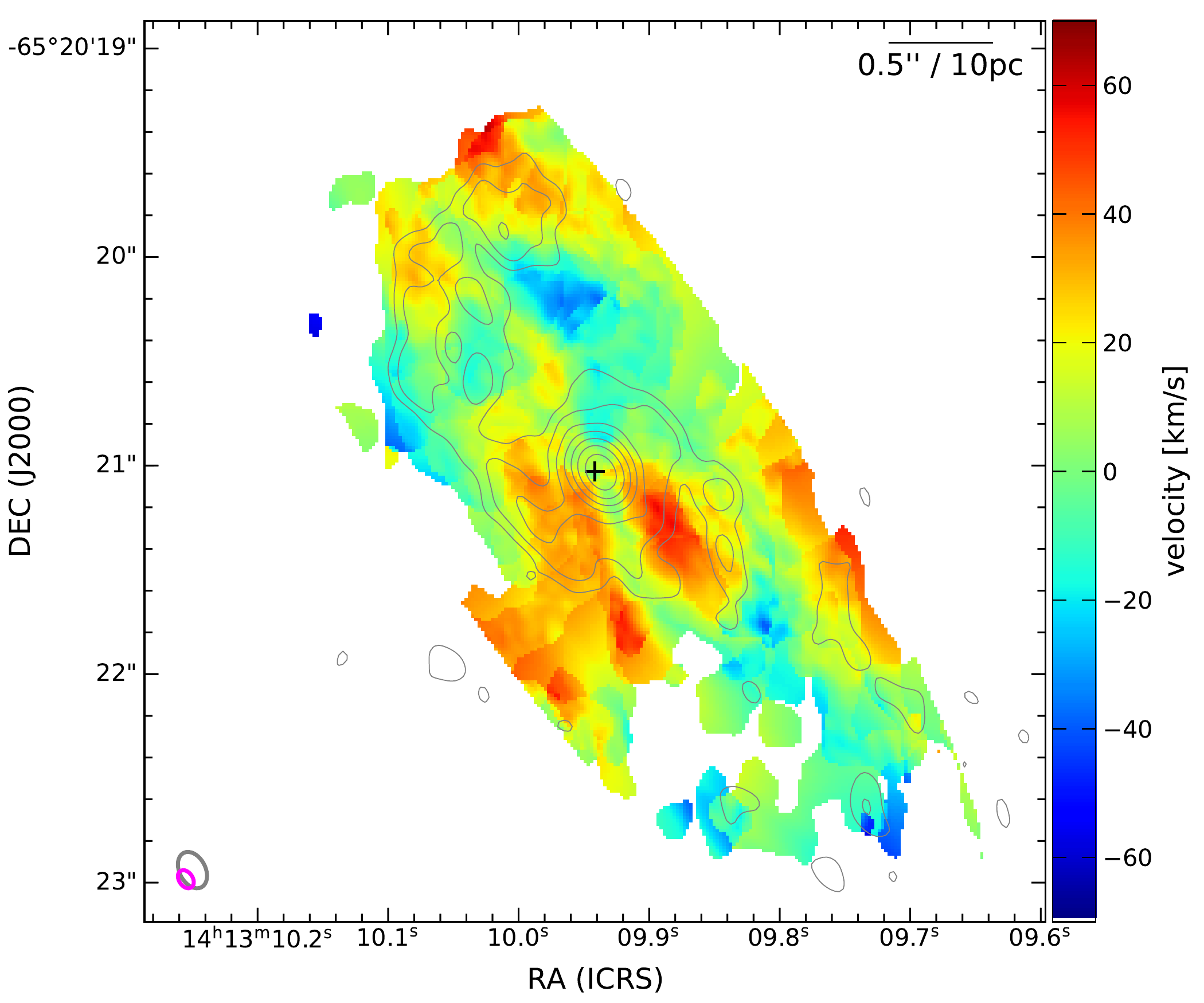}}{\Large\par}
\par\end{centering}
\begin{centering}
{\Large{}\includegraphics[width=0.49\textwidth]{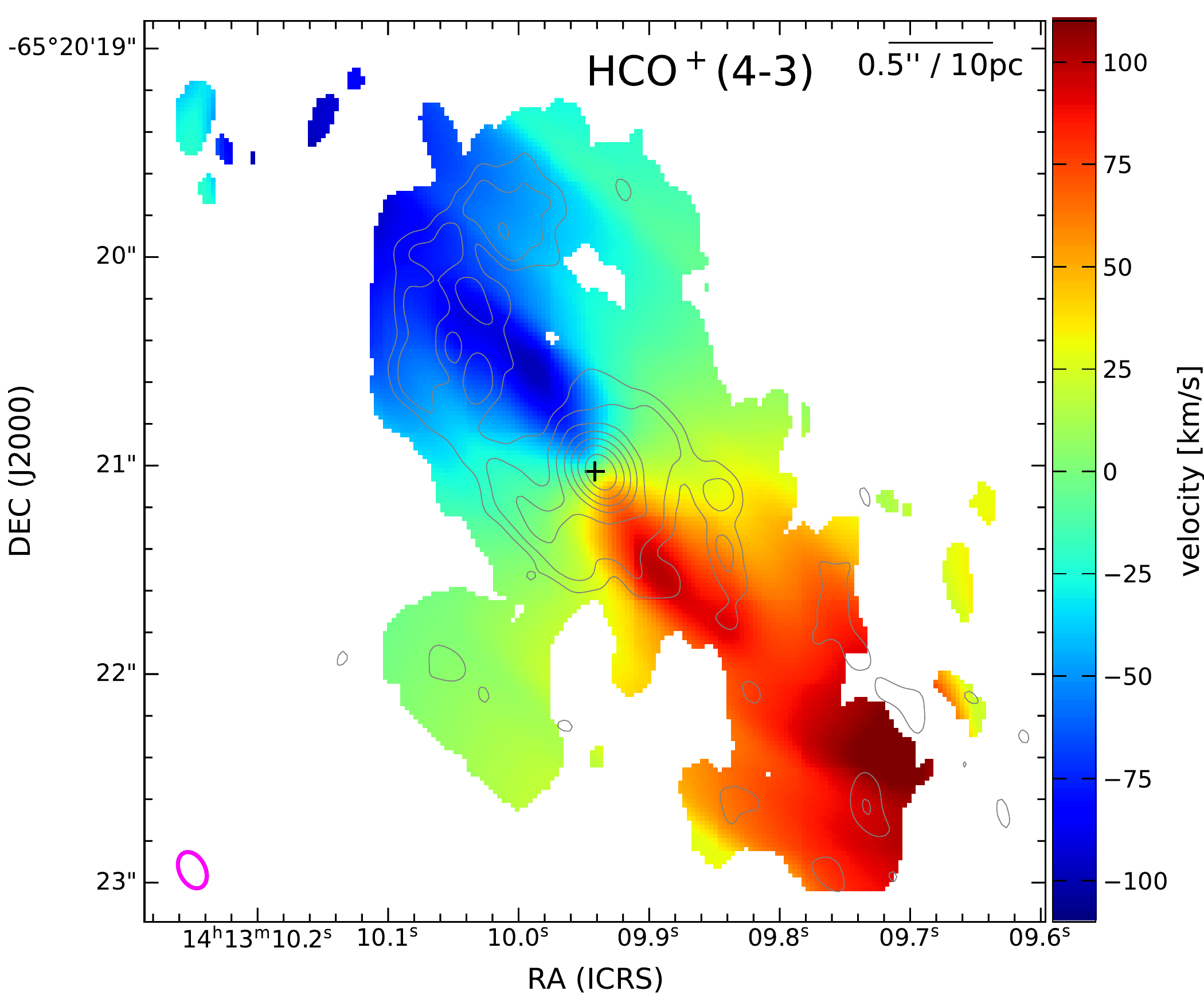}\hfill{}\includegraphics[width=0.49\textwidth]{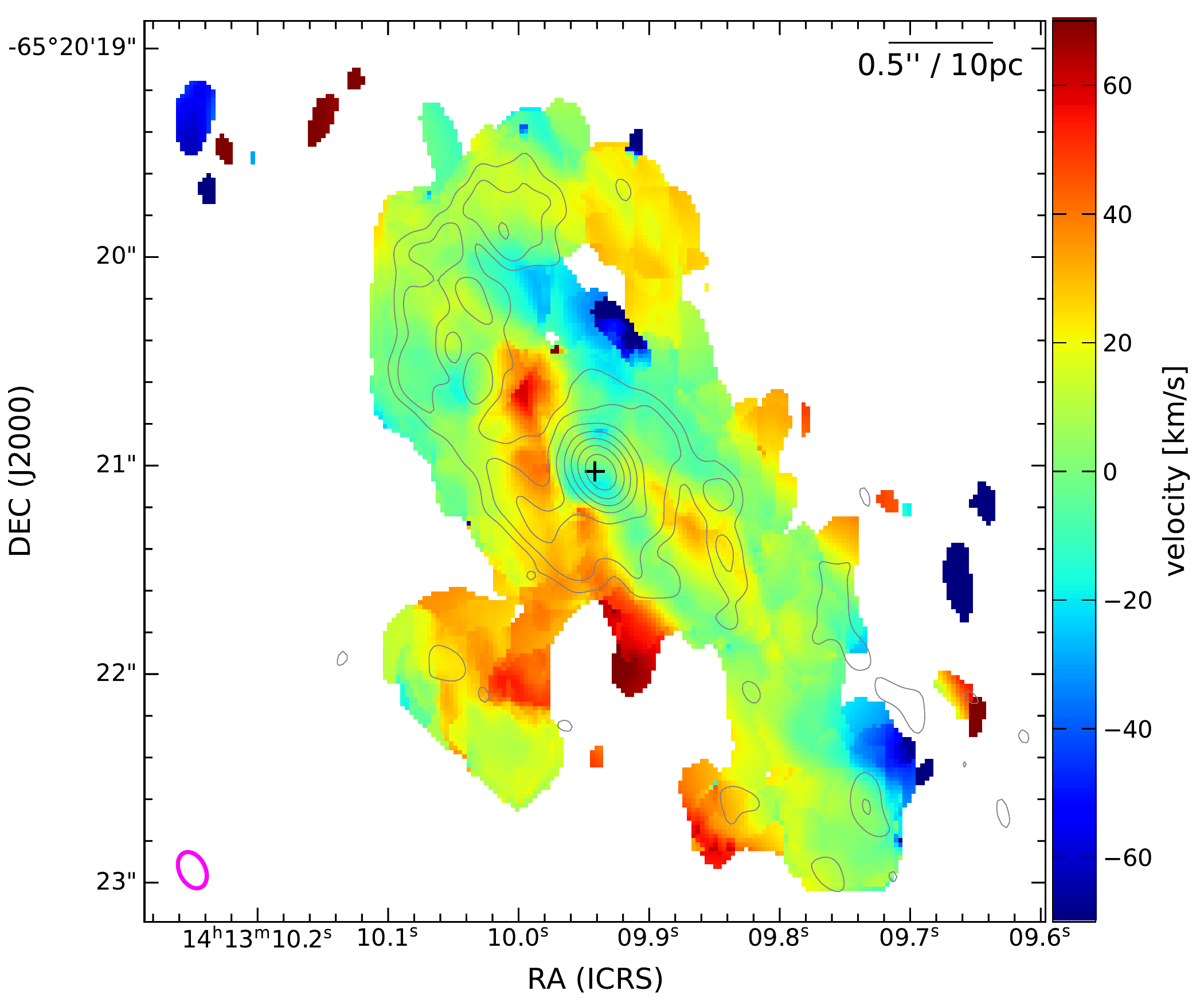}}{\Large\par}
\par\end{centering}
\caption{Kinematic modelling with $^{\mathrm{3D}}\mathrm{BAROLO}$ for CO(3$-$2)
(top row), CO(6$-$5) (middle row) and $\mathrm{HCO}^{+}$(4$-$3)
(bottom row). For each emission line the model velocity field (left
column), and the residual velocity field (right column) are shown.
\label{fig:3dbarfit-maps}}
\end{figure*}
\begin{figure*}
\includegraphics[width=1\textwidth]{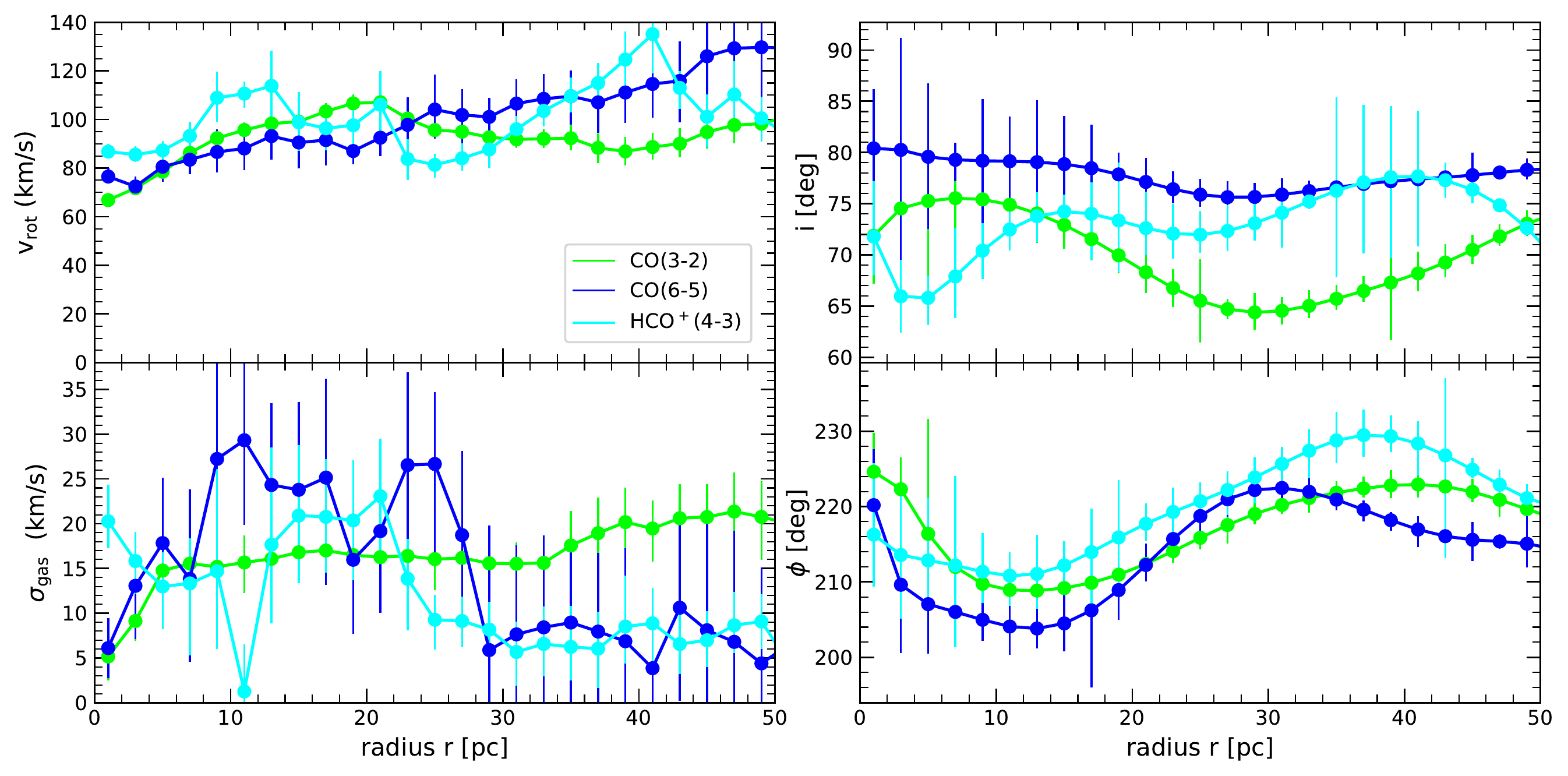}

\caption{Parameter results for the kinematic modelling with $^{\mathrm{3D}}\mathrm{BAROLO}$
for the three fitted emission lines: CO(3$-$2) (green), CO(6$-$5)
(blue) and $\mathrm{HCO}^{+}$(4$-$3) (cyan). \label{fig:3dbarfit-params}}
\end{figure*}
The kinematics of the nuclear material can give hints on the physical
mechanism that keeps the obscuring material geometrically thick. To
investigate this, we model the emission line kinematics for the three
transitions with the highest signal to noise and completeness, CO(3$-$2),
CO(6$-$5) and $\mathrm{HCO}^{+}$(4$-$3), using $^{\mathrm{3D}}\mathrm{BAROLO}$
\citep{DiTeodoro2015}. $^{\mathrm{3D}}\mathrm{BAROLO}$ fits 3D titled
ring models to spectroscopic data cubes of emission lines such as
those obtained with ALMA. With such a simple model, we aim to obtain
rough estimates of the kinematical properties of the gaseous material,
considering that the observed kinematics are clearly dominated by
rotation. We acknowledge that more complex kinematics, such as streaming
motions, lopsidedness and extra-planar gas, will not be captured by
this model. The following parameters were fitted for each ring: the
inclination angle $i$ ($90^{\circ}$ for edge-on); the position angle
$\phi$ of the major axis on the receding half of the ring, counted
anticlockwise from the north; the rotational velocity $v_{\mathrm{rot}}$;
and the velocity dispersion of the gas $\sigma$; that is in total
four parameters. The software also allows for a fit of the spatial
coordinates of the centre as well as for the systemic velocity $v_{\mathrm{sys}}$.
However we used the centre as defined by the continuum peaks at $345\,\mathrm{GHz}$
and $691\,\mathrm{GHz}$ (the latter being identical with the former
after correction for the shift between both bands) and set $v_{\mathrm{sys}}=0$
because our data cubes are already corrected for redshift (see Sect.~\ref{subsec:datareduction}).
Attempts to also fit for these parameters do not result in a significantly
improved fit, nor in a significant change of the parameter values
with respect to the fixed values. Due to the inhomogeneous line brightness
distribution, we chose to normalise the face-on surface density pixel
by pixel such that the line intensity maps of the model and the observations
are the same. This permits for a non-axisymmetric intensity distribution
and avoids that untypical regions, such as areas with strong and clumpy
emission or holes affect the global fit \citep[see][]{DiTeodoro2015}.
Furthermore, $\chi^{2}$ minimisation was used because it produces
significantly more consistent results between rings than the other
options.

The fitting was carried out in two steps. In the first step, all four
free parameters were fitted independently. In a second step, the inclination
and the position angle are regularised using a Bezier function and
only the rotation velocity and velocity dispersion are free parameters
for the final fit. Uncertainties on the parameters are estimated by
the software via a Monte Carlo method. The fits for the Circinus galaxy
turn out to be robust and consistent for variations of the radial
sampling (ring radii separated by $\delta r=0.1\,\mathrm{arcsec}$,
corresponding to $2\,\mathrm{pc}$) and of the initial parameter guesses.

Our fitted velocity maps and velocity residuals are shown in Fig.~\ref{fig:3dbarfit-maps}
and the radial dependency of the parameters is displayed in Fig.~\ref{fig:3dbarfit-params}.
We find the four fitted parameters to be consistent within uncertainties
for all three emission lines, confirming that they trace the same
general kinematical structure dominated by rotation. The rotational
velocity increases very slowly from $v_{\mathrm{rot}}=80\,\mathrm{km}\,\mathrm{s}^{-1}$
at the centre to $v_{\mathrm{rot}}=110\,\mathrm{km}\,\mathrm{s}^{-1}$
at a radius of $50\,\mathrm{pc}$. We find a relatively constant inclination
of the rings of $i=75\pm5^{\circ}$, which is in between the lower
inclinations found on larger, and the almost edge-on disk found on
smaller scales \citep[and references therein]{Curran2008}. The residual
maps as well as direct comparison of the model cubes to the measured
data reveal deviations from the rotation pattern up to $\Delta v=30\,\mathrm{km}\,\mathrm{s}^{-1}$
in a patchy structure, indicating significant non-circular motions
in the circumnuclear disk. Including a radial velocity component in
the fit does not significantly reduce these residuals and the fitted
radial velocity component is consistent with $v_{\mathrm{rad}}=0\,\mathrm{km}\,\mathrm{s}^{-1}$.
This means the deviations are not caused by uniform in- or outflowing
motions on scales of the disk, consistent with the result found for
molecular hydrogen \citep{Hicks2009}. The residual bulk motions are
larger than the typical velocity dispersion of the ring fit model
and they are larger than the velocity dispersion of a single velocity
component in the data. We see this as evidence for the kinematics
of the gas on scales of a few tens of parsecs not being dominated
by the random motion of small clouds, but rather by in- or outflowing
motions of larger filamentary structures or clouds similar to what
is seen in hydrodynamical simulations of the torus \citep[e.g.][]{Schartmann2009,Wada2016}.

\subsubsection{Disk scale height}

We try to estimate the disk scale height for the molecular gas via
two methods. In the first we assume a self-gravitating (thin) disk,
for which the thickness of the disk is given as $h(r)=\sigma^{2}/\left(2\pi G\Sigma\right)$,
where $\Sigma$ is the mass surface density of the disk of the model
\citep[e.g][]{Hicks2009}. With Keplerian rotation $v_{\mathrm{rot}}=\left(GM(r)/r\right)^{0.5}$
and assuming a roughly homogeneous distribution of the material in
the disk, that is $\Sigma(r)=M(r)/\left(\pi r^{2}\right)$ we obtain
$h/r=0.5\sigma^{2}/v_{\mathrm{rot}}^{2}$ and hence $h/r<0.1$ for
the dense molecular material on scales of ten parsecs.

In the second method, we assume vertical hydrostatic equilibrium and
hence $h/r\sim\sigma/v_{\mathrm{rot}}$ for which we obtain a less
stringent limit of $h/r<0.3$. Both values are much too low to collimate
the ionising radiation to a cone with an opening angle of $\sim90^{\circ}$
for which $h/r\sim1$ would be required, that is $\sigma\gtrsim v_{\mathrm{rot}}$,
which is clearly not the case. We conclude that the collimation of
the ionising radiation is not carried out by the material in the circumnuclear
disk, but rather on scales $<5\,\mathrm{pc}$. Our results are in
rough agreement with those found by \citet{Izumi2018}, although we
find somewhat lower values for the velocity dispersion in our higher
resolution data and hence infer lower scale heights. Also our data
do not confirm the increase to $h/r>0.4$ at $5\,\mathrm{pc}$ seen
by \citet{Izumi2018}. Our molecular lines trace the (cold) dense
mid-plane of the disk. A much higher velocity dispersion, $\sigma>50\,\mathrm{km}\,\mathrm{s}^{-1}$,
is measured in molecular hydrogen, \citep{Hicks2009}, implying much
higher scale heights for the hot molecular gas component.

\begin{figure*}
{\Large{}\includegraphics[width=0.49\textwidth]{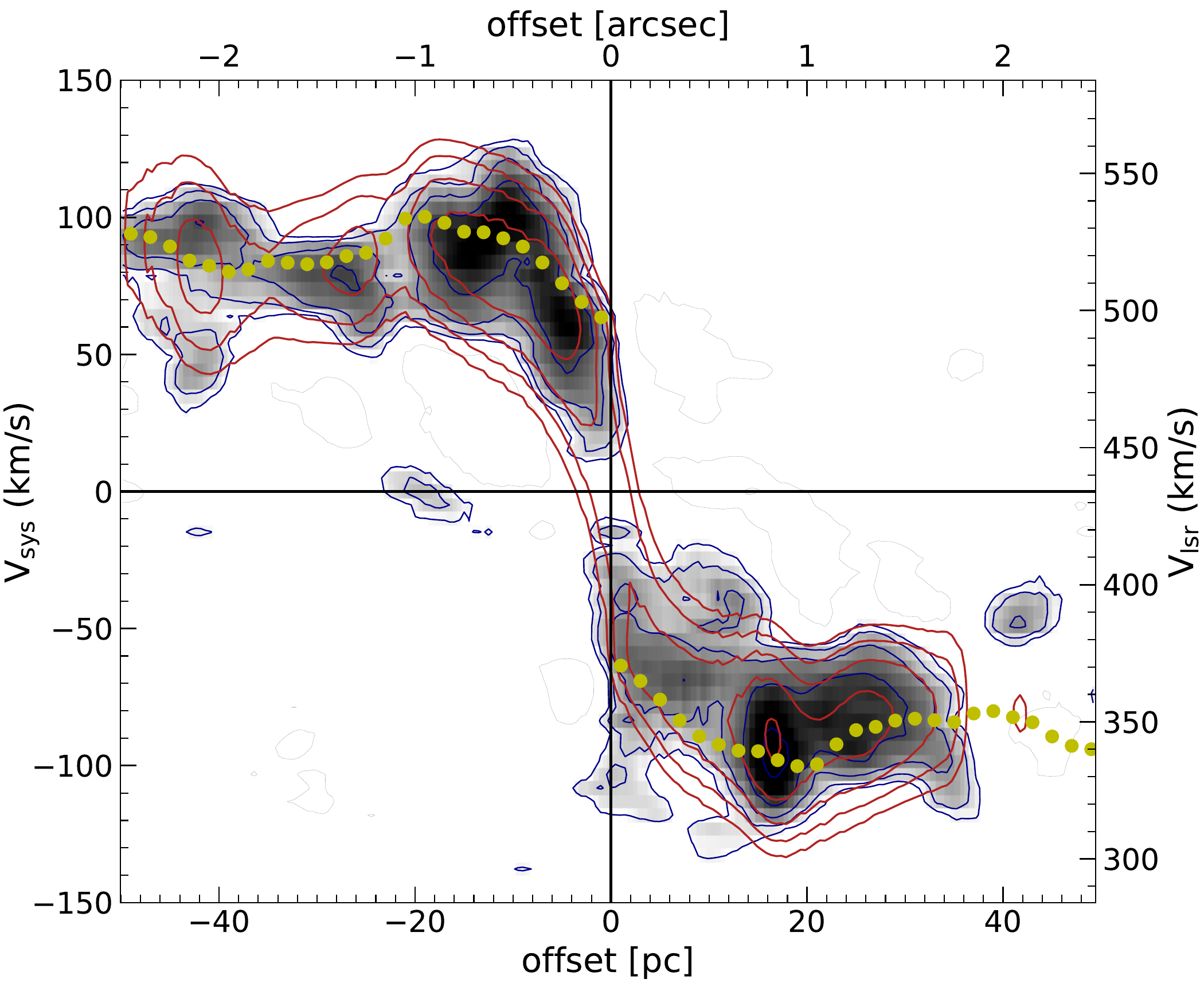}\hfill{}\includegraphics[width=0.49\textwidth]{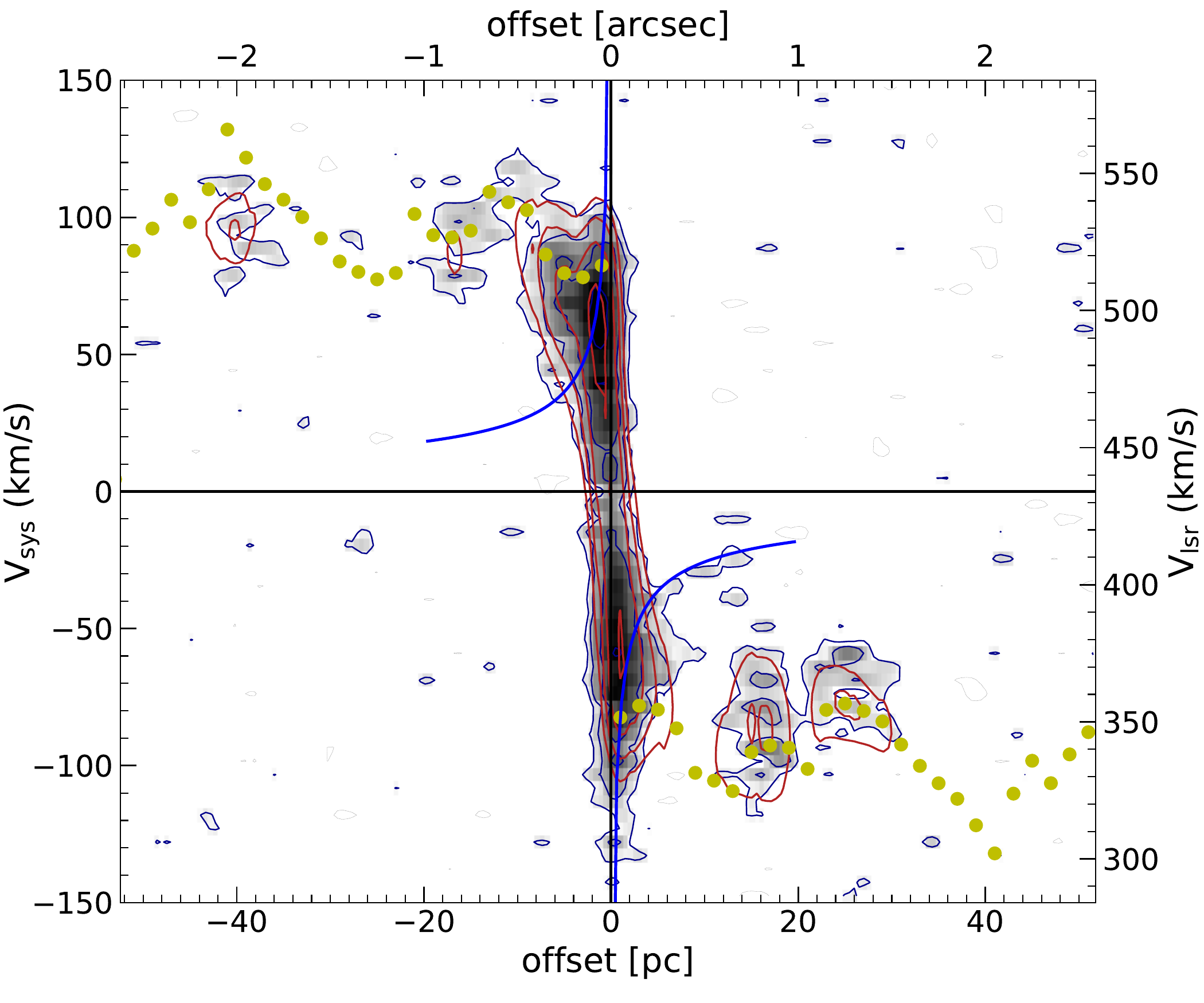}}{\Large\par}

\caption{Position velocity diagrams along the major axis ($\phi\sim220^{\circ}$)
for CO(3$-$2) (left) and $\mathrm{HCO}^{+}$(4$-$3) (right). The
data are shown in grey-scale and blue contours; the model with $^{\mathrm{3D}}\mathrm{BAROLO}$
is overplotted in red contours, the values of the modelled line-of-sight
velocities $v_{\mathrm{los}}=\sin i\times v_{\mathrm{rot}}$ are plotted
as yellow filled circles. For $\mathrm{HCO}^{+}$(4$-$3), the expected
Keplerian rotation curve for a $(1.7\pm0.3)\times10^{6}\,M_{\odot}$
black hole \citep{Greenhill2003a} is indicated in blue. \label{fig:3dbarfit-pvdiag}}
\end{figure*}

\subsubsection{Possible signs for a molecular outflow}

In the inner ${\sim}10\,\mathrm{pc}$ to the south east of the nucleus,
we find mainly redshifted residual velocities, while to the north-west
the residual velocities are mainly blue-shifted; best seen for CO(3$-$2),
top right panel of Fig.~\ref{fig:3dbarfit-maps}. This was already
noted in Sect.~\ref{subsec:results-kinematics} and may be a sign
for a molecular outflow in polar direction emanating from the nucleus.
The north-western side of this outflow would be slightly oriented
out of the plane of the sky towards the observer, similar to the ionisation
cone. However any signature of such an outflow is very weak and its
velocity, $v_{\mathrm{out}}<40\,\mathrm{km}\,\mathrm{s}^{-1}$, is
significantly lower than the rotational velocity. This is in line
with the only extremely weak detection of an outflow in CO(1$-$0)
on scales of $330$ to $650\,\mathrm{pc}$ \citep{Zschaechner2016}.
On the other hand, there are several signs for a stronger polar outflow
from the AGN in the Circinus galaxy: water masers trace a wide-angle
outflow out to $\sim1\,\mathrm{pc}$ \citep{Greenhill2003a}, \citet{Curran1999}
found an outflow in CO(2$-$1) with velocities up to $190\,\mathrm{km}\,\mathrm{s}^{-1}$
at $\pm500\,\mathrm{pc}$ from the nucleus along the rotation axis
and an outflow of ionised gas is clearly observed in the prominent
ionisation cone on scales between $2$ and $900\,\mathrm{pc}$ with
velocities of up to $v_{\mathrm{out}}=300\,\mathrm{km}\,\mathrm{s}^{-1}$
\citep[e.g.][]{Marconi1994,Veilleux1997,Fischer2013}. This discrepancy
can be explained, if the outflow on parsec scales is mainly in an
ionised gas phase and not in a molecular phase. We hence conclude
that cold molecular gas is not well suited to trace the outflow in
the Circinus galaxy on these scales.

This result is in contrast to those found for various other nearby
AGNs, where the kinematics show clear signs of molecular outflows
on pc scales, for example the high velocity molecular outflows or
jets in \object{NGC~1068} \citep{Gallimore2016,Impellizzeri2019}
and \object{NGC~1377} \citep{Aalto2016,Aalto2020}, or the AGN-driven
molecular outflows in \object{NGC~613} \citep{Audibert2019} and
\object{NGC~3227} \citep{AlonsoHerrero2019}. These outflows are
seen as evidence for the disk-wind scenario for the obscuring torus
in AGNs. Our results are more similar to those found for \object{NGC~4388},
\object{NGC~5506} and \object{NGC~5643}, where the kinematics are
dominated by rotation and only from the residual velocities or kinematic
fits there are much less clear signs for outflows \citep{GarciaBurillo2021,AlonsoHerrero2018}.
Similar to the case of the Circinus galaxy, the outflows are much
clearer identified in the ionised gas components than in the molecular
gas in these sources \citep{Fischer2013,RodriguezArdila2017}. 

\subsubsection{Self-absorption\label{subsec:self-absorption}}

\begin{figure}
\begin{centering}
{\Large{}\includegraphics[width=0.485\textwidth]{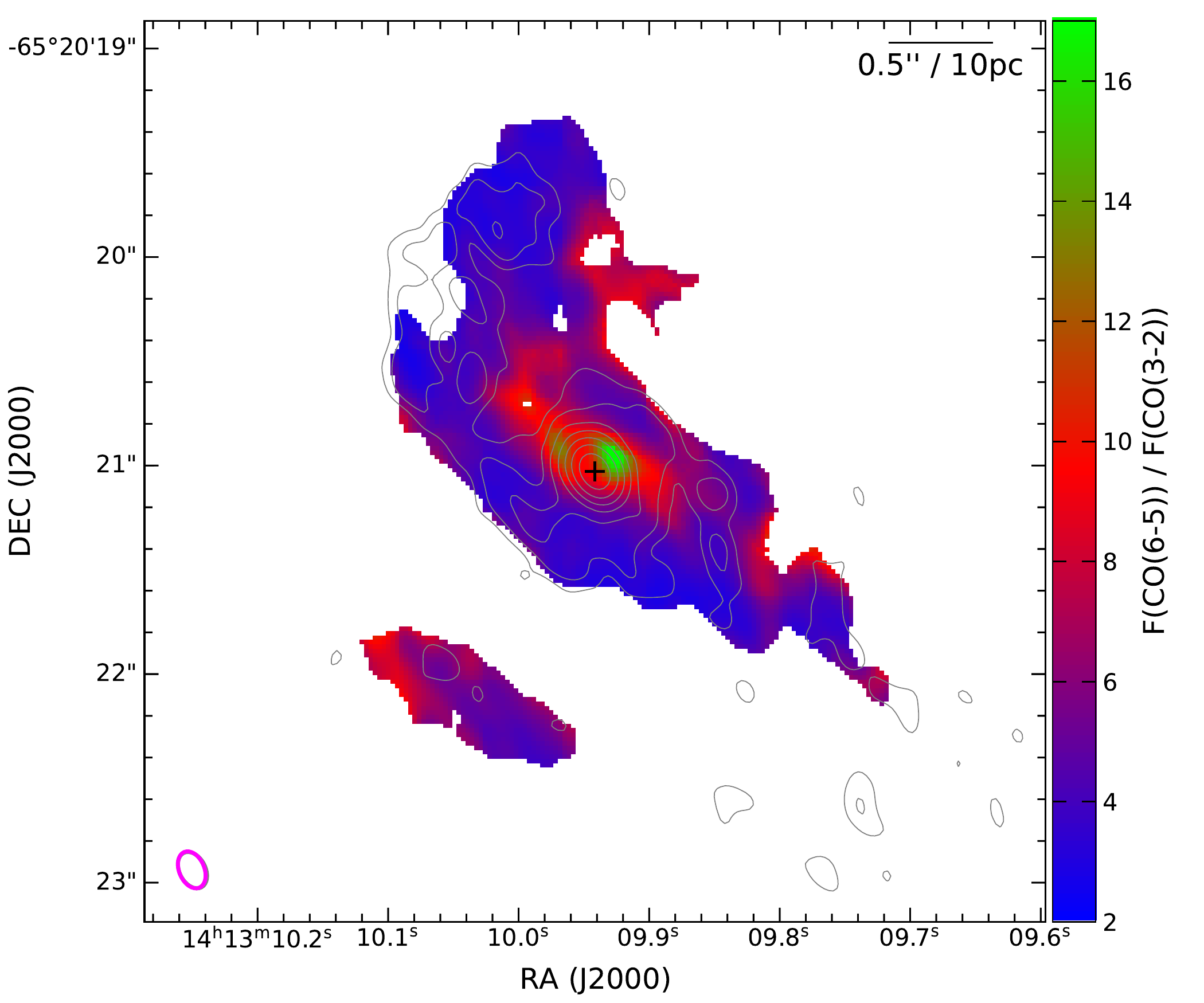}}{\Large\par}
\par\end{centering}
\begin{centering}
{\Large{}\includegraphics[width=0.485\textwidth]{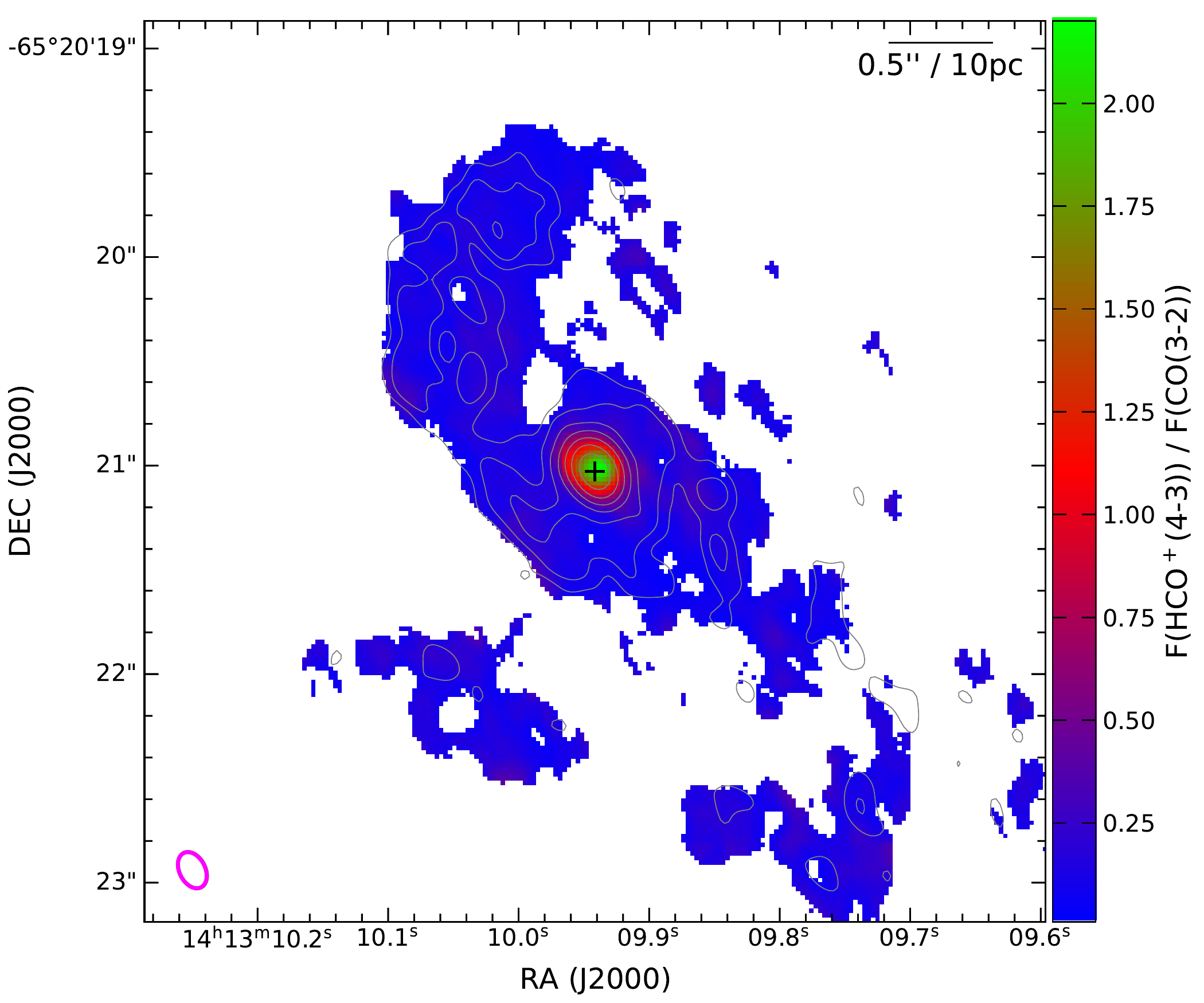}}{\Large\par}
\par\end{centering}
\begin{centering}
{\Large{}\includegraphics[width=0.485\textwidth]{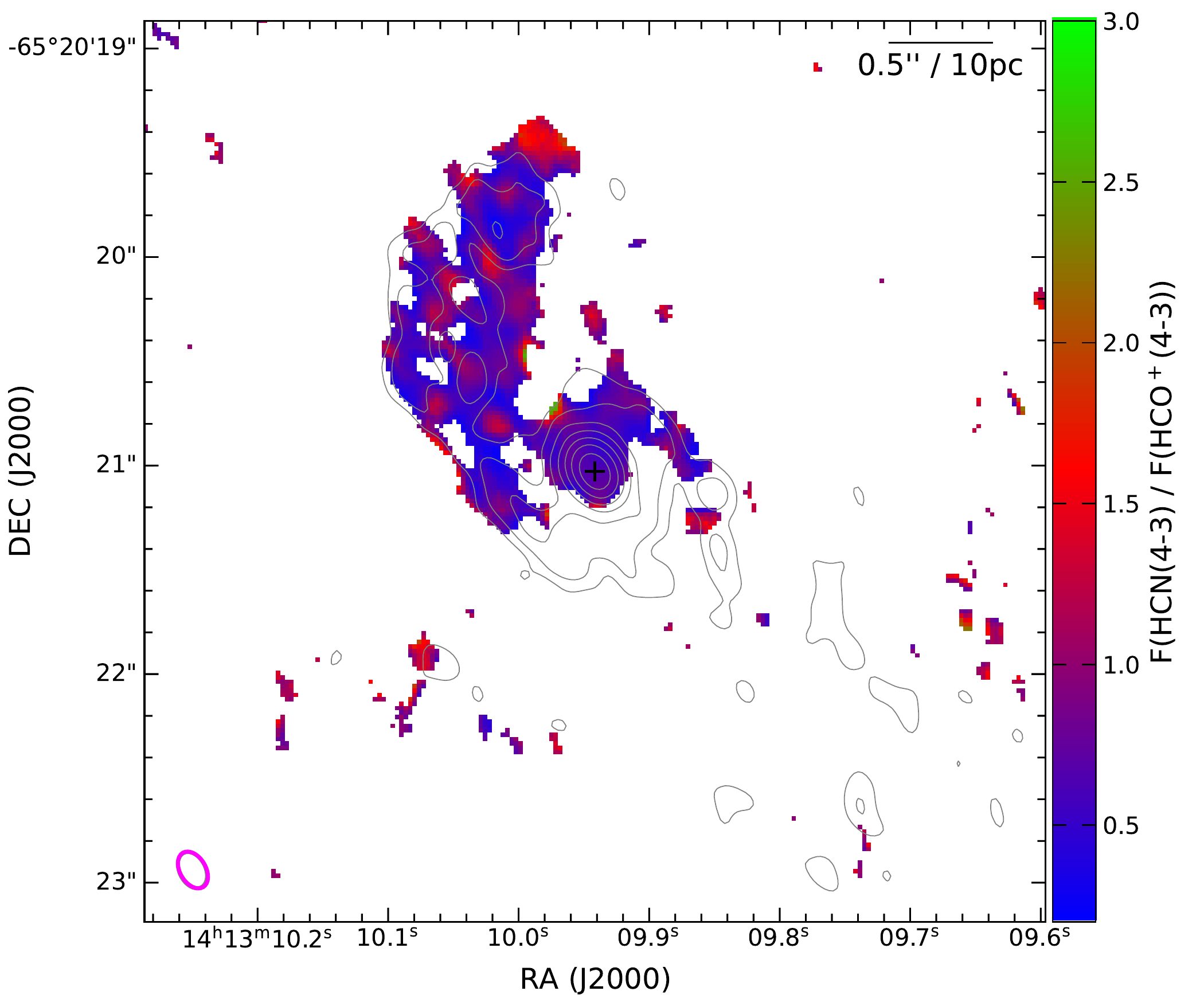}}{\Large\par}
\par\end{centering}
\caption{Emission line ratio maps for CO(6$-$5)/CO(3$-$2) (top panel), $\mathrm{HCO}^{+}$(4$-$3)/CO(3$-$2)
(middle panel) and HCN(4$-$3)/$\mathrm{HCO}^{+}$(4$-$3) (bottom
panel). For the CO(6$-$5)/CO(3$-$2) map, the CO(6$-$5) data were
smoothed to the resolution of the CO(3$-$2) data, while for the HCN(4$-$3)/$\mathrm{HCO}^{+}$(4$-$3)
map, the $\mathrm{HCO}^{+}$(4$-$3) was truncated to velocities $<15\,\mathrm{km}\,\mathrm{s}^{-1}$
in order to match the velocity coverage of the HCN(4$-$3) line.\label{fig:emission-line-ratios}}
\end{figure}
In Sect.~\ref{subsec:results-kinematics}, we found that the nuclear
spectrum has a double peaked profile, with varying depth of the dip
at systemic velocities and a broader profile for $\mathrm{HCO}^{+}$(4$-$3)
than for CO(3$-$2). The tilted ring fit completely fails to reproduce
this spectral profile at the centre, especially the dip at the systemic
velocity is not reproduced. Although often associated with a rotating
disk, a double horned spectral profile is in fact only obtained for
a truncated disk, which is clearly not the case for the Circinus nucleus.
We therefore see the observed profile as evidence for radiative transfer
effects and more precisely for significant amounts of self-absorption
\citep[see also][]{Elitzur2012} as well as absorption against the
bright background continuum \citep[e.g. ][]{Scoville2017}. Self-absorption
is strongest for CO(3$-$2), where we expect a lot of foreground material
in the galactic disk; it is less strong for higher gas density species
such as $\textrm{HCO}^{+}$. This can also be clearly seen in the
position velocity diagrams (see Fig.~\ref{fig:3dbarfit-pvdiag}),
where there is a clear hole in the CO(3$-$2) diagram at the position
of the nucleus and for zero velocity.

\subsubsection{Dynamical mass and black hole sphere of influence}

\begin{figure*}
\begin{minipage}[b]{12cm}%
\centering

\includegraphics[clip,width=12cm]{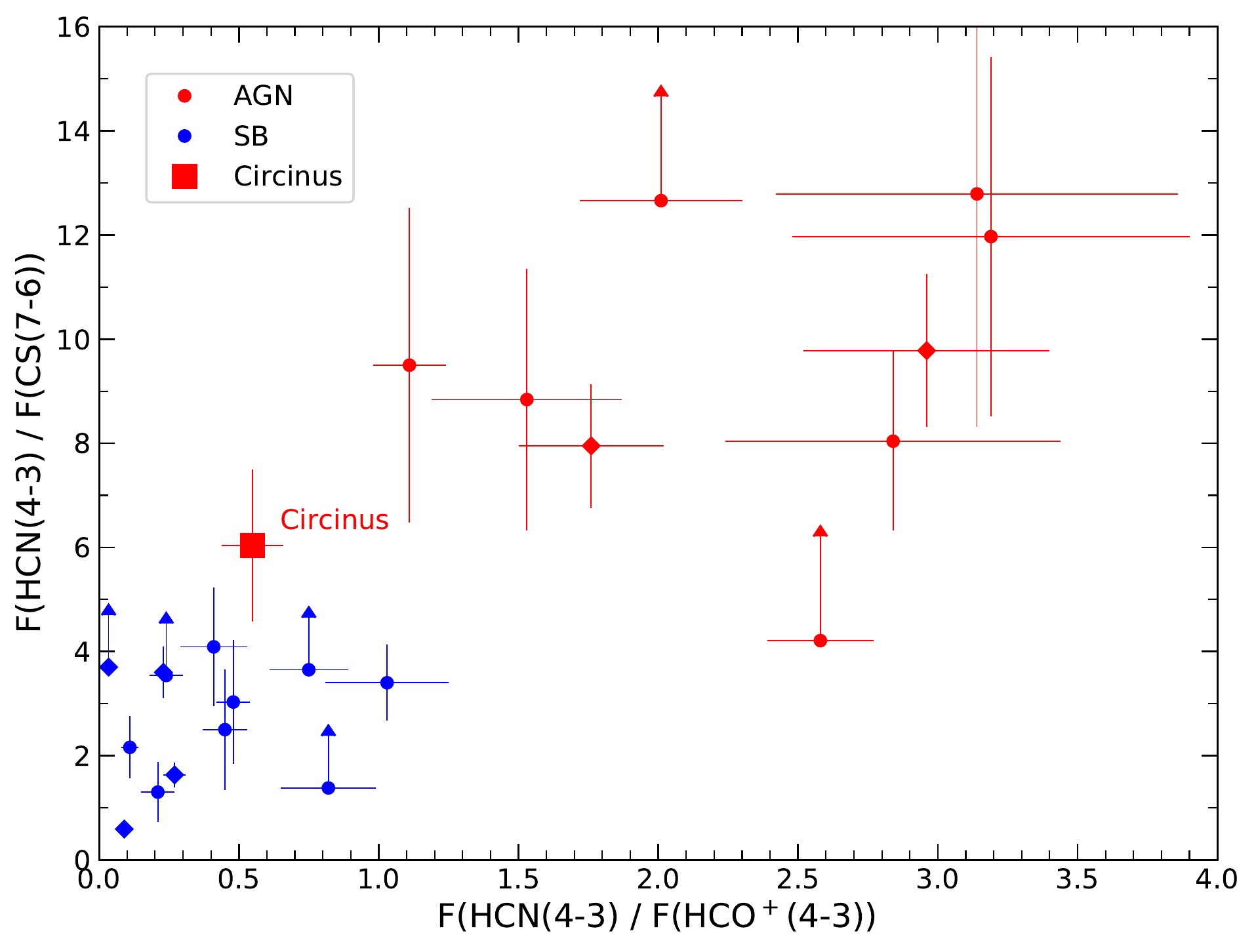}%
\end{minipage}\hfill{}%
\begin{minipage}[b]{6cm}%
\caption{High resolution (spacial resolution $<500\,\mathrm{pc}$) submillimetre
HCN diagram with the HCN(4$-$3) to CS(4$-$3) ratio plotted as a
function of HCN(4$-$3)/$\mathrm{HCO}^{+}$(4$-$3) following \citet{Izumi2016},
filled circles. Additional data points for \protect\object{NGC~613}
\citep{Audibert2019} and \protect\object{NGC~1808} \citep{Audibert2021}
are shown by diamonds. The location of the Circinus nucleus derived
from our measurements is shown by the large red box.\label{fig:hcn-diagram}}
\end{minipage}
\end{figure*}
From $v_{\mathrm{rot}}\sim80\,\mathrm{km}\,\mathrm{s}^{-1}$ at $r\sim2\,\mathrm{pc}$
(the radius of the innermost ring at $100\,\mathrm{mas}$) we obtain
a dynamical mass of $\sim3\times10^{6}\,M_{\odot}$, which is less
than a factor two higher than the mass of the black hole, $(1.7\pm0.3)\times10^{6}\,M_{\odot}$
determined from the maser disk \citep{Greenhill2003a}. This means,
we just fall short of starting to directly resolve the region kinematically
dominated by the supermassive black hole. In fact, $\mathrm{HCO}^{+}$(4$-$3)
and to some degree also CO(6$-$5) show significantly higher velocities
at the nucleus, up to $130\,\mathrm{km}\,\mathrm{s}^{-1}$, especially
on the blue-shifted side. We overplotted the expected line-of-sight
velocities due to Keplerian rotation around the black hole in the
position velocity diagram of $\mathrm{HCO}^{+}$(4$-$3), right panel
of Fig.~\ref{fig:3dbarfit-pvdiag}, and it seems that the innermost
velocities at the centre are influenced by the black hole potential.
Higher resolution observations will be necessary to probe this region
in more detail. 

\subsection{Excitation of the gas}

\subsubsection{The hole in the CO(3$-$2) emission}

The apparent deficit in CO(3$-$2) emission at the location of the
nucleus was already noticed by \citet{Izumi2018} and attributed to
the destruction of CO in the X-ray dominated region (XDR) surrounding
the nucleus. However, because we clearly detect CO(6$-$5) at the
nucleus, this hypothesis can be ruled out to a certain degree. Only
for the highest velocities at the nuclear position, which start following
the Keplerian field of the black hole and are only observed in $\mathrm{HCO}^{+}$(4$-$3),
there might be indications of destruction of CO. 

We rather suggest that most of the CO molecules have been excited
to higher transitions and that CO(3$-$2) is partially self-absorbed
or absorbed against the background continuum (see Sect.~\ref{subsec:self-absorption}).
The former is supported by the increase of the CO excitation observed
already on larger scales in the Circinus galaxy \citep{Zhang2014},
where the high-$J$ CO ($J\gtrsim\mathrm{\textrm{4-3}}$) is likely
dominated by the emission from the nucleus. Similar CO excitation
has also been found in \object{NGC~1068} from Herschel observations
\citep{Spinoglio2012}. Further high resolution observations of several
CO transitions will be required to obtain a spectral line energy distribution
and confirm this scenario.

In addition to the excitation, also absorption most likely leads to
a stronger suppression of the CO(3$-$2) emission at the nucleus than
for CO(6$-$5). This is especially the case for close to systemic
velocities, because we expect most of the CO at larger distances from
the nucleus, but on the line of sight towards the nucleus, to be in
a lower excited state, making CO(3$-$2) more absorbed than CO(6$-$5).

\subsubsection{Emission line ratios}

In order to get a first idea of the chemical and physical state of
the gas, we show in Fig.~\ref{fig:emission-line-ratios} the molecular
line ratio maps for CO(6$-$5) to CO(3$-$2), $\mathrm{HCO}^{+}$(4$-$3)
to CO(3$-$2) and HCN(4$-$3) to $\mathrm{HCO}^{+}$(4$-$3). For
the CO(6$-$5) to CO(3$-$2) ratio map, the higher resolution CO(6$-$5)
data from band 9 were smoothed and regridded to match the properties
of the CO(3$-$2) data in band 7, before calculating the ratio map.
For the HCN(4$-$3) to $\mathrm{HCO}^{+}$(4$-$3) ratio map, the
$\mathrm{HCO}^{+}$(4$-$3) line was truncated to velocities $<15\,\mathrm{km}\,\mathrm{s}^{-1}$
in order to match the velocity coverage of the HCN(4$-$3) line; the
ratio map therefore only covers the north-eastern part of the nucleus.

The line ratio of CO(6$-$5) to CO(3$-$2) is typically $<8$ in the
extended emission; only in an elongated, disk-like region within $\sim10\,\mathrm{pc}$
of the nucleus, the ratio clearly increases to $>8$. This is much
higher than the expected ratio under the local thermal equilibrium
(LTE) and the Rayleigh-Jeans limit, which is 4 from the $\left(j_{6}/j_{3}\right)^{2}$
scaling \citep[e.g. ][]{Narayanan2014}, indicating a deviation from
LTE conditions. Despite that the CO(6$-$5) emission peaks at the
nucleus, the highest ratio is, however, not directly found at the
nucleus itself, but slightly to the north-west with a secondary peak
to the north-east. This is mainly caused by the `finger' of emission
in CO(3$-$2) at the location of the nucleus (see Sect.~\ref{subsec:CO-line-emission}).
Some of the smaller-scale structures in the ratio map may, however,
also be influenced by the low signal to noise of the CO(6$-$5) data.
Nevertheless we find that CO(6$-$5) is clearly enhanced with respect
to CO(3$-$2) in the inner $\sim10\,\mathrm{pc}$ region. Taken at
face value, this trend indicates an increase in excitation and hence
in gas temperature towards the nucleus, as has been discussed in the
previous section. It is also possible that the increase in the CO(6$-$5)
line is simply due to an increase of gas density. Also for \object{NGC~1068}
\citep{Viti2014} and \object{Centaurus~A} \citep{Espada2017} an
increase by a factor of a few of the CO(6$-$5)/CO(3$-$2) line ratio
has been observed towards the nucleus on scales of tens of parsec.
For \object{NGC~1068}, the ratio displays a clear peak at the nuclear
position, while for \object{Centaurus~A} the situation is less clear
due to the low resolution and low signal to noise of the data.

In contrast, the $\mathrm{HCO}^{+}$(4$-$3) to CO(3$-$2) ratio clearly
shows a very strong enhancement towards the nucleus: While the ratio
is $<0.4$ outside of the central beam, it rises to values of $\sim2$
at the centre. This relative enhancement towards the nucleus hence
confirms that this molecule is a better tracer for very dense molecular
material than CO. We note that the $\mathrm{HCO}^{+}$(4$-$3) to
CO(3$-$2) ratio is between 0.01 and 0.11 in the nuclear region of
\object{NGC~1068} \citep{Viti2014}, and that no enhancement of the
ratio is observed towards the very nucleus in that galaxy. For \object{NGC~1068},
the highest value of this ratio is actually found in the so-called
east knot, $\sim60\,\mathrm{pc}$ ($0.9\,\mathrm{arcsec}$) to the
east of the location of the supermassive black hole, where also the
CO(3$-$2) emission peaks. 

Finally we find a HCN(4$-$3)/$\mathrm{HCO}^{+}$(4$-$3) ratio of
around 1, with values ranging between 0.5 and 1.5 in a patchy structure,
possibly due to the low signal to noise of the HCN line. We hence
do not see any clear enhancement of the HCN/$\textrm{HCO}^{+}$ ratio
in the nucleus as has been reported for other galaxies hosting AGNs
\citep[e.g. ][]{Kohno2005,Krips2008,Imanishi2014,Imanishi2018b}.
Our value around unity is similar to that found in the starburst rings
of \object{NGC~1068} and \object{NGC~1097} at $\sim1\,\mathrm{kpc}$
distance from their nuclei \citep{Viti2014,Martin2015}, while the
HCN/$\textrm{HCO}^{+}$ ratio (in the $J=$ 4$-$3 and 1$-$0 transitions,
respectively) increases to values of $1.5$ at the nucleus and to
values between 2 and 3 in the circumnuclear disk. Also for \object{NGC~613}
and \object{NGC~1808}, the HCN(4$-$3) line is found to be two to
three times brighter than $\mathrm{HCO}^{+}$(4$-$3) in the nuclear
region, while at larger distances, $>50\,\mathrm{kpc}$, in surrounding
star-forming rings lower values of $\ll1$ are found for this line
ratio \citep{Audibert2019,Audibert2021}. 

To investigate this further, we show in Fig.~\ref{fig:hcn-diagram}
the so-called submillimetre HCN diagram, with the HCN(4$-$3) to CS(4$-$3)
ratio plotted as a function of the HCN(4$-$3) to $\textrm{HCO}^{+}$(4$-$3)
ratio following \citet{Izumi2016}. Enhanced HCN(4$-$3) to $\mathrm{HCO}^{+}$(4$-$3)
and/or HCN(4$-$3) to CS(7$-$6) intensity ratios have been proposed
as an extinction-free energy diagnostic tool for the nuclear regions
of galaxies \citep{Izumi2013,Izumi2016}. The HCN enhancement can
be explained by either a truly enhanced abundance of HCN compared
to $\mathrm{HCO}^{+}$ and CS in AGNs in contrast to starburst galaxies,
that is an astrochemical effect dominated by X-ray activities raised
by AGN \citep[e.g. ][]{Meijerink2007,Harada2013}, or by a systematically
enhanced gas densities in AGNs. For Circinus, we find $R_{\textrm{HCN(4–3)}/\textrm{HCO}^{+}\textrm{(4–3)}}=0.55\pm0.11$
and $R_{\textrm{HCN(4–3)}/\textrm{CS(4–3)}}=6.0_{-1.5}^{+4.1}$,
when integrating the fluxes in a $1\,\mathrm{arcsec}$ aperture centred
on the nucleus and limiting the integration of all lines to the restricted
velocity range of the HCN line. The nucleus of Circinus thus has a
comparatively low HCN(4$-$3) to $\mathrm{HCO}^{+}$(4$-$3) ratio,
in agreement with that found in starburst regions. Only the HCN(4$-$3)
and CS(4$-$3) ratio is slightly higher than for starburst regions.
The reason for this lack in AGN-driven enhancement may be due to a
relatively weak AGN radiation field in contrast to the starburst intensity.
Another explanation may be that the X-rays from the AGN are obscured
by the molecular torus, due to the particular edge-on point of view,
assuming an exceptionally high optical depth in the HCN line in the
mid-plane of the torus.

In conclusion, we find some indications for CO excited to higher transitions
as well as increased emission from dense gas tracers, especially $\mathrm{HCO}^{+}$(4$-$3)
in the nucleus of the Circinus galaxy in line with the AGN activity.
However, Circinus does not show the HCN enhancement found in several
other AGNs. More detailed studies of the molecular composition and
excitation using multiple molecules and transitions will be the subject
of a future publication.

\section{Conclusions\label{sec:conclusions}}

Here we present ALMA band 7 ($345\,\mathrm{GHz}$) and band 9 ($690\,\mathrm{GHz}$)
continuum as well as molecular emission line observations of the nearby
Seyfert 2 nucleus in the Circinus galaxy, with unprecedented spatial
resolutions of ${\sim}190\,\mathrm{mas}$ (${\sim}3.8\,\mathrm{pc}$)
and ${\sim}110\,\mathrm{mas}$ (${\sim}2.2\,\mathrm{pc}$), respectively.
Our results can be summarised as follows:
\begin{itemize}
\item The continuum emission shows a central peak plus an S-shaped spiral
structure out to a radius of $40\,\mathrm{pc}$. The central peak
is essentially unresolved at $345\,\mathrm{GHz}$, with a size of
$<2\,\mathrm{pc}$; at $691\,\mathrm{GHz}$ the central peak is slightly
resolved with a size of $3.2\,\mathrm{pc}\times2.2\,\mathrm{pc}$. 
\item The extended emission at $r>6\,\mathrm{pc}$ from the nucleus has
spectral indices of $3.3<\alpha<4.3$, which implies that this emission
is the Rayleigh-Jeans tail of typical interstellar dust. The nucleus
on the other hand has a much lower spectral index of $\alpha\sim2$.
From an analysis of the infrared to radio SED of the Circinus nucleus,
we infer that the nuclear continuum emission at $691\,\mathrm{GHz}$
is dominated to $\gtrsim95\%$ by dust emission, while at $345\,\mathrm{GHz}$
there is a significant contribution most likely from free-free, or
from synchrotron emission. This additional contribution explains both
the lower spectral index as well as the more compact emission at $345\,\mathrm{GHz}$.
\item No similarity between the morphology of the dust emission at submillimetre
wavelengths and at mid-infrared wavelengths is observed. This implies
that at these two wavelength ranges we are probing two completely
different dust components: a relatively warm dust component with $T_{\mathrm{dust}}\sim300\,\mathrm{K}$
in polar direction for which the SED drops sharply towards the submillimetre
regime; as well as a cold dust component with $T_{\mathrm{dust}}\lesssim100\,\mathrm{K}$
in an equatorial plane, roughly following the distribution of the
molecular gas.
\item We detect four emission lines in the band 7 data, CO(3$-$2), $\mathrm{HCO}^{+}$(4$-$3),
HCN(4$-$3) and CS(4$-$3), as well as one emission line, CO(6$-$5),
in the band 9 data.
\item All emission lines show a similar morphology as the continuum emission,
that is extended emission in an S-shaped structure and a central emission
peak. The CO(3$-$2) transition is an exception: it does not show
a peak at the position of the AGN, but rather a hole (deficit) up
to $r\sim5\,\mathrm{pc}$ from the nucleus. This deficit can be explained
by CO being excited to higher transitions as well as CO(3$-$2) being
partially self-absorbed or absorbed against the background continuum.
We argue that no special XDR chemistry is required to explain the
lack of CO(3$-$2) at the nucleus.
\item The kinematics of the molecular gas is dominated by rotation, with
some perturbation by the spiral arms and by filaments of less dense
material above or below the disk with distinct velocity components.
While the Circinus galaxy hosts a well known ionised outflow in polar
direction, we only find very weak signs for a molecular outflow in
our ALMA data.
\item The velocity dispersion is much lower than the rotational velocity,
which implies that the molecular gas is located a relatively thin
disk with $h/r<0.3$. This is much too low to collimate the ionising
radiation to a cone with an opening angle of $\sim90^{\circ}$. The
collimation must hence occur already on scales of $<5\,\mathrm{pc}$,
that is the structure playing the role of the `obscuring torus'
of the unified scheme of AGNs is compact and yet unresolved by the
ALMA observations.
\item While we do find indications for CO to be excited to higher transitions
and for increased emission from dense gas tracers, especially $\mathrm{HCO}^{+}$(4$-$3),
at the position of the nucleus, the Circinus galaxy does not show
the HCN enhancement found in several other AGNs. The Circinus nucleus
is rather located in between the loci of starbursts and AGNs in the
submillimetre HCN diagram, somewhat questioning the robustness for
this diagram to identify AGNs.
\end{itemize}
Our findings support the most recent radiative transfer calculations
of the obscuring structures in AGNs, which find a similar two-component
structure. Further observations with even higher angular resolution
are, however, required to resolve the region responsible for the collimation
of the ionisation cone and the material kinematically dominated by
the supermassive black hole. 
\begin{acknowledgements}
We thank the anonymous referee for his valuable suggestions to improve
the manuscript, which were provided in a record time. We are further
grateful to Adele Plunkett for sharing her python routines, to Marko
Stalevski for providing the SEDs of his Circinus torus models. Z.-Y.
Zh. acknowledges the support of NSFC (grants 12041305, 12173016),
the Program for Innovative Talents, Entrepreneur in Jiangsu, China,
and the science research grants from the China Manned Space Project
with NO.CMS-CSST-2021-A08.

This paper makes use of the following ALMA data: ADS/JAO.ALMA\#2012.1.00479.S.
ALMA is a partnership of ESO (representing its member states), NSF
(USA) and NINS (Japan), together with NRC (Canada), MOST and ASIAA
(Taiwan), and KASI (Republic of Korea), in cooperation with the Republic
of Chile. The Joint ALMA Observatory is operated by ESO, AUI/NRAO
and NAOJ.
\end{acknowledgements}

\bibliographystyle{aamod}
\bibliography{circalma}

\end{document}